\documentclass[useAMS,usenatbib]{mnras}

\usepackage{graphicx}
\usepackage[T1]{fontenc}
\usepackage{aecompl}
\usepackage{amsfonts}
\usepackage[]{amsmath,amssymb}
\usepackage{color}
\usepackage{xspace}
\usepackage{textcomp}
\usepackage[flushleft]{threeparttable}
\usepackage[dvipsnames]{xcolor}

\newcommand{\msun}{M_\odot}
\graphicspath{{figures/}}

\newcommand{\mytab}{\begin{table}[htb]}
\newcommand{\myfig}{\begin{figure}[htbp]}

\newcommand{\ud}{\mathrm{d}}

\def\lsim{ \lower .75ex \hbox{$\sim$} \llap{\raise .27ex \hbox{$<$}} }
\graphicspath{{figures/}}

\title[]{A comparative study of satellite galaxies in Milky Way-like galaxies from HSC, DECaLS and SDSS
}

\author[Wang et al.]{Wenting~Wang$^{1,2}$\thanks{wenting.wang@sjtu.edu.cn}, Masahiro~Takada$^{3}$, Xiangchong~Li$^{3}$, Scott G. Carlsten$^{4}$,\newauthor
Ting-Wen~Lan$^{5,3}$, Jingjing~Shi$^{3}$, Hironao~Miyatake$^{6,7}$, Surhud~More$^{8}$, \newauthor
Rachael L. Beaton$^{4}$, Robert~Lupton$^{4}$, Yen-Ting Lin$^{9}$, Tian~Qiu$^{3}$, Wentao~Luo$^{3}$ \\
{}$^{1}$ Department of Astronomy, Shanghai Jiao Tong University, Shanghai 200240, China \\
{}$^{2}$ Shanghai Key Laboratory for Particle Physics and Cosmology, Shanghai 200240, China\\
{}$^{3}$ Kavli IPMU (WPI), UTIAS, The University of Tokyo, Kashiwa, Chiba 277-8583, Japan\\
{}$^{4}$ Department of Astrophysical Sciences, Princeton University, 4 Ivy Lane, Princeton, NJ08544, USA\\
{}$^{5}$ Department of Astronomy and Astrophysics, UCO/Lick Observatory, University of California, 1156 High Street, Santa Cruz,\\
         \hspace{2.5mm}CA 95064, USA\\
{}$^{6}$ Institute for Advanced Research, Nagoya University, Furo-cho, Nagoya 464-8601, Japan\\
{}$^{7}$ Division of Particle and Astrophysical Science, Graduate School of Science,Nagoya University, Furo-cho, Nagoya 464-8602, Japan\\
{}$^{8}$ The Inter-University Centre for Astronomy and Astrophysics, Post bag 4, Ganeshkhind, Pune 411007, India \\
{}$^{9}$ Institute  of Astronomy  and  Astrophysics,  Academia  Sinica,  Taipei 10617, Taiwan\\
}

\begin{document}

\maketitle

\begin{abstract}
We conduct a comprehensive and statistical study of the luminosity functions (LFs) for satellite galaxies,
by counting photometric galaxies from HSC, DECaLS and SDSS around isolated central galaxies (ICGs) and paired 
galaxies from the SDSS/DR7 spectroscopic sample. Results of different surveys show very good agreement. 
The satellite LFs can be measured down to $M_V\sim-10$, and for central primary galaxies as small as $8.5<
\log_{10}M_\ast/\msun<9.2$ and $9.2<\log_{10}M_\ast/\msun<9.9$, which implies there are on average 3--8
satellites with $M_V<-10$ around LMC-mass ICGs. The bright end cutoff of satellite LFs and the satellite 
abundance are both sensitive to the magnitude gap between the primary and its companions, indicating 
galaxy systems with larger magnitude gaps are on average hosted by less massive dark matter haloes. By 
selecting primaries with stellar mass similar to our MW, we discovered that i) the averaged satellite 
LFs of ICGs with different magnitude gaps to their companions and of galaxy pairs with different colour 
or colour combinations all show steeper slopes than the MW satellite LF; ii) there are on average
more satellites with $-15<M_V<-10$ than those in our MW; iii) there are on average 1.5 to 2.5 satellites 
with $M_V<-16$ around ICGs, consistent with our MW; iv) even after accounting for the large scatter
predicted by numerical simulations, the MW satellite LF is uncommon at $M_V>-12$. Hence the MW and its 
satellite system are statistically atypical of our sample of MW-mass systems. In consequence, our MW is 
not a good representative of other MW-mass galaxies. Strong cosmological implications based on only MW 
satellites await additional discoveries of fainter satellites in extra-galactic systems. Interestingly, 
the MW satellite LF is typical among other MW-mass systems within 40~Mpc in the local Universe, perhaps 
implying the Local Volume is an under-dense region.
\end{abstract}   


\section{Introduction}
\label{sec:intro}

In the structure formation paradigm of the $\Lambda$ cold dark matter ($\Lambda$CDM) scenario, galaxies 
form in the centres of an evolving population of dark matter haloes \citep{1978MNRAS.183..341W}. Dark 
matter haloes grow in mass and size through both mergers with other haloes and the smooth accretion of 
diffuse matter \citep[e.g.][]{2011MNRAS.413.1373W}. Smaller haloes and their own central galaxies fall 
into larger haloes and become so-called ``subhaloes'' and ``satellites'' of the galaxy in the centre 
of the host halo. 

Compared with the extra-galactic satellites around more distant central galaxies, the satellite galaxies 
in our Milky Way (hereafter MW) can be observed in great details. We are not only able to observe the MW 
satellites down to much fainter magnitudes, but also it is possible to measure their full 3-dimensional 
velocities as well as internal dynamics through observations of bright individual member stars. The 
MW system with its associated satellite galaxies thus offer an ideal environment to closely study 
satellite-properties, which, in turn, helps to probe the small mass end of galaxy formation, constrain the 
underlying dark matter distribution, and test the standard cosmological model on small scales. 

With the observed abundance and properties of the MW satellites, a few so-called challenges to the standard 
cosmological model have been raised and are still under debates. The missing satellite problem \citep[e.g.][]{
1999ApJ...522...82K,1999ApJ...524L..19M} is one example. Twenty years ago, it was pointed out that the 
observed number of satellite galaxies in the MW is significantly lower than the predicted number of 
dark matter subhaloes by numerical simulations. In the last decade, more faint satellite galaxies have been 
discovered in the Local Group \citep[e.g.][]{2007ApJ...656L..13I,2008A&A...477..139L,2008ApJ...684.1075M,
2007ApJ...659L..21Z,2009MNRAS.398.1757W,2010ApJ...712L.103B}, but the number of observed satellites is still 
much smaller than the predicted number of subhaloes in cold dark matter simulations. To explain the problem, 
some studies invoke the model of warm dark matter, which predicts much less surviving small substructures 
\citep[e.g.][]{2014MNRAS.439..300L}. On the other hand, this problem can be possibly explained by galaxy 
formation physics \cite[e.g.][]{2000ApJ...539..517B,2002MNRAS.333..156B}, such as photoionization and supernova 
feedback, which inhibit the star formation in small haloes, \citep[e.g.][]{2000ApJ...539..517B,2002MNRAS.333..156B,
2002MNRAS.333..177B,2002ApJ...572L..23S}, predicting that a significant number of small subhaloes do not 
host a galaxy. Moreover, it was estimated that there could be at least a factor of three to five times 
more faint satellites in the MW to be discovered \citep[e.g.][]{2008ApJ...686..279K,2008ApJ...688..277T,
2009AJ....137..450W,2014ApJ...795L..13H}. More recently, with deep imaging surveys, more satellite 
galaxies or candidates in our MW and M31 are being discovered and reported \citep[e.g.][]{2015ApJ...807...50B,
2015ApJ...813..109D,2015ApJ...804L..44K,2015ApJ...805..130K,2018PASJ...70S..18H,2019PASJ...71...94H}. 

Another problem, known as the ``too big to fail'' problem,  was raised by \cite{2011MNRAS.415L..40B}. Our MW 
has nine classical dwarf spheroidal satellites that have maximum circular velocities, $V_\mathrm{max}$, smaller 
than 30~km/s. Only the Large and Small Magellanic Clouds (LMC and SMC) have $V_\mathrm{max}$ greater than 60~km/s, 
while the Sagittarius dwarf in our MW has $V_\mathrm{max}$ likely in between 30 and 60~km/s before infall. 
However, the most massive dark matter subhaloes predicted by $\Lambda$CDM simulations of MW-like systems are 
found to have $V_\mathrm{max}$ larger than those of MW satellites. It is thus hard to explain why the most 
massive subhaloes predicted by cold dark matter simulations do not match the properties of the most massive 
observed satellites. Proper mechanisms are required to explain how the central density of the most massive 
subhaloes in simulations can be reduced in order to match those of observed satellites, and such mechanisms 
include supernova feedback and subhalo disruption by the massive host halo \citep[e.g.][]{2019MNRAS.487.1380G}. 
Alternatively, the problem can be explained if the virial mass of our MW is smaller than that has been assumed. 
For example, \cite{2012MNRAS.424.2715W} and \cite{2014MNRAS.445.1820C} reported that the MW satellites are 
consistent with $\Lambda$CDM predictions at the 10\% confidence level if the MW halo has a virial mass smaller 
than $1.3\times10^{12}\msun$.

A lighter-mass MW-like halo, however, would have difficulties predicting the existence of the two massive 
satellite galaxies, LMC and SMC. They are very likely accreted by our MW, not as two individual satellites, 
but rather as a pair given their similarity in phase space \citep{2006ApJ...652.1213K}, in good consistency 
with simulation results \citep[e.g.][]{2008ApJ...686L..61D}. For galaxies with LMC stellar mass, the typical host 
halo mass is $\sim2\times 10^{11}\msun$ before being accreted by a more massive host halo and becoming a 
satellite. The host halo mass of the SMC is approximately a factor of 2 to 3 smaller than that of the LMC. More 
recent studies suggest that the LMC could be as massive as $\sim3\times 10^{11}\msun$ at infall \cite[e.g.][]{2019MNRAS.483.2185C}.
Looking for subhaloes that are similar to the mass, Galactocentric distance, and the orbital velocity of 
the LMC in numerical simulations, \cite{2011MNRAS.414.1560B} concluded that the fraction of dark matter haloes 
smaller than $M_{200}\sim1.25\times10^{12}\msun$ hosting LMC-like subhaloes is very low ($<5\%$), and over 90\% 
of haloes hosting LMC-like subhaloes have $M_{200}>1.7\times10^{12}\msun$. In addition, although subhaloes 
with LMC or SMC-like masses are commonly found in MW-like haloes, it is rare to find MW-like haloes to host 
both LMC and SMC-like subhaloes. Only $\sim$2.5\% of lighter-mass MW-like haloes with LMC analogous also 
have LMC-SMC pairs. Similar conclusions have been reached by, e.g., \cite{2011ApJ...743..117B,2013ApJ...770...96G} 
and \cite{2017MNRAS.464.3825P}. In addition, \cite{2011ApJ...733...62L} looked at MW-MC-like systems in SDSS, 
and they claimed that galaxies with luminosity similar to the MW have only 3.5\% probability of hosting both 
LMC and SMC-like satellites within a projected distance of 150~kpc.

The $V_\mathrm{max}$ distribution of classical satellites in our MW is also atypical. \cite{2014MNRAS.445.2049C}
investigated the $V_\mathrm{max}$ distribution of massive subhaloes in MW-like hosts, by looking 
for haloes hosting at most three subhaloes with $V_\mathrm{max}\ge 30$km/s and at least two subhaloes with 
$V_\mathrm{max}\ge 60$km/s. They found that such cases are rare in $\Lambda$CDM simulations, with 
at most 1\% of haloes of any mass having a similar distribution.

Thus, MW-like systems can be predicted and do exist in $\Lambda$CDM simulations, but some of the properties of 
the MW-system are statistically uncommon compared with simulated systems of similar central stellar mass or 
luminosity. However, we can not robustly rule out or claim challenges to the standard cosmology model with 
just the single case of our MW without determining statistical constraints on the properties of the satellite 
system. It is thus important to look at other extra-galactic satellites around MW-like galaxies, and investigate 
how differently the satellites in our MW are compared to the satellites of other galaxies. Such a comparative 
study can help to assess the statistical significance of previous cosmological implications based on MW satellites. 

The MW satellite luminosity function can be measured down to $M_V\sim 0$ \citep[e.g.][]{2018MNRAS.479.2853N}. 
The satellites of more distant extra-galactic galaxies, however, are more difficult to study to such faint 
magnitudes. In the very local universe (within $\sim$ 10 to 40~Mpc), efforts have been made to the detection 
and confirmation of faint satellites in other host galaxies. For example, \cite{2018ApJ...865..125T} investigated 
a few nearby MW-mass galaxies within $\sim$20~Mpc. With statistical background subtraction, their satellite 
luminosity functions can be measured down to $M_V\sim-9$, which show a large diversity. \cite{2020arXiv200602443C} 
measured the surface brightness fluctuation distances \citep[e.g.][]{2009ApJ...694..556B,2018ApJ...856..126C,
2019ApJ...879...13C} for dwarf satellites of 10 massive galaxies within 12~Mpc, and they reported that the MW 
satellite luminosity function is remarkably typical. Moreover, the SAGA \citep[Satellites Around Galactic Analogs;][]{
2017ApJ...847....4G,2020arXiv200812783M} survey measures the distribution of satellites around $\sim$100 MW-like 
systems within 40~Mpc, and the number of satellites per host shows large diversities. The MW satellite luminosity 
function is typical among these systems, though the satellites of MW and M31 were found to be redder than those 
in other MW-like systems within 40~Mpc.

In order to measure the intrinsic luminosities and distances of satellite galaxies beyond 40~Mpc and still at 
low redshifts, usually spectroscopic observations are required\footnote{Photometric redshifts suffer from 
large relative errors at low redshifts.}. However, spectroscopic surveys have bright flux limits, and thus 
studies based on pure spectroscopic data are often limited to a few brightest satellites. For example, 
the SAGA survey can reach $M_V\sim-12$ at the distance of 40~Mpc. At the redshift of $z\sim0.1$, the 
distance is about 400~Mpc, corresponding to $\sim$5 magnitudes brighter than the limit at 40~Mpc. 

Instead of looking at only spectroscopic companions, efforts have been devoted to measure the radial 
profiles and luminosity functions of photometric satellites around spectroscopic host galaxies at $z\sim 0.1$ 
\citep[e.g.][]{2011AJ....142...13L,2011ApJ...734...88W,2011MNRAS.417..370G,2012MNRAS.424.2574W,2012ApJ...760...16J,
2014MNRAS.442.1363W,2015MNRAS.449.2576C,2016MNRAS.459.3998L,2019arXiv191104507T} and also to intermediate 
and high redshifts \citep[e.g.][]{2013ApJ...772..146N,2014ApJ...792..103K}, based on statistical background 
subtraction approaches to remove the contamination by foreground and background galaxies. The significantly 
fainter flux limits of photometric surveys can help to study satellites that are much fainter than those 
accessible by spectroscopic observations. Using the photometric catalogues of the Sloan Digital Sky Survey 
(SDSS), the satellite luminosity function of galaxies at $z\sim0.1$ can be measured down to $M_V\sim -14$, 
which covers the bright end\footnote{As we will show in Section~\ref{sec:MWsat}, $M_V\sim -14$ is about 4 
and 3 magnitudes fainter than the LMC and SMC. All other MW satellites are fainter than $M_V\sim -14$.} 
of MW satellites \citep[e.g.][]{2011MNRAS.417..370G,2012MNRAS.424.2574W}. 

More recently, there are a few on-going and future deep imaging surveys such as the DESI Legacy Imaging 
Survey \citep{2019AJ....157..168D}, the Hyper Suprime-Cam Subaru Strategic Program \citep[HSC;][]{2018PASJ...70S...4A} 
and the future Vera Rubin Observatory Legacy Survey of Space and Time \citep[LSST;][]{2008arXiv0805.2366I}. These 
surveys are promising to extend previous studies based on SDSS down to fainter magnitudes, though follow-up 
spectroscopic surveys are still required to identify central galaxies in their footprints. Fortunately, 
the on-going HSC survey and the DESI Legacy Imaging Survey \citep{2019AJ....157..168D} already have part of 
their footprints overlapping with the SDSS spectroscopic galaxies. In this study, we select central primary 
galaxies similar to the MW from the SDSS spectroscopic Main galaxy sample and investigate their satellite 
luminosity function (hereafter LF) using photometric sources from HSC and DECaLS\footnote{One of the three 
parts of DESI Legacy Imaging Survey.} (the Dark Energy Camera Legacy Survey). We also repeat the same
analysis with SDSS imaging data, in order to cross check the consistency of our measurements in 
different surveys. We will show whether the LF of satellites in our MW is typical compared with the 
averaged satellite LF around extra-galactic central galaxies with similar properties.

The structure of this paper is organised as follows. We introduce how we select central primary galaxies 
and the photometric source catalogues of HSC, DECaLS and SDSS used to construct satellite galaxies in 
Section~\ref{sec:data}. Our methodology of satellite counting, background subtraction and the projection 
effect are described in Section~\ref{sec:method}. Results are presented in Section~\ref{sec:results}, for 
the measured satellite LFs centred on primary galaxies selected in many different ways. We discuss and 
conclude in the end (Sections~\ref{sec:disc} and \ref{sec:concl}).

For observational results, we adopt the first-year Planck cosmology \citep{2014A&A...571A..16P} as 
our fiducial cosmological model, with the values of the Hubble constant $H_0=67.3\ \mathrm{km~s^{-1}/Mpc}$, 
the matter density $\Omega_m=0.315$ and the cosmological constant $\Omega_\Lambda=0.685$.

\section{data}
\label{sec:data}

In order to investigate the extra-galactic satellite LFs, we need to first select samples of 
primary galaxies that sit in the centre of dark matter haloes. We make the selection in a few 
different ways. In brevity, we select isolated central galaxies that are brighter than or brighter 
by at least 1 magnitude than all companions. Then, we select galaxy pairs similar to our Local 
Group system. These primary galaxies are selected from the SDSS spectroscopic Main galaxy sample 
\citep{{2005AJ....129.2562B}}, and the satellite counts are made from the photometric galaxies 
in either the HSC, DECaLS and SDSS footprints. In the following, we introduce how we select central 
primaries and the photometric source catalogues of different surveys used to construct satellites. 

\subsection{Isolated central galaxies}
\label{sec:isogal}

To identify a sample of primary galaxies that are highly likely sitting in the centre of dark 
matter haloes (purity), we select the brightest galaxies within given projected and line-of-sight 
distances. The parent sample used for the selection is the NYU Value Added Galaxy Catalogue 
\citep[NYU-VAGC;][]{2005AJ....129.2562B}, which is based on the spectroscopic Main galaxy sample 
from the seventh data release of the Sloan Digital Sky Survey \citep[SDSS/DR7;][]{2009ApJS..182..543A}. 
The sample includes galaxies in the redshift range between $z=0.001$ and $z\sim0.4$, which is flux 
limited down to an apparent magnitude of $\sim$17.7 in SDSS $r$-band, with most of the objects
below redshift $z=0.25$. Stellar masses in VAGC were estimated from the K-corrected galaxy 
colours by fitting the stellar population synthesis model \citep{2007AJ....133..734B} assuming 
a \cite{2003PASP..115..763C} initial mass function.

We adopt two different sets of isolation criteria: i) Galaxies that are the brightest within the projected 
virial radius, $R_{200}$, of their host dark matter haloes\footnote{$R_{200}$ is defined to be the 
radius within which the average matter density is 200 times the mean critical density of the universe.
The virial radius and velocity are derived through the abundance matching formula between stellar 
mass and halo mass \citep{2010MNRAS.404.1111G}, and based on mock catalogues it was demonstrated 
that the choice of three times virial velocity along the line of sight is a safe criterion that 
identifies all true companion galaxies.} and within three times the virial velocity along the 
line of sight. ii) Galaxies that are at least one magnitude brighter than all companions projected 
within the virial radius and within three times the virial velocity along the line of sight. In 
addition, galaxies selected in i) and ii) should not be within the projected virial radius (also 
three times virial velocity along the line of sight) of another brighter galaxy.

The SDSS spectroscopic sample suffers from the fiber-fiber collision effect that two fibers cannot 
be placed closer than 55$\arcsec$. As a result, galaxies in dense regions, such as those within 
galaxy groups and clusters, are spectroscopically incomplete. Moreover, for criterion ii), 
the companions used for the selection of isolated central galaxies fainter than $r=16.7$ can be 
fainter than the flux limit of the SDSS spectroscopic sample ($r=17.7$), and thus do not have 
spectroscopic redshift measurements\footnote{\cite{2012MNRAS.424.2574W} adopted a flux cut of $r<16.7$ 
for isolated primaries. In this study, we keep the flux limit of $r=17.7$ to maximise our sample 
size, while we use photometric redshift (photoz) probability distribution of photometric companions 
to compensate the selection.}. Hence to avoid the case when a galaxy has a brighter companion but 
this companion does not have available redshift and is hence not included in the SDSS spectroscopic 
sample, we use the SDSS photometric catalogue to make further selections. The photometric catalogue 
is the value-added Photoz2 catalogue \citep{2009MNRAS.396.2379C} based on SDSS/DR7, which provides 
photometric redshift probability distributions of SDSS galaxies. We further discard galaxies that 
have a photometric companion satisfying the magnitude requirement, whose redshift information is 
not available but is within the projected separation of the above selection criterion, and the 
photoz probability distribution of the photometric companion gives a larger than 10\% of probability 
that it shares the same redshift as the central galaxy.

By adopting the flux limit of $r=17.7$ for primaries might induce a luminosity bias in the analysis,
because when the selection is made near the faint limit of the spectroscopic sample, the faintest 
primaries will only have photometrically-identified companions. To ensure that our selection does not 
include any additional bias, we have explicitly repeated our analysis in this paper by adopting a flux 
limit of $r<16.7$ for primaries, so that all companions used for the selection of isolated central 
galaxy are above the flux limit of SDSS spectroscopic observations. Except for one of the conclusions 
which we will discuss in Section~\ref{sec:sigblueblue}, all the other conclusions of this paper is 
robust against changes in the flux limit of central primaries. 

Throughout the paper, primaries selected by adopting criteria i) and ii) are referred as ICG1 and ICG2 
correspondingly. ICG1s are only mildly isolated. As tested against a mock galaxy catalogue based on the 
semi-analytical model of \cite{2010MNRAS.404.1111G}, the completeness of ICG1s among all true halo 
central galaxies is above 90\%. The purity is above 85\%, which reaches $>$90\% at $\log_{10}M_\ast/\msun
>11.5$. The readers can check more details in Figure~2 of \cite{2019MNRAS.487.1580W}. ICG2s are more 
strongly isolated, with a larger magnitude gap (at least one magnitude) to the companions. Thus, the 
completeness of ICG2s is lowered to between 60\% and 70\% at $\log_{10}M_\ast/\msun>11.2$, between 
70\% and 80\% at $10.8<\log_{10}M_\ast/\msun<11.2$, between 80\% and 90\% at $10<\log_{10}M_\ast/\msun<10.8$ 
and is still above 90\% at the small mass end. The purity fraction slightly increases to above 87\%, 
which reaches $>$95\% at $\log_{10}M_\ast/\msun>11.5$. The purity and completeness fractions predicted 
by the Illustris TNG-100 simulation \citep{2019ComAC...6....2N} are similar though noisier. The 
comparison between results based on ICG1s and ICG2s will help to determine whether the magnitude 
gap between the central primary and the companions can affect the satellite LF.

In this study, we will calculate the LF for satellites projected within the halo virial radius 
of ICGs. Instead of using the abundance matching formula to estimate $R_{200}$ for each individual 
ICG in a given stellar mass bin, we adopt the mean $R_{200}$ based on ICG1s selected from the mock 
galaxy catalogue of \cite{2011MNRAS.413..101G}. Note the $R_{200}$ of ICG2s and primaries in pairs 
(see the next subsection) can be a bit different, but to ensure fair comparisons, we adopt the 
same $R_{200}$ for central primaries selected in different ways. The values of $R_{200}$ in these 
different bins are provided in Table~\ref{tbl:r200}. In addition, we will also choose a stellar 
mass bin of $10.63<\log_{10}M_\ast/\msun<10.93$ for primaries with stellar mass similar to our MW 
\citep{2015ApJ...806...96L}, and for this particular bin, we fix its virial radius to be 260~kpc. 
The choice of 260~kpc is motivated by the distance of Leo I (see Table~\ref{tbl:MWsat}), and the 
virial radius of MW-mass ICG1s in the mock galaxy catalogue of \cite{2011MNRAS.413..101G} is close 
to 260~kpc. Table~\ref{tbl:r200} also provides the total number of ICG1s and ICG2s in each bin. 
Note the numbers of central primary galaxies in HSC are more than 10 times smaller, while the 
numbers of primaries in SDSS are slightly more than 50\% the numbers in DECaLS due to better 
overlap, as primaries are selected from the SDSS spectroscopic Main galaxy sample. We avoid 
repeatedly showing the numbers of primaries in the footprints of all the three surveys.

%%%%%%%%%%%%%%%%%%%%%%%%%%%%%%%%%%%%%%%%%%%%%%%%%%%
\begin{table}
\caption{The average halo virial radii ($R_{200}$) for primary galaxies grouped by stellar mass, 
which are estimated using ICG1s selected from a mock galaxy catalogue of Guo et al. (2011).
In each stellar mass bin, the numbers of ICG1s and ICG2s in the DECaLS footprint are provided. 
}
\begin{center}
\begin{tabular}{lccc}\hline\hline
$\log M_\ast/\msun$ & \multicolumn{1}{c}{$R_\mathrm{200}$ [kpc]} & $N$ of ICG1 & $N$ of ICG2 \\ \hline
11.4-11.7 & 758.65 & 10188 & 3785  \\
11.1-11.4  & 459.08 & 39621 & 19649  \\
10.8-11.1 & 288.16 & 73755 & 49509  \\
10.5-10.8 &  214.80 & 75366 & 59374 \\
10.2-10.5 &  173.18  & 50795 & 43066 \\
9.9-10.2 & 142.85  & 27873 & 24545 \\
9.2-9.9 &  114.64 &  26919  &  24523  \\
8.5-9.2 & 82.68 &  9787 &  9037  \\
10.63-10.93 & 260.00 & 79840 & 59537 \\
\hline
\label{tbl:r200}
\end{tabular}
\end{center}
\end{table}
%%%%%%%%%%%%%%%%%%%%%%%%%%%%%%%%%%%%%%%%%%%%%%%%%%%

\subsection{Galaxy pairs}
\label{sec:galaxy_pair}

Our MW has a massive companion, M31. The separation between the MW and M31 is $\sim$700 to 800~kpc. 
The virial radius of our MW is estimated to be in between 200 and 300~kpc in most previous
studies\citep[e.g.][]{2010ApJ...720L.108G,2015ApJ...806...54E}. We have introduced how the sample 
of ICGs are selected above. ICG1s have high completeness. They are more representative of the 
whole population of halo central galaxies and can contain galaxy pairs like our Local Group 
system. ICG1s can thus help to investigate the satellite LF around a general population of 
halo central galaxies. On the other hand, the more strictly selected ICG2s are at least one magnitude 
brighter than all companions within the virial radius, and thus the magnitude gap between ICG2s
and their satellites is more comparable to the condition of our MW. Galaxy pairs like the MW and 
M31 can be included in the sample of ICG2s as well. In addition to ICG1s and ICG2s, here we also 
select a sample of galaxy pairs analogous to our MW and M31, in order to investigate 
whether their satellites have different LFs than ICGs and whether the satellite LFs depend on the 
properties of the other massive companion galaxy in the pair. 

For galaxies with $10.2<\log_{10}M_\ast/\msun<11.1$ from the NYU Value Added Galaxy Catalogue, we identify 
pairs whose projected separations are in between twice the virial radius\footnote{Virial radius calculated 
through the abundance matching formula of \cite{2010MNRAS.404.1111G}.}, $R_{200}$, of the more massive 
galaxy in the pair and 1,500~kpc. The line-of-sight distances are required to be smaller than seven 
times the mean virial velocity of the two galaxies in the pair. We require the mass ratio (large versus 
small value) of galaxies in the pair should be less than a factor of two\footnote{The stellar mass 
of M31 and MW are about $1-5\times10^{11}\msun$ and $6\times 10^{10}\msun$, respectively \citep[e.g.][]{
2012A&A...546A...4T,2015ApJ...806...96L,2015IAUS..311...82S}.}. In addition, centred on the middle point 
of the two galaxies\footnote{Just geometrical middle point of both projected and line-of-sight directions, 
not weighted by mass.}, all other companions projected within 800~kpc, and within seven times the virial 
velocity of the more massive primary galaxy along the line of sight, should be at least one magnitude 
fainter than the fainter primary galaxy in the pair. The SDSS photometric catalogue with photo-$z$ 
probability distributions \citep{2009MNRAS.396.2379C} is adopted for further selections as well, in 
order to compensate missing brighter companions due to fiber-fiber collisions. 

In this paper, we will pick up galaxies in such pairs whose stellar masses are in the range 
of $10.63<\log_{10}M_\ast/\msun<10.93$, i.e., sharing a similar stellar mass range as our MW. However, 
when computing satellite counts around a given galaxy in a pair whose stellar mass is in this range, 
we choose not to make requirements on the stellar mass of its companion. In other words, though the
stellar mass ratio between the two galaxies in the pair is less than a factor of two (sharing 
similar stellar mass), the stellar mass of the companion might still be out of the range of 
$10.63<\log_{10}M_\ast/\msun<10.93$. Our choice helps to maximise the available sample size. 
Besides, in this study we will investigate paired galaxies with different colour or colour combinations. 
We divide galaxies into red and blue populations by drawing a colour division line of $^{0.1}(g-r)=0.065
\log_{10}M_\ast/\msun+0.1$ over the colour-magnitude diagram of SDSS spectroscopic galaxies.
The number of red and blue primary galaxies in pair with $10.63<\log_{10}M_\ast/\msun<10.93$ are 
893 and 738, respectively. The numbers of galaxies in red-red, blue-blue and red-blue pairs are 524, 
341 and 766, respectively. Note the global colour of our MW very likely lies on this colour 
division line or in the green valley region, which we will discuss later in Section~\ref{sec:disc}.

\begin{figure} 
\includegraphics[width=0.49\textwidth]{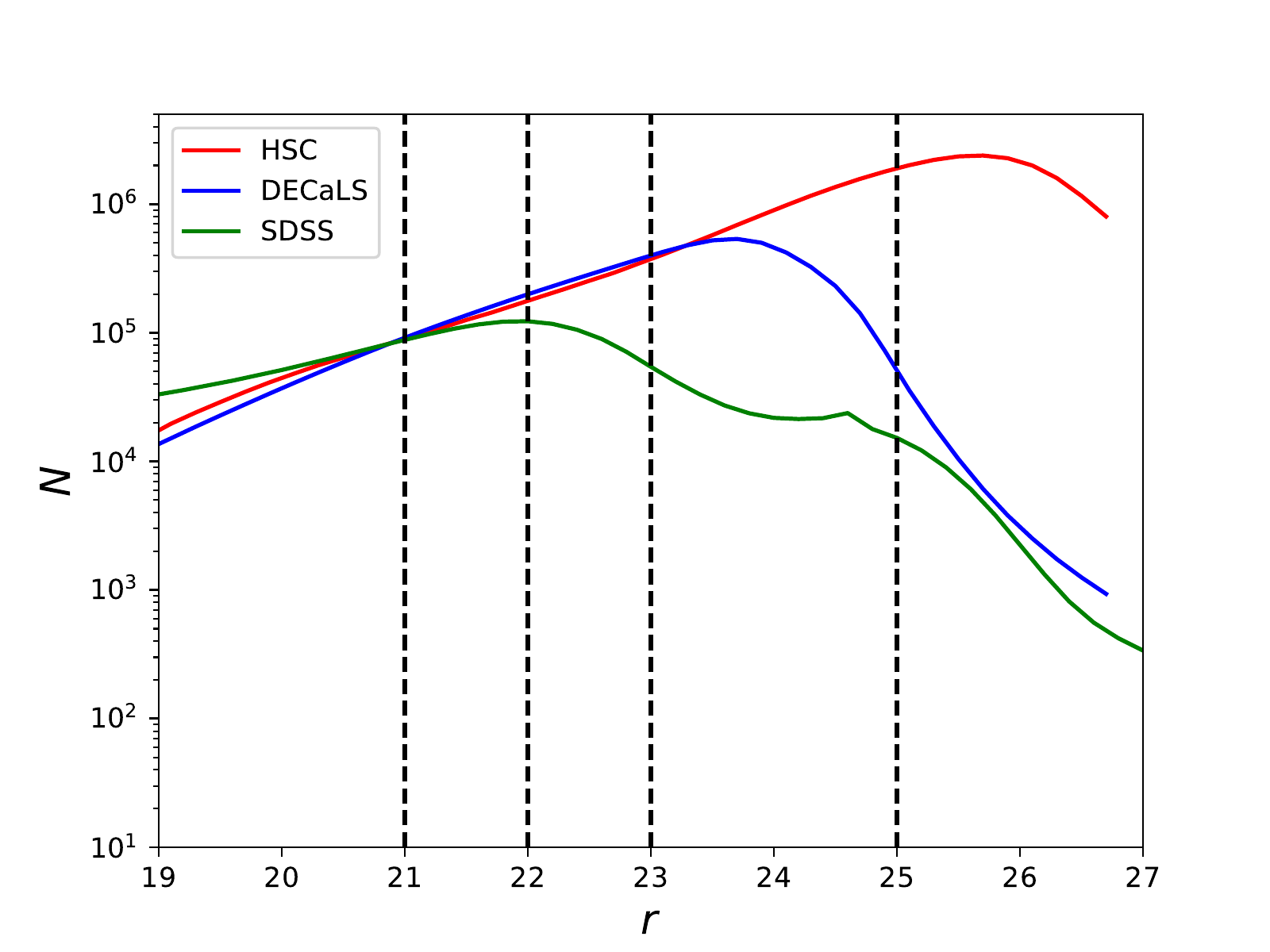}
\caption{Number counts of all photometric sources in HSC, DECaLS and SDSS, as functions of $r$-band 
apparent magnitudes. Vertical dashed lines mark a few typical values of apparent magnitudes, i.e., 
$r=21, 22, 23$ and 24.
}
\label{fig:Nm}
\end{figure}

\subsection{HSC photometric sources}
\label{sec:step}

The HSC survey \citep{2018PASJ...70S...4A} is based on the prime-focus camera, the Hyper Suprime-Cam 
\citep{2012SPIE.8446E..0ZM,2018PASJ...70S...1M,2018PASJ...70S...2K,2018PASJ...70S...3F} 
on the 8.2-m Subaru telescope. It is a three-layered survey, aiming for a wide field of 1,400 deg$^2$ 
with a depth of $r\sim26$, a deep field of 26 deg$^2$ with a depth of $r\sim27$ and an ultra-deep 
field of 3.5 deg$^2$ with one magnitude fainter. In this work we use the wide field data.
HSC photometry covers five bands, namely HSC-$grizy$. The transmission and wavelength range for 
each of the HSC $gri$-bands are almost the same as those of SDSS \citep{2018PASJ...70...66K}. 

The HSC data reduction pipeline is an specialised version of the LSST \citep{2017ASPC..512..279J,
2019ApJ...873..111I} pipeline code. Details about the HSC pipeline are available in the pipeline 
paper \citep{2018PASJ...70S...5B}, and we just briefly introduce the main steps in the following. 

Data reduction is at first achieved on individual exposure basis. The sky background is estimated 
and subtracted before source detections, and the detected sources are used to calibrate the zero 
point and a gnomonic world coordinate system for each CCD by matching to external SDSS and Pan-STARRS1
\citep[PS1;][]{2012ApJ...756..158S,2012ApJ...750...99T,2013ApJS..205...20M,2016arXiv161205560C} sources. 
A secure sample of non-blended stars are used to construct the PSF model. Then a step called joint 
calibration is run to refine the astrometric and photometric calibrations, requiring that the 
same source appearing on different locations of the focal plane during different visits should give 
consistent positions and fluxes \citep{2018PASJ...70S...5B}. The HSC pipeline then resamples images 
to the pre-defined output skymap using a 5th-order Lanczos kernel\citep[e.g.][]{2011PASP..123..596J,
Turkowski:1990:FCR:90767.90805}. All resampled images are combined together (coaddition). 

From the coadd images, objects are detected, deblended and measured. A maximum-likelihood detection 
algorithm is run independently on the coadd image for each band. Background is estimated and subtracted 
in this step once again. The detected footprints and peaks of sources are merged across different bands 
to remove spurious peaks and maintain detections that are consistent over different bands. These 
detected peaks are deblended and a full suite of source measurement algorithm is run on all objects, 
yielding independent measurements of position and source parameters in each band. A reference band of 
detection is then defined for each object based on both the signal-to-noise and the purpose of maximising 
the number of objects in the reference band. Finally, the measurements of sources are run again with the 
position and shape parameters fixed to the values in the reference band to achieve the ``forced'' 
measurements, which brings consistency across bands and enables computing object colours 
using the magnitude difference in different bands.

In this paper, we use detected photometric galaxies from the S19A internal data release. The HSC 
database provides the extendedness flag to classify whether a detected source is more likely to 
be a point source or an extended galaxy. We thus use primary sources which are classified as 
extended. The type of magnitude for these extended HSC sources is the so-called cModel magnitude, 
based on fitting composite model combining Exponential and de Vaucouleurs profiles to source images. 
We exclude sources with any of the following pixel flags set as true in $g$, $r$ and $i$-bands: 
bad, crcenter, saturated, edge, interpolatedcenter or suspectcenter. We also limit ourselves to 
footprints which have reached full depth in $g$ and $r$-bands, i.e., number of exposures in $g$ 
and $r$ greater than or equal to four. The S19A bright star masks are created based on stars from 
{\it Gaia} DR2, and sources within the ghost, halo and blooming masks of bright stars in $i$-band 
have been excluded. The total area is a bit more than 450 square degrees.

As shown by Figure~\ref{fig:Nm}, the number counts of HSC sources keep rising to $r\sim25$. 
In fact, as have been pointed out by \cite{2018PASJ...70S...8A}, the completeness of photometric 
sources in HSC is very close to 1 at $r\sim25$. Thus throughout this paper, unless otherwise 
specified, we adopt a flux limit of $r<25$ for HSC photometric galaxies. However, also according 
to  \cite{2018PASJ...70S...8A}, the quality of star-galaxy separation is very good at $r\sim23$, 
but the completeness fraction of galaxies can drop to as low as 60--70\% at $r\sim25$, which depends 
on seeing. Note we do not worry about cases when stars are mis-classified as galaxies, because stars 
are not correlated with our sample of central primary galaxies, and thus are not expected to bias 
our results, though the contamination by stars might increase the level of foreground contamination 
and hence make the results noisier. On the other hand, if galaxies are mis-classified as stars, 
our results might be affected. Hence for results based on HSC in this study, we will also try 
a few different cuts in flux, including $r<21$, $r<23$ and $r<25$, to ensure the robustness 
of our results.

\subsection{DECaLS photometric sources}

The DESI Legacy Imaging Surveys are imaging the sky in three optical bands ($g$, $r$ and $z$) and four infrared bands, which 
comprise 14,000 deg$^2$ of sky area, bounded by $-18\deg<\mathrm{Dec.}<84\deg$ in celestial coordinates and $|b|>18\deg$ in 
Galactic coordinates \citep{2019AJ....157..168D}. The surveys are comprised of 3 imaging projects, the Beijing-Arizona Sky 
Survey \citep[BASS;][]{2017PASP..129f4101Z}, the Mayall $z$-band Legacy Survey (MzLS) and the Dark Energy Camera Legacy 
Survey (DECaLS).  The survey footprints are observed at least once, while most fields are observed twice or more times. In 
fact, the depth varies over the sky. In this paper, we mainly focus on the sky region with $\mathrm{Dec.}<33\deg$, i.e., 
the DECaLS footprint, which covers $\sim$9,000~deg$^2$. This southern footprint also includes data from the Dark Energy 
Survey \citep[DES;][]{2011AJ....141..185R}, which covers $\sim$5,000~deg$^2$ and is based on the same instrument as DECaLS, 
i.e., the Dark Energy Camera \citep[DECam;][]{2015AJ....150..150F} at the 4-m Blanco telescope at the Cerro Tololo 
Inter-American Observatory.

Data used in this study are downloaded from the eighth data release of the DESI Legacy surveys. All data from the Legacy Surveys 
are first processed through the NOAO Community Pipelines \citep{2019AJ....157..168D}. Briefly, after processing raw CCD images, 
the astrometric calibration is achieved against sources with known coordinates from external reference star samples. The world 
stars are from {\it Gaia} DR1 for data releases made later than the third release of the DESI Legacy surveys. A word coordinate 
system is determined for each CCD using these reference stars. The photometric calibration is based on PS1 DR1 sources. The zero 
point for each CCD is determined independently. 

The photometric source product is constructed by \textsc{tractor}\footnote{\url{https://github.com/dstndstn/tractor}}. \textsc{tractor} 
generates an inference-based model of the sky which best fits the real data. Here we briefly introduce the main post-processing
steps. Basically, the survey footprint is divided into 0.25$\times$0.25$\deg^2$ regions, which is referred to as ``bricks''. 
For a given brick, all CCDs overlapping with it can be found, and for each CCD, an initial sky background is estimated without 
masking sources. After subtracting the initial sky, sources are detected and masked, and the sky background is estimated again 
based on remaining pixels. The PSF for each CCD is determined using PSFEx \citep{2011ASPC..442..435B}. Five independent stacks 
using sky subtracted and PSF convolved CCD images are then made for source detections, which include weighted sums of all CCDs 
in $g$, $r$ and $z$-bands, i.e., a weighted sum of all three bands to optimise a ``flat'' SED with zero AB magnitude colour and 
a weighted sum of all three bands to optimise a ``red'' SED with $g-r=1$ and $r-z=1$. Sources are then detected using the five 
stacks with a threshold of $6\sigma$. For each detected source, \textsc{tractor} models its pipeline-reduced images from different 
exposures and in multiple bands simultaneously. This is achieved by fitting parametric profiles including a delta function (for 
point sources), a de Vaucouleurs law, an exponential model or de Vaucouleurs plus exponential to each image simultaneously. The 
model is assumed to be the same for all images and is convolved with the corresponding PSF in different exposures and bands
before fitting to each image. The best fit is  achieved by minimising the residuals of all images. \textsc{tractor} also outputs 
the quantity which can be used to distinguish extended sources (galaxies) from point sources (stars). 

Starting from the DESI Legacy Survey database sweep files, we remove all sources with TYPE classified as ``PSF'', and require 
BITMASK not containing any of the following: BRIGHT, SATUR\_G (saturated), SATUR\_R, SATUR\_I, ALLMASK\_G\footnote{ALLMASK\_X 
denotes a source that touches a pixel with problems in all of a set of overlapping X-band images. Explicitly, such pixels 
include BADPIX, SATUR (saturated) , INTERP (interpolated), CR (hit by cosmic ray) or EDGE (edge pixels).}, ALLMASK\_R, ALLMASK\_I. 
According to Figure~\ref{fig:Nm}, the average number counts of DECaLS sources keep rising to $r\sim23$. However, given the 
variation of the depth over the sky, unless otherwise specified, we adopt a flux cut of $r<22.5$ to define a safe flux limited 
sample throughout our analysis in this study. We only use the regions with depth deeper than 22.5 in $r$. This is achieved by 
at first selecting bricks with at least three exposures in both $g$ and $r$-bands, and we also require the $r$-band GALdepth 
of the bricks to be deeper than 22.5. We only include galaxies within these selected bricks. For each galaxy, its GALdepth 
is additionally required to be deeper than $r=22.5$, otherwise this galaxy is excluded. In addition, to ensure the robustness 
of our results, we will test another two different choices of flux limits, $r<21$ and $r<23$. For DECaLS, the completeness of
objects classified as extended galaxies is above 90\% at $r<22.5$.

\subsection{SDSS photometric sources}
SDSS used a dedicated wide-field 2.5m telescope \citep{2000AJ....120.1579Y,2006AJ....131.2332G} to image the sky in the drift scan 
mode with five optical filters, $ugriz$ \citep{1996AJ....111.1748F,1998AJ....116.3040G}. The sample of SDSS photometric galaxies 
used to construct satellite counts in this study is exactly the same as those used in \cite{2012MNRAS.424.2574W} and \cite{2014MNRAS.442.1363W}. 
As have been discussed extensively in the Appendix of \cite{2012MNRAS.424.2574W}, by using internal tests based on the data itself, 
external tests using mock catalogues based on both directly projecting the simulation box and light-cone mocks considering a series 
of realistic observational effects, the measured satellite luminosity and stellar mass functions are robust against sample completeness, 
star-galaxy separation, projection effects, background subtraction and K-corrections. In this study, we repeat our analysis with 
our new primaries, and we will make detailed comparisons among results based on SDSS, DECaLS and HSC, to validate the robustness 
of our results based on surveys with different observing strategy, data reduction, resolution and depth. 

The sample of photometric galaxies is downloaded from Casjobs of SDSS DR8. Specifically, we downloaded sources which are 
classified as galaxies in the survey’s primary object list, and that do not have any of the flags: BRIGHT, SATURATED, SATURCENTER 
or  NOPETROBIG\footnote{NOPETROBIG means the Petrosian radius appears to be larger than the outermost point of the extracted 
radial profile.} set. This follows the selection for the DR7 photoz2 catalogue used above when selecting primary galaxies. 
SDSS DR8 has included an improved algorithm of background subtraction than previous releases, and has eliminated the number 
of spurious sources which exist in DR7 \citep[e.g.][]{2005MNRAS.361.1287M,2011ApJS..193...29A}. We only use the sources within 
the acception masks of the DR7 VAGC catalogue, which are stored as spherical polygons. Note using later data releases will 
not increase the signal, because our spectroscopic primary galaxies are selected from the DR7 Main galaxies, and the overlapping 
footprints with releases later than DR8 are not increased. Throughout the paper, we use the Petrosian magnitude\footnote{We have 
repeated the calculation by using SDSS model magnitudes, and the satellite LFs based on Petrosian or model magnitudes are very 
similar.} for SDSS, which self-consistently defines the colour of galaxies within the same aperture size crossing different 
bands. The flux limit we adopt for SDSS is the same as \cite{2012MNRAS.424.2574W}, i.e., $r<21$. As the readers can also see 
from Figure~\ref{fig:Nm}, number counts of SDSS sources keep rising to $r\sim21$, and as has been shown in the Appendix of 
\cite{2012MNRAS.424.2574W}, the completeness of SDSS extended galaxies is very close to 1 above this flux limit.

%%%%%%%%%%%%%%%%%%%%%%%%%%%%%%%%%%%%%%%%%%%%%%%%%%%
\begin{table*}
\caption{Galactocentric positions and $V$-band absolute magnitudes of six MW satellites brighter than $M_V\sim -10$.
Riley et al. (2019) defined the spherical coordinate system in the following way. The $x$-axis points from 
the Sun to the Galactic centre, the $y$-axis points towards the direction of Galactic longitude, $l=90$~deg, and the 
$z$-axis points towards the north Galactic pole. $\theta$ is defined from the $z$-axis, and $\phi$ is defined 
from the $x$-axis so that the Galactic rotation is in the $-\phi$ direction.}
\begin{center}
\begin{tabular}{lcccc}\hline\hline
Satellite      & $r_\mathrm{GC}$ [kpc] & $\theta$ [deg]         & $\phi$ [deg]           & $M_V$ \\ \hline
 Sagittarius I & $18.3_{-2.0}^{+2.0}$  & $110.6_{-0.6}^{+0.8} $ & $8.2_{-0.3}^{+0.3}$    & $-$13.5 \\
 LMC           & $50.3_{-1.9}^{+2.0}$  & $123.3_{-0.0}^{+0.0}$  & $-90.7_{-0.5}^{+0.4}$  & $-$18.1 \\
 SMC           & $61.3_{-3.8}^{+4.2}$  & $136.9_{-0.1}^{+0.1}$  & $-66.8_{-0.7}^{+0.6}$  & $-$16.8 \\
 Sculptor      & $84.0_{-1.5}^{+1.5}$  & $172.5_{-0.1}^{+0.1}$  & $-119.7_{-0.1}^{+0.9}$ & $-$11.1 \\
 Fornax        & $149.5_{-9.0}^{+8.6}$ & $153.9_{-0.1}^{+0.1}$  & $-129.1_{-0.4}^{+0.3}$ & $-$13.4 \\
 Leo I         & $261.9_{-9.3}^{+9.2}$ & $41.7_{-0.0}^{+0.0}$   & $-135.8_{-0.1}^{+0.1}$ & $-$12.03 \\
\hline
\label{tbl:MWsat}
\end{tabular}
\end{center}
\end{table*}
%%%%%%%%%%%%%%%%%%%%%%%%%%%%%%%%%%%%%%%%%%%%%%%%%%%

\subsection{Milky Way satellites}
\label{sec:MWsat}
The $V$-band luminosity, R.A., Dec. and distances of the MW satellites are taken from \cite{2018MNRAS.479.2853N}, \cite{2019MNRAS.486.2679R} 
and \cite{2020ApJ...894...10L}. As the readers will see, the MW satellites which can be used to compare with extra-galactic satellites from 
HSC, DECaLS and SDSS photometric sources introduced above are all brighter than $M_V\sim -10$. These include LMC, SMC, Fornax, Leo I, 
Sculptor and Sagittarius-I. We provide in Table~\ref{tbl:MWsat} the Galactocentric distances, positions and $V$-band absolute magnitudes of 
these satellites. 

\subsection{Illustris TNG-100 and L-Galaxies}
\label{sec:simdata}

We will use the hydro-dynamical Illustris TNG-100 simulation, and the L-Galaxies semi-analytical 
mock galaxy catalogue to aid our analysis. In the following we briefly introduce them.

The TNG series of simulations include a comprehensive model for galaxy formation under the standard cosmological context. 
The fiducial cosmological parameters are the 2015 Planck cosmology \citep{2016A&A...594A..13P} ($H_0=67.74\ \mathrm{
km~s^{-1}/Mpc}$, $\Omega_m=0.3089$ and $\Omega_\Lambda=0.6911$). It self-consistently solves for the coupled evolution of dark 
matter, gas, stars, and black holes from early times to $z=0$ \citep{2019ComAC...6....2N,2018MNRAS.480.5113M}, which produced 
reasonable and quantitative agreement with the colour distribution, clustering, satellite abundance and stellar mass distribution 
of observed galaxies \citep[e.g.][]{2018MNRAS.475..624N,2018MNRAS.475..676S,2018MNRAS.475..648P}. The box size of TNG-100 is 
75~$h^{-1}$Mpc, and the particle mass of dark matter is $7.5\times10^{6}\msun$, which corresponds to a resolution limit of 
about $7.5\times10^{8}\msun$ in halo mass. The satellite LFs of TNG-100 tend to be incomplete beyond $M_V\sim-16$ due to the 
resolution limit.

The semi-analytical galaxy formation code, L-Galaxies, describes the physical processes of galaxy formation analytically, 
by tracing the halo merger histories of the Millennium \citep{2005Natur.435..629S} and Millennium-II \citep{2009MNRAS.398.1150B} 
simulations. The 2015 L-galaxies mock galaxy catalogue \citep{2015MNRAS.451.2663H} rescales \citep{2010MNRAS.405..143A} the 
original simulation to the first-year Planck cosmology \citep{2014A&A...571A..16P} ($H_0=67.3\ \mathrm{km~s^{-1}/Mpc}$, 
$\Omega_m=0.315$ and $\Omega_\Lambda=0.685$). Compared with early versions of the Munich semi-analytical models \citep{2011MNRAS.413..101G,
2013MNRAS.428.1351G}, it has included a few modifications in the treatment of baryonic processes, in order to reproduce observations 
on the abundance and passive fractions of galaxies from $z=3$ to $z=0$. In this study, we use the L-Galaxies mock galaxy catalogue 
based on Millennium-II, which has a higher resolution limit than Millennium, a simulation box size of 100~$h^{-1}$Mpc, and the dark 
matter particle mass of $6.9\times10^{6}h^{-1}\msun$. In addition, L-Galaxies models the evolution of orphan galaxies whose dark 
matter haloes have been entirely disrupted, by tracing the most bound dark matter particles after disruption. Over the luminosity 
range probed in this paper ($M_V<-10$), the L-Galaxies satellite LFs based on Millennium-II agree well with real observations.

\section{Methodology}
\label{sec:method}

\subsection{Satellite luminosity function}

Our methods of counting photometric satellites around spectroscopic primary galaxies and computing the intrinsic luminosities 
of satellites are based on the approach of \cite{2012MNRAS.424.2574W}. For each primary galaxy in a given stellar mass bin, we 
at first count all its photometric companion galaxies down to a certain flux limit and projected within the halo virial radius, 
$R_{200}$ (see Table~\ref{tbl:r200}). The physical scale is calculated based on the redshift and angular diameter distance of 
the central primary. However, without redshift information and accurate distance measurements for these photometric companions,
the companion counts can only be recorded as a function of apparent magnitude and observed-frame colour. In addition, the total 
companion counts not only include true satellites, but also include contamination by fore/background sources. 

We obtain the intrinsic luminosities and rest-frame colours for companions in the following way. For each companion, we employ 
the empirical K-correction of \cite{2010PASP..122.1258W} to estimate its rest-frame colour by using the observed colour and 
also assuming that the companion is at the same redshift as the primary. This is a reasonable approximation, because physically 
associated satellite galaxies are expected to share very similar redshifts as the central primary. For fore/background galaxies, 
their K-correction is wrong, but we will subtract the fore/background counts later. The distance modulus correction is also based 
on the redshift of the central primary. After obtaining the absolute magnitudes and rest-frame colours. A conservative red end 
cut of $^{0.1}(g-r)<0.065\log_{10}M_\ast/\msun+0.35$ is made to the companions to reduce the number of background sources which 
are too red to be at the same redshift of the primary, and hence increase the signal\footnote{In order to test whether we might 
have excluded some extremely red satellite galaxies, we calculate the LF using galaxies redder than this colour cut. We find 
that the signals are very close to zero. Sometimes the signals can be positive, but the amplitudes are significantly less than 
1\textperthousand~of the satellite LFs measured in this paper. Thus even if a small number of extremely red satellites are lost 
due to this red end cut, our results are unlikely to have been significantly affected. }. The colour cut is drawn from the colour 
distribution of SDSS spectroscopic Main galaxies.

To ensure the completeness of satellite number counts in different luminosity bins, for each primary, we convert the corresponding 
flux limit of a given survey to a K-corrected absolute magnitude, $M_{r,{\rm lim}}$, using the redshift of the primary and a 
colour chosen to be on the red envelope of the intrinsic colour distribution for galaxies at that redshift. For a given luminosity 
bin, primaries are allowed to contribute to the final companion counts only if $M_\mathrm{r,lim}$ is fainter than the fainter bin 
boundary. As a result, the numbers of actual central primaries contributing to different luminosity bins can vary. The fainter 
the luminosity, the less number of primaries can contribute the counts, and their redshifts are lower. In the end, the total 
counts are divided by the total number of primaries which actually contribute to the satellite counts in each bin, which provides 
the averaged and complete satellite LF per primary galaxy. 

To subtract fore/background contamination, we repeat exactly the same procedures using a sample of random primaries, which 
are assigned the same redshift and stellar mass distributions as true central primaries, but their coordinates have been 
randomised within the survey footprint. The averaged companion counts per primary around these random primaries are subtracted 
from the counts around real primaries.  

Due to the survey boundary and masks of bad pixels and bright stars, we should estimate the completeness of the projected area
around primaries. For HSC and DECaLS, this is achieved by using their photometric random samples provided in the database. We 
apply exactly the same selection and masks to random points. The completeness of the projected area is estimated as 

\begin{equation}
    f_\mathrm{complete}=\frac{\mathrm{number\ of\ actual\ random\ points}}{\mathrm{area}\times \mathrm{(surface\ density\ of\ random\ points)}}.
\end{equation}

For SDSS, the completeness fraction is estimated through the acception masks/spherical polygons of the DR7 VAGC catalogue. 
Around each primary, we generate random points by ourselves, and $f_\mathrm{complete}$ is defined as the ratio between the 
number of random points within the spherical polygons and all random points in the projected area around each primary. Our 
actual companion counts around both real and random primaries are divided by $f_\mathrm{complete}$ for incompleteness corrections. 

\subsection{Inner radius cut and projection effects}

\begin{figure*} 
\includegraphics[width=0.98\textwidth]{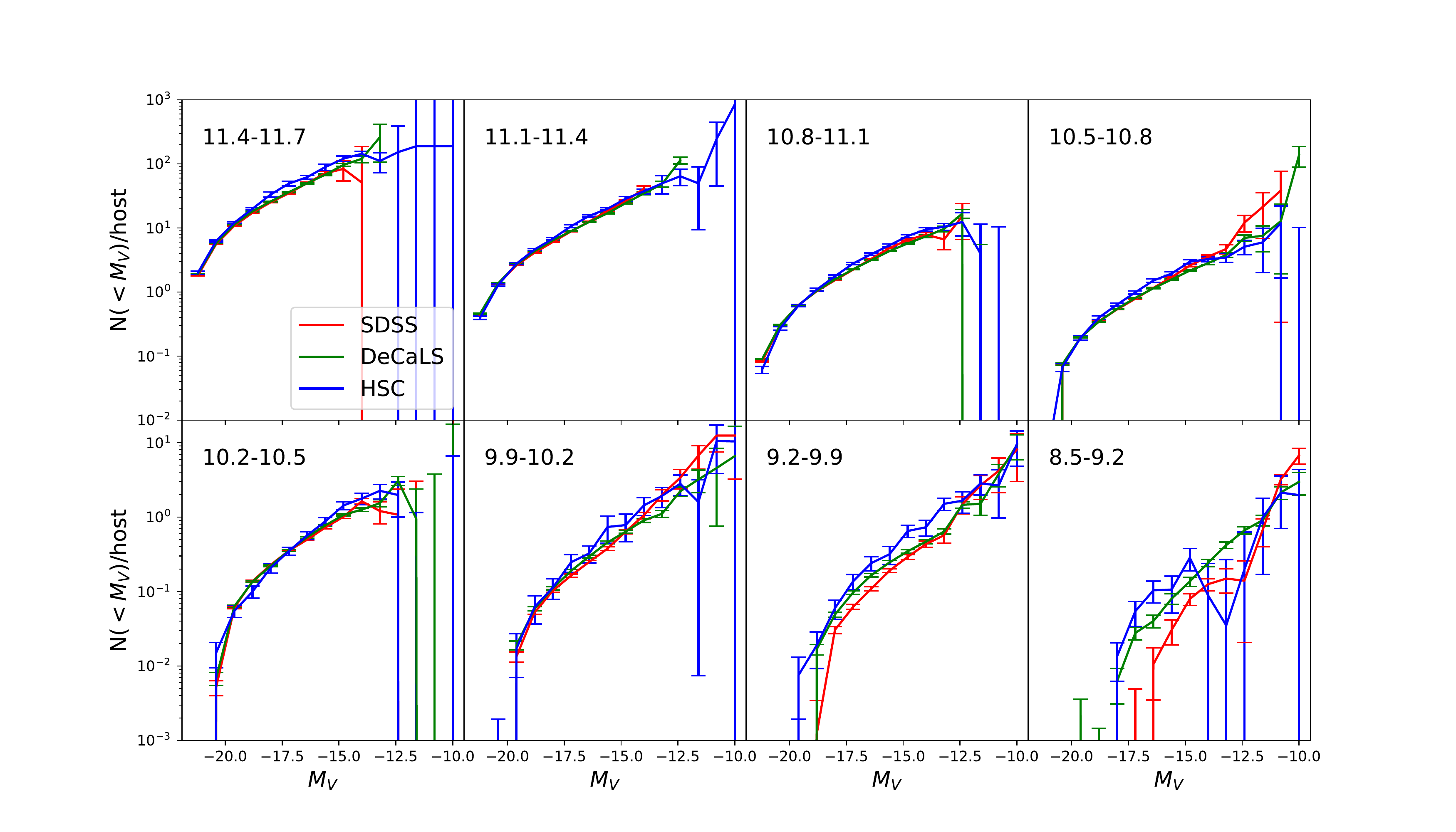}
\caption{Cumulative $V$-band luminosity functions (LFs) for satellite galaxies projected within the halo virial radius 
to isolated central galaxies (ICG1s) in HSC, DECaLS and SDSS (see the legend). In addition, an inner radius cut of 
$r_p>30$~kpc has been adopted for all three surveys to avoid failures in source deblending. Flux limits are adopted 
to be $r<21$, $r<22.5$ and $r<25$ for SDSS, DECaLS and HSC, respectively. The text in each panel indicates the log 
stellar mass range of primaries. Errorbars are 1-$\sigma$ scatters based on  100 boot-strap subsamples.
}
\label{fig:LFcum}
\end{figure*}

As have been introduced in Section~\ref{sec:data}, HSC, DECaLS and SDSS have very different survey depth and  
resolution. The deeper a survey is, the more fainter sources and the more extended low surface brightness 
structures can be detected, and thus the observed field is more crowded. As a result, deep surveys such as HSC suffer 
from more serious source blending issues than other shallower surveys. Details about direct comparisons among 
the three surveys are provided in Appendix~\ref{app:blending}. As the readers can see from Figure~\ref{fig:blending}, 
there are far more faint sources detected in HSC and around the central primary galaxies. Some sources are real, 
which are failed to be detected in both SDSS and DECaLS. However, some sources are in fact fake detections, which 
are in the photometric region of the central galaxy itself and are mistakenly deblended to be companions. 

To avoid such deblending failures on small scales to the central primaries, we adopt an inner radius cut of 
$r_p>30$~kpc for all the results in the main text of this paper. We also provide results in Appendix~\ref{app:blending} 
showing the evidences supporting the choice of $r_p>30$~kpc. We also found that failures of source deblending 
is more frequent around blue primaries, because blue galaxies have rich substructures such as star-forming regions 
along the spiral arms, which are more easily to be mistakenly deblended as companion sources. Thus the inner radius 
cut is very important in order to properly avoid the deblending failures and ensure consistencies among HSC, DECaLS 
and SDSS. 

The inner radius cut not only excludes fake sources, but real sources within $r_p\sim$30~kpc are also excluded. 
This, however, does not affect the fair comparison among different surveys if we adopt the same cut for all of them. 
In addition, to ensure fair comparison with the MW satellite LF, we need to properly consider the projection 
effect. This is because extra-galactic satellites are observed in projection, whereas the MW satellites are observed 
in 3-D. We choose to project the observed 3-D positions of MW satellites along 600 randomly selected ``line-of-sight'' 
directions. For each direction of projection, we calculate the satellite LF after excluding satellites projected within 
30~kpc. In the end, we calculate the mean MW satellite LF based on all 600 projections, and the 1-$\sigma$ scatter 
among these different projections is adopted to represent the uncertainties. The inner radius cut of 30~kpc always 
excludes Sagittarius-I in Table~\ref{tbl:MWsat} from our analysis, due to its short Galactocentric distance of 
$r_{\rm GC}=18.3\pm2.0$~kpc. The other MW satellites we use for the comparison are at distances greater than 30~kpc, 
although the projection would make satellites appear to be closer than this distance cut in some cases.

In the end, we note that the MW satellite LF is measured in $V$-band. However, we do not have $V$-band data for any of
the three surveys used in this study. To ensure fair comparisons, we convert the $r$-band magnitudes to $V$-band based 
on the transformations provided by \cite{2007AJ....133..734B} and by using the $g-r$ colours of satellites.

\section{Results}
\label{sec:results}

\begin{figure*}
\includegraphics[width=0.98\textwidth]{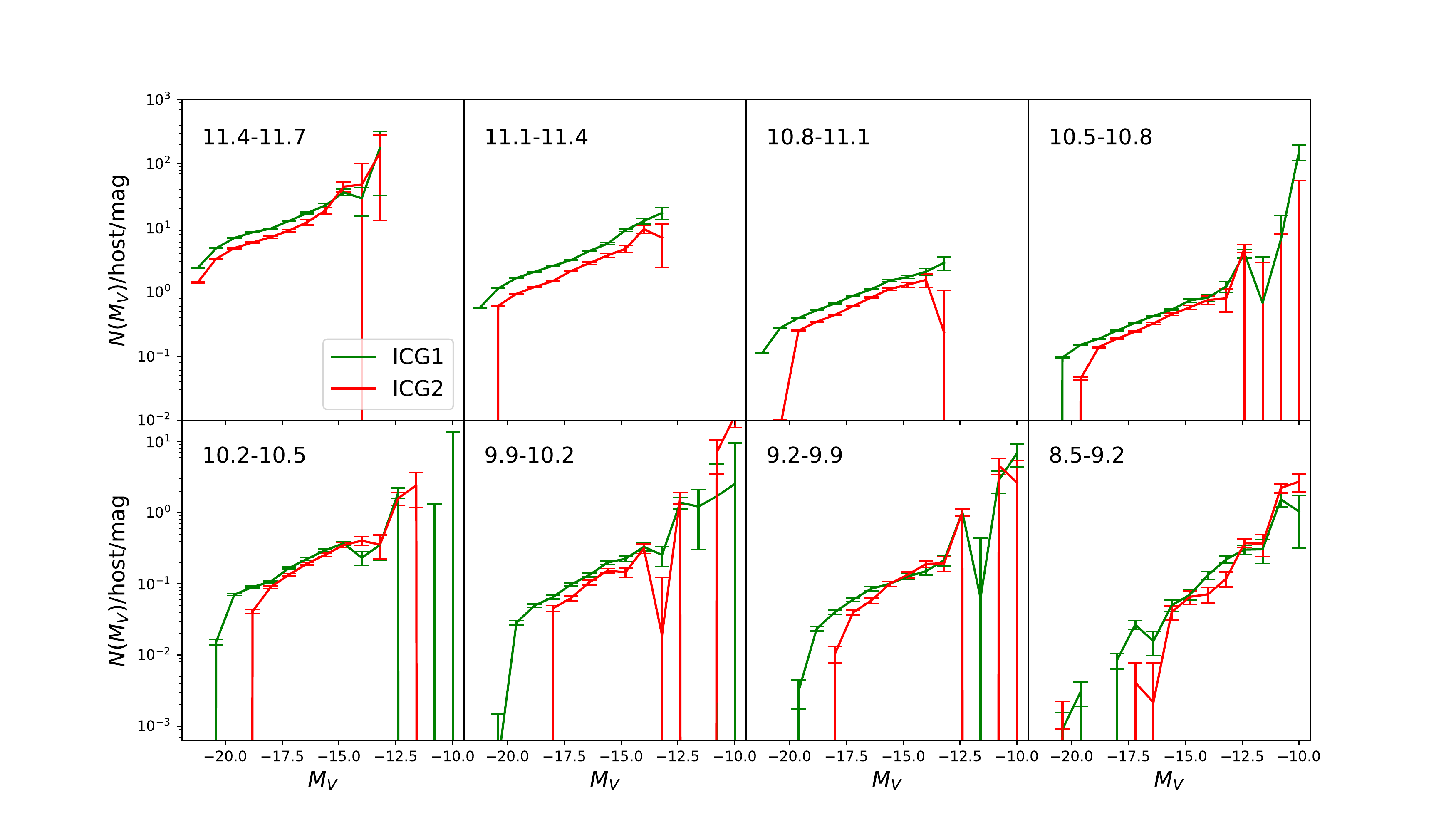}
\caption{Differential $V$-band LFs for satellite galaxies projected between 30~kpc and the halo virial radius, and 
centred on two different populations of isolated central galaxies (ICG1 and ICG2). The results are based on DECaLS. 
Companions of ICG1s are required to be fainter than ICG1s, while companions of ICG2s are at least one magnitude 
fainter. Due to the selection, ICG2s have less satellites, which could be an indication of smaller host halo mass. 
Errorbars are 1-$\sigma$ scatters based on 100 boot-strap subsamples.
}
\label{fig:LFcum_diffICG}
\end{figure*}

In this section we present our satellite LF measurements. First, we show results for all primary galaxies grouped 
into eight stellar mass bins, and we investigate whether the satellite LF depends on the magnitude gap between 
the central primary galaxy and its satellites, using ICG1s and ICG2s. Then we move on to present results focusing 
on isolated central galaxies (or galaxy pairs) sharing similar properties as our MW (or as the MW and M31), and 
compare the measurements with the MW satellite LF. 

\subsection{Satellites of all primary galaxies}
\label{sec:allprim}

\begin{figure*} 
\includegraphics[width=0.98\textwidth]{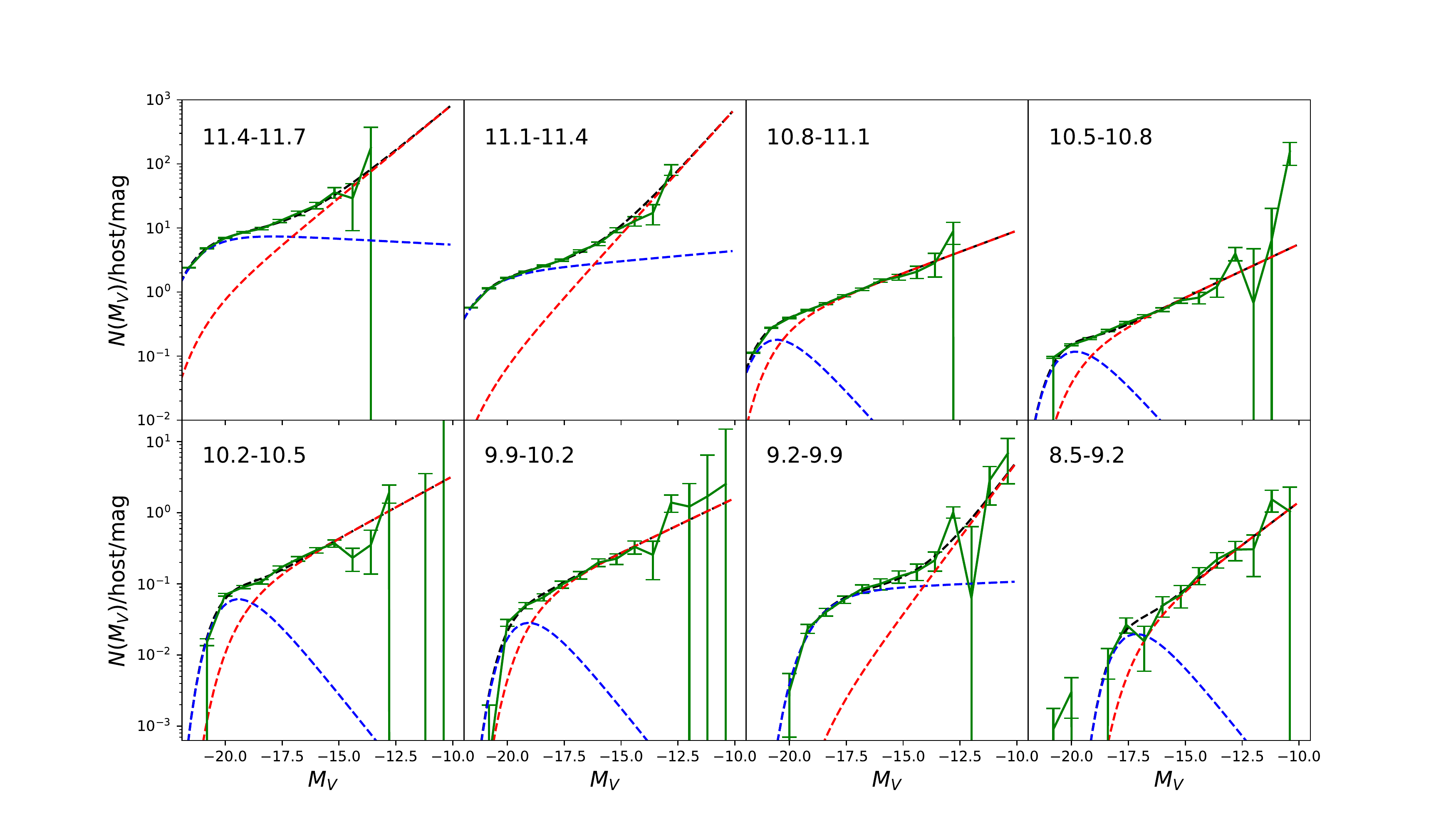}
\caption{Differential $V$-band LFs for satellite galaxies projected between 30~kpc and the halo virial radius to ICG1s 
in DECaLS. Errorbars are 1-$\sigma$ scatters based on 100 boot-strap subsamples. Double Schecter functions are fit to 
the LFs. Black dashed curves show the best fits, while the two components are plotted as blue and red dashed curves. 
}
\label{fig:LFdiff}
\end{figure*}

%%%%%%%%%%%%%%%%%%%%%%%%%%%%%%%%%%%%%%%%%%%%%%%%%%%
\begin{table*}
\caption{The best-fit double Schecter function parameters to satellite LFs around ICG1s
in Figure~\ref{fig:LFcum_diffICG}.}
\begin{center}
\begin{tabular}{lccccc}\hline\hline
$\log M_\ast/\msun$ & $M_0$ & $\Phi_{\ast,1}$ & $\Phi_{\ast,2}$ & $\alpha$ & $\beta$  \\ \hline
11.4-11.7 & $-$21.27$\pm$0.10 & 9.662$\pm$0.567 & 0.477$\pm$0.703 & $-$0.953$\pm$0.111 &  $-$1.730$\pm$0.251  \\
11.1-11.4  &  $-$21.42$\pm$0.10  & 2.038$\pm$0.163 & 0.026$\pm$0.035 & $-$1.081$\pm$0.064  & $-$1.979$\pm$0.444  \\
10.8-11.1 &  $-$20.56$\pm$0.05  & 0.532$\pm$0.045 & 0.393$\pm$0.017 & $-$0.000$\pm$0.251 & $-$1.332$\pm$0.018 \\
10.5-10.8 &  $-$19.83$\pm$0.04  & 0.346$\pm$0.024 & 0.141$\pm$0.012 & $-$0.000$\pm$0.028 & $-$1.416$\pm$0.034 \\
10.2-10.5 &  $-$19.42$\pm$0.06  &0.180$\pm$0.021  & 0.080$\pm$0.010  & $-$0.000$\pm$0.040 & $-$1.437$\pm$0.047 \\
%9.2-10.2 &  $-$18.81$\pm$0.17  & 0.065$\pm$0.025 & 0.062$\pm$0.012  & $-$0.000$\pm$0.169 & $-$1.284$\pm$0.081 \\
9.9-10.2 &  $-$19.06$\pm$0.19  & 0.084$\pm$0.028 & 0.072$\pm$0.013  & $-$0.000$\pm$0.287 & $-$1.381$\pm$0.071 \\
9.2-9.9  &  $-$18.81$\pm$0.29  & 0.088$\pm$0.026 & 0.001$\pm$0.002  & $-$-1.035$\pm$0.121 & $-$2.064$\pm$0.138 \\
8.5-9.2 & $-$17.15 $\pm$0.38  & 0.058$\pm$0.036 & 0.029$\pm$0.015  & $-$0.000$\pm$0.144 & $-$1.601$\pm$0.159 \\
\hline
\label{tbl:LFpara}
\end{tabular}
\end{center}
\end{table*}
%%%%%%%%%%%%%%%%%%%%%%%%%%%%%%%%%%%%%%%%%%%%%%%%%%% 

Figure~\ref{fig:LFcum} shows the LFs for satellite galaxies projected between 30~kpc and the halo virial radius to ICG1s 
in HSC, DECaLS and SDSS, and ICG1s are grouped into eight stellar mass bins (see the text in different panels). Throughout 
this paper, the errorbars for extra-galactic satellite LFs are based on the 1-$\sigma$ scatters of 100 boot-strap subsamples, 
which reflect the errors on the mean satellite population of different primaries. We are able to measure the satellite LFs 
around primaries as small as $8.5<\log M_\ast/\msun<9.2$. The satellite LFs based on HSC, DECaLS and SDSS are very similar 
to each other in all of the panels. Considering the very different observational mode, data reduction, depth and image 
resolution, the consistency among these surveys is very encouraging.

There are, however, still some very small and delicate differences. At the bright end, the difference might be partly due to 
the small number of bright galaxies and hence relatively large sample variance. At fainter magnitudes, the difference could 
be caused by many reasons. First of all, different flux limits are adopted for the three surveys. In fact, we have explicitly 
tested that after adopting the same flux limit of $r<21$ for all surveys, the results in in Figure~\ref{fig:LFcum} remain 
almost unchanged. We avoid repeatedly showing the figure with $r<21$, but the readers can still find part of the tests 
based on a stellar mass bin similar to our MW in Section~\ref{sec:mwlikeprim}. 

Secondly, we can see HSC measurements tend to show slightly higher amplitudes than both SDSS and DECaLS in almost all of 
the panels. This is at least partly due to the deeper surface brightness depth of HSC, and thus more low surface brightness 
satellites can be detected. While the integrated flux of a faint satellite is still above the survey flux limit, it might 
not have been detected because it is diffuse and has low surface brightness \citep[e.g.][]{2018ApJ...856...69D,
2020ApJ...891..144C}. We provide more evidences in Appendix~\ref{app:blending}. 

In addition, given the different image resolution or PSF size, the ability for different survey pipelines to deblend sources 
can vary. HSC is able to disentangle multiple sources which have very small angular separations, whereas SDSS might treat 
them as one single source. This could cause some very delicate differences in both satellite number counting and magnitudes.

Moreover, the photometric system, data reduction steps and depth of different surveys are not the same, and the magnitudes 
are defined and calculated in different ways. As a result, even for the same source, its magnitudes in HSC, DECaLS and SDSS can 
differ. However, the filter system between SDSS and HSC is very similar, and as have been checked in \cite{2020arXiv200412899Q}, 
the mean magnitude difference between the same source in SDSS and HSC is negligible, and thus we do not expect the difference 
in magnitudes can significantly affect our results. We provide in Appendix~\ref{app:blending} (Figure~\ref{fig:testmag}) more 
detailed comparisons by using matched sources between different surveys.

Lastly, we have excluded photometric sources with a few bad-quality photometric flags setting up to be true, including 
bad pixels, saturated, edge, cosmic rays and so on (see Section~\ref{sec:data} for details). However, it is difficult to 
have exactly the same selection by photometric flags for different surveys. We have tested that after removing the 
selection by photometric flags, and the change in measured satellite LFs is smaller than 1\%. 

For primaries more massive than $\log M_\ast/\msun\sim10.8$, we can robustly push down to $M_V\sim-14$ or $-13$. Results 
based on HSC can push even fainter (close to $M_V\sim-10$ in the two most massive bins), but the measurements are quite 
noisy due to the small HSC footprint. For smaller primaries, we can push close to $M_V\sim-10$. This is because
massive bright primaries are biased to have higher redshifts\footnote{We can observe bright galaxies out to larger 
distances, and at nearby distances, the number of bright galaxies is small due to the low volume density and the small 
volume in the nearby universe.}, but for fainter luminosity bins, only those satellites around more nearby primaries 
are complete and are allowed to contribute to the satellite counts (see Section~\ref{sec:method} for details). Thus 
we do not have enough number of nearby massive bright primaries contributing to the number counts of intrinsically 
faint satellites. On the contrary, smaller and fainter primaries have lower redshift distributions, and thus we are 
able to push down to even fainter magnitudes for satellites around them. 

Previous measurements of satellite LFs based on SDSS can go as faint as $M_r\sim-14$ \citep[e.g.][]{2012MNRAS.424.2574W} 
or about $M_V\sim-14$ as well \citep[e.g.][]{2011MNRAS.417..370G}, i.e., at most $\sim$8 magnitudes fainter than the 
central primaries. Very recently, after carefully considering the varying survey depths, new measurements of satellite LFs 
based on DECaLS can also reach $M_r\sim-14$ \citep{2019arXiv191104507T}. Note the measured satellite LFs in \cite{2012MNRAS.424.2574W}
stopped at the stellar mass of $\log_{10}M_\ast/\msun=10.2$ for primaries. In this study, we have managed to extend the 
previous measurements down to much fainter satellites ($M_V\sim-10$) and around smaller primaries ($9.9<\log_{10}M_\ast/
\msun<10.2$, $9.2<\log_{10}M_\ast/\msun<9.9$ and $8.5<\log_{10}M_\ast/\msun<9.2$). 

The stellar mass of the LMC is likely to be in between the two lowest stellar mass bins in our analysis. The LMC 
should have its own satellites before merging into our Galaxy. Many efforts have been devoted to potentially determine 
how many satellites does the LMC have before infall \citep[e.g.][]{2015MNRAS.453.3568D,2016MNRAS.461.2212J,2017MNRAS.465.1879S,
2018MNRAS.476.1796S}, which still tend to show large uncertainties. As have been pointed out by \cite{2017MNRAS.472.1060D}, 
it is useful to look at the population of satellites around isolated galaxies with LMC stellar mass, and these isolated 
LMC-like systems are also important for the study of environmental effects on the evolution of dwarf galaxies 
\citep[e.g.][]{2015ApJ...807...49W}. Our measurements predict an average number of 3--8 satellites brighter than 
$M_V\sim-10$ of LMC-mass ICGs.

Unfortunately, despite the much deeper survey depths of $r<25$, HSC does not seem to be able to push significantly 
fainter than DECaLS ($r<22.5$) or SDSS ($r<21$). This is mainly due to the much smaller footprint of HSC (a bit more than 
450 square degrees) than the other two surveys. The footprints of SDSS and DECaLS are above 8,000 and 9,000 square degrees, 
respectively\footnote{Considering the footprint overlapping with spectroscopic central galaxies from the SDSS Main sample 
galaxies, the effective footprint of DECaLS is smaller than SDSS.}. However, it is still very encouraging that even given the 
more than ten times smaller footprint, HSC is still able to achieve comparable or slightly better performance. Hence it is 
promising to wait for the completion of the HSC mission and use the full 1,400 square degrees of the planned footprint to 
push even fainter. Besides, our results based on HSC can be regarded as a pioneer study of the future LSST survey, which is 
designed to have similar depth but much larger footprint (about 23,000 sq. deg.) than HSC. Our results based on HSC 
have demonstrated the power of such deep surveys to help revolutionise our understanding towards faint extra-galactic 
satellites.

Compared with ICG1s, ICG2s are selected with more strict criteria, i.e., its companions should be at least 
one magnitude fainter. Now we move on to compare satellite LFs measured around ICG1s and ICG2s. The results are shown 
in Figure~\ref{fig:LFcum_diffICG}. We only show the differential LFs based on DECaLS, because of its larger footprint 
than HSC and greater depth than SDSS, though we have explicitly checked that HSC and SDSS show very similar results. 
Green curves are based on ICG1s, while red curves are based on ICG2s, and due to the larger magnitude gap between 
the primaries and their companions, the bright end cut off of satellite LFs around ICG2s becomes more 
significant by definition. In addition, we also see that the LFs around ICG2s tend to have lower amplitudes than 
those around ICG1s, and this is true over the wide luminosity range probed here. 

The difference between ICG1s and ICG2s is very interesting. Although we have only changed the magnitude gap between 
primaries and their companions in our selection, the magnitude gap affects not only the bright end, but also 
the overall abundance of satellite galaxies at different magnitudes. Many previous studies have tried to link satellite 
abundance or total satellite luminosity to host halo mass \citep[e.g.][]{2012MNRAS.424.2574W,2013MNRAS.428..573S,
2014MNRAS.442.1363W,2019arXiv191104507T}, and such links have been proved by \cite{2016MNRAS.457.3200M} through 
direct weak lensing measurements. Thus the lower amplitudes for ICG2s in Figure~\ref{fig:LFcum_diffICG} imply that 
ICG2s are likely hosted by less massive haloes. This is also supported by studies based on numerical 
simulations, which claim that the host halo mass depends on the magnitude gap between central and satellite galaxies 
\citep[e.g.][]{2015ApJ...804...55L}. About the nature of why such magnitude gaps can be linked to satellite abundance 
or halo mass, it might be related to the assembly history of haloes, which we will investigate in future studies. 

We fit the following double Schecter functions to satellite LFs in DECaLS (green curves in Figure~\ref{fig:LFcum} and 
\ref{fig:LFcum_diffICG})
\begin{equation}
    \Phi(L)\ud L= \left\{\Phi_{\ast,1} \left[ \frac{L}{L_0} \right]^\alpha +  \Phi_{\ast,2}\left[ \frac{L}{L_0} \right]^\beta \right\}\exp{\left(-\frac{L}{L_0}\right)}\ud L,
    \label{eqn:LF}
\end{equation}
and because the relation between luminosity and absolute magnitude is $\frac{L}{L_0}=10^{-0.4(M-M_0)}$, 
Equation~\ref{eqn:LF} can be expressed in terms of absolute magnitude $M$ as

\begin{align}
    \Phi(M)\ud M &= 0.4\ln{10} \nonumber \\ 
    &\hspace{-2em}\times \left[\Phi_{\ast,1} 10^{-0.4(M-M_0)(\alpha+1)}+\Phi_{\ast,2} 10^{-0.4(M-M_0)(\beta+1)} \right]\nonumber \\
    &\times \exp{\left[-10^{-0.4(M-M_0)}\right]} \ud M. 
\end{align}
The best fits are shown in Figure~\ref{fig:LFdiff}, in which we present the differential satellite LFs instead
of cumulative ones. The best-fit parameters are provided in Table~\ref{tbl:LFpara}. Except for the stellar mass 
bin of $9.2<\log_{10}M_\ast/\msun<9.9$, which has a faint end slope slightly steeper than $-2$, the best-fit 
faint end slopes of all the other panels are shallower than $-2$. In addition, there are some indications 
in the top right panel that the faint end tends to show some signs of up-turning. The measurements at the 
faint end are quite noisy, and thus we avoid drawing a very strong conclusion in this paper. However, the faint 
end slopes of satellite LFs have very important cosmological implications. For example, the predicted number of 
substructures by warm and cold dark matter models only vary at the small mass end \citep[e.g.][]{2014MNRAS.439..300L}, 
and if the satellite LF continues to rise sharply at the faint end, the rich number of small satellites can 
provide important clues to distinguish different dark matter models. In addition, the faint end slopes of satellite 
LFs contain information about the formation history of galaxies at the early Universe and can be used to constrain 
models of galaxy formation \citep[e.g.,][]{2017MNRAS.464.3256L,2016MNRAS.459.3998L}. Future deep and wide photometric 
surveys such as LSST, combined with our method of counting photometric satellites around spectroscopic primaries, 
is thus very promising and powerful to help improving the statistics at the faint end and hence can potentially 
revolutionise our understanding towards the nature of extremely faint and small satellites, though one has to 
very carefully deal with possible systematics at such faint magnitudes. 

\begin{figure} 
\includegraphics[width=0.49\textwidth]{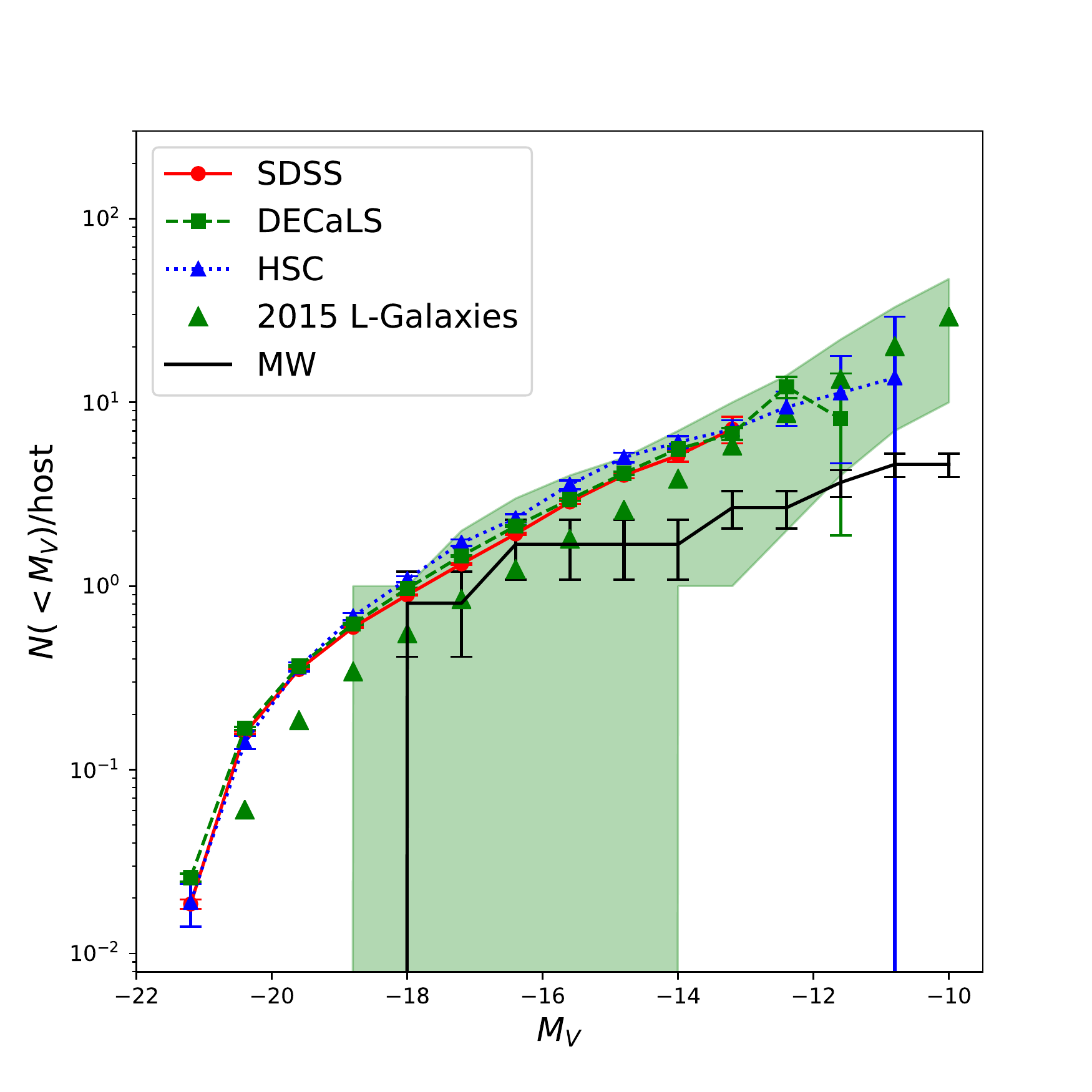}
\caption{Cumulative $V$-band LFs for satellite galaxies projected between 30 and 260~kpc to MW-mass ICG1s ($10.63<
\log_{10}M_\ast/\msun<10.93$). Results based on SDSS, DECaLS and HSC are overplotted with each other for comparisons 
(see the legend). Flux limits are adopted to be $r<21$, $r<22.5$ and $r<25$ for SDSS, DECaLS and HSC, respectively. 
Black solid histogram shows the cumulative LF for MW satellites after projection and with the same inner radius 
cut of $r_p>30$~kpc. Errorbars for HSC, DECaLS and SDSS results are 1-$\sigma$ scatters based on 100 boot-strap 
subsamples, which reflect the error on the averaged satellite LF. Errorbars for the MW satellite LF are 
calculated from the 1-$\sigma$ scatters among 600 different projections. Green triangles show predictions by 
the 2015 L-Galaxies model, and the green shaded region associated with it shows the scatter, i.e. 16th 84th 
percentiles, in the LFs of individual systems.}
\label{fig:LFcum_MW}
\end{figure}

\subsection{Satellites of Milky-Way-like primary galaxies}
\label{sec:mwlikeprim}

\begin{figure*} 
\includegraphics[width=0.49\textwidth]{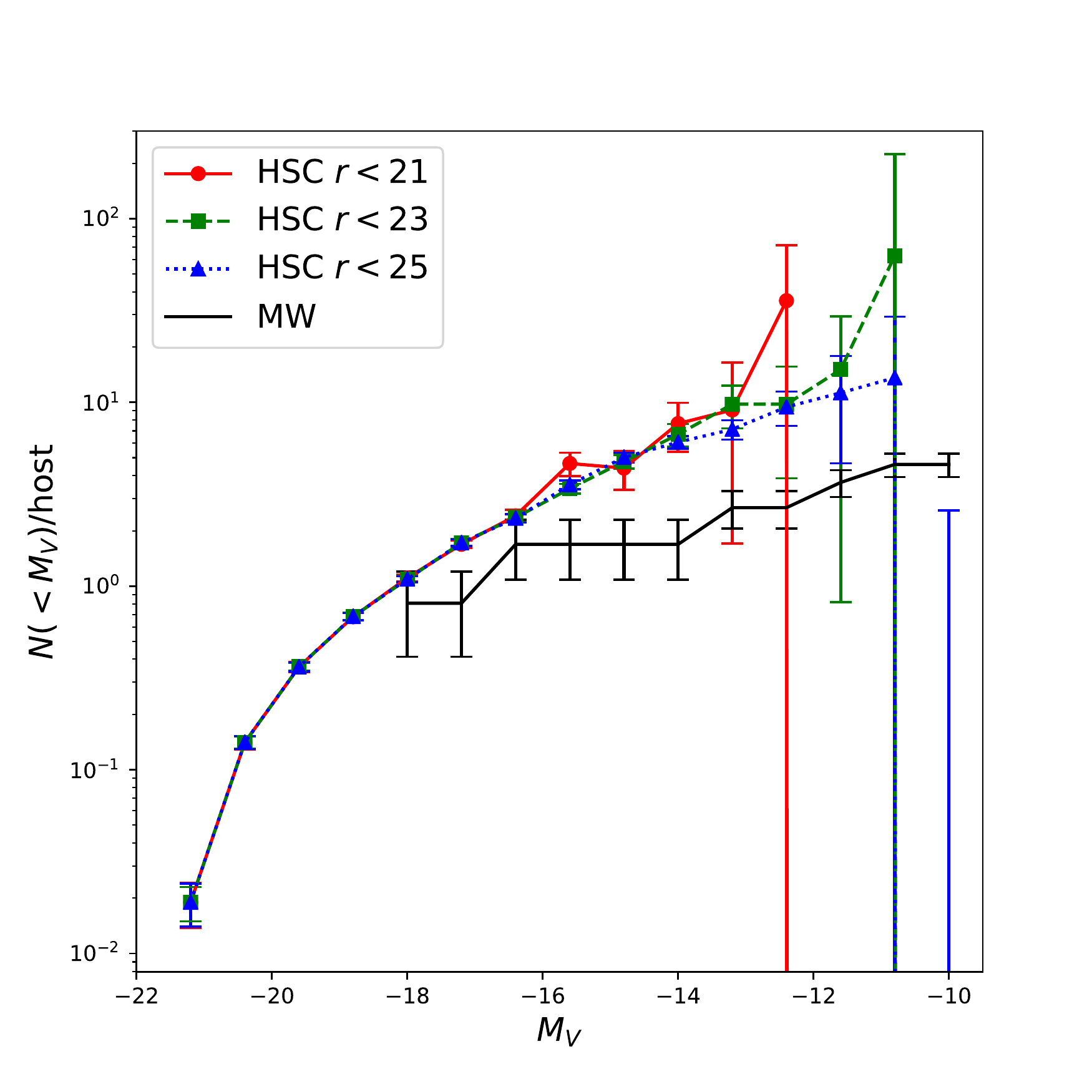}%
\includegraphics[width=0.49\textwidth]{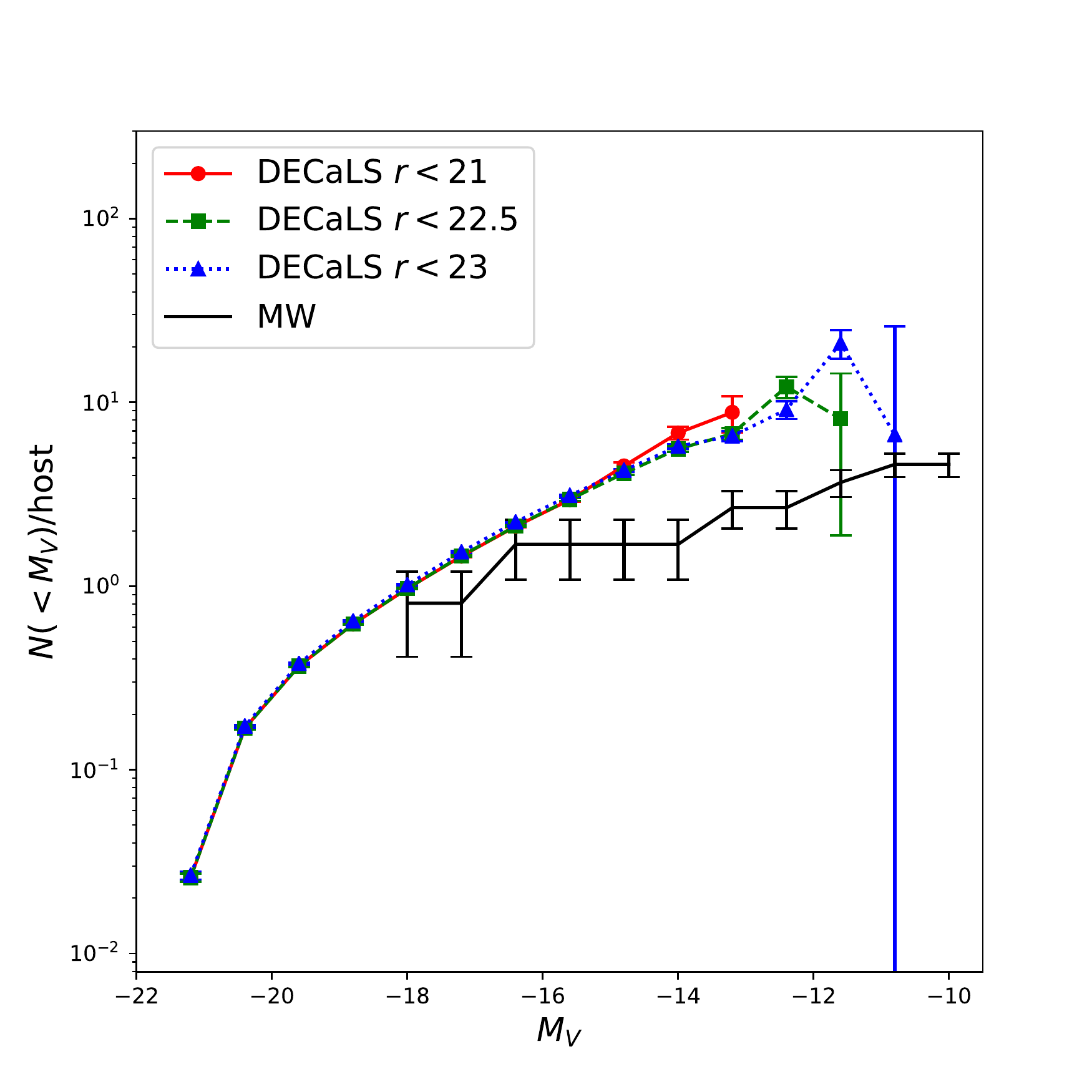}
\caption{{\bf Left:} Cumulative $V$-band LFs for satellite galaxies projected between 30 and 260~kpc 
to MW-mass ICG1s ($10.63<\log_{10}M_\ast/\msun<10.93$). The results are based on HSC. Three different 
flux limits of $r<21$, $r<23$ and $r<25$ are adopted. LFs based on different flux limits show very 
good agreement with each other, which validates our method and the sample completeness of HSC. {\bf 
Right:} Similar to the left panel, but shows results based on DECaLS. Three different flux limits of 
$r<21$, $r<22.5$ and $r<23$ are adopted. In both panels, the black solid histogram shows the cumulative 
LF for MW satellites after projection and with the same inner radius cut of $r_p>30$~kpc. Errorbars 
are calculated in the same way as Figure~\ref{fig:LFcum_MW}. }
\label{fig:LFcum_MW_fluxlimit}
\end{figure*}
\begin{figure*} 
\includegraphics[width=0.49\textwidth]{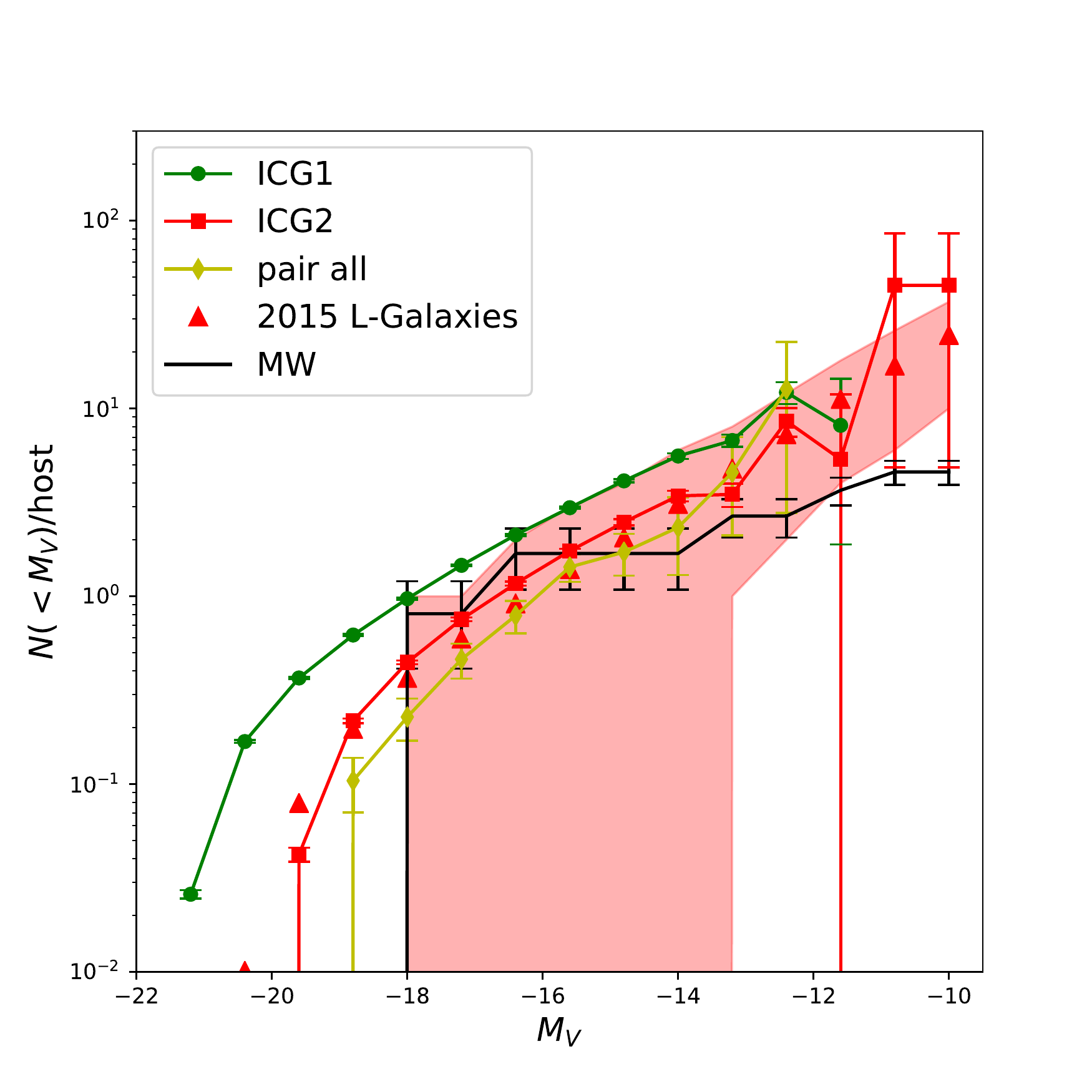}
\includegraphics[width=0.49\textwidth]{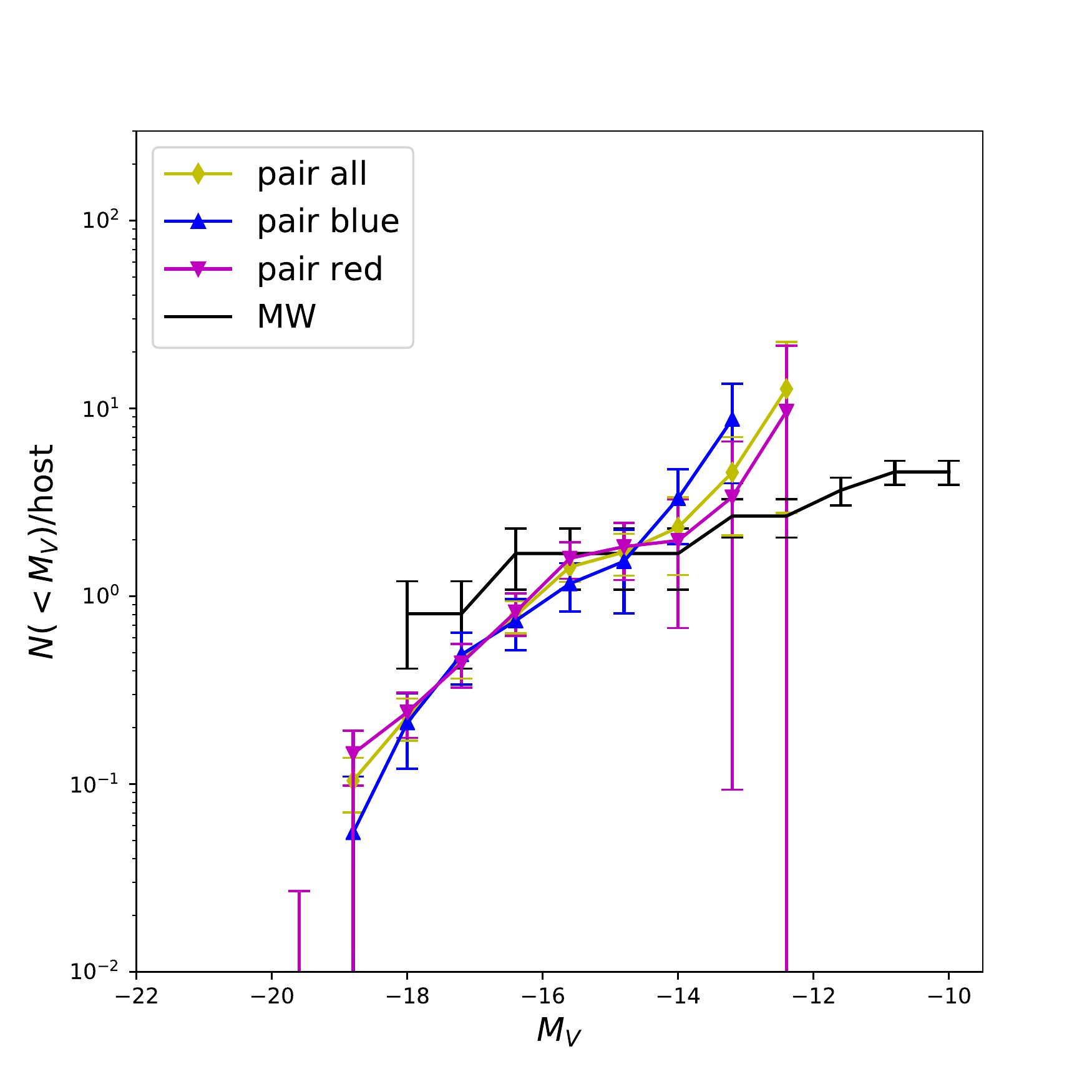}
\caption{{\bf Left:} Cumulative $V$-band LFs for satellite galaxies projected between 30 and 260~kpc 
to ICG1s, ICG2s and primary galaxies in pair (see the legend). All different types of primary galaxies 
are required to have similar stellar mass as our MW, i.e., $10.63<\log_{10}M_\ast/\msun<10.93$. In addition, 
although upon selecting galaxy pairs, the mass ratio between the two galaxies in pair is required to 
be smaller than a factor of two, the stellar mass of the companion is not necessarily in the range of 
$10.63<\log_{10}M_\ast/\msun<10.93$. For such cases, only one galaxy in the pair is used for this plot, 
while the other is not. This maximises our sample size. Red triangles show the predicted satellite 
LFs of ICG2s by the 2015 L-Galaxies model, and the red shaded region associated with it shows the 
scatter, i.e. 16th 84th percentiles in the LFs of individual systems. {\bf Right:} Similar to the left 
plot, but shows the satellite LFs of all, red and blue primary galaxies in pair (see the legend). The 
yellow curve is exactly the same in both plots. For red or blue central galaxies in pair, no restrictions 
are made to the colour of the other companion. Results in both plots are are based on DECaLS. The black 
solid histogram shows the cumulative LF for MW satellites after projection and with the same inner radius 
cut of $r_p>30$~kpc. Errorbars are calculated in the same way as Figure~\ref{fig:LFcum_MW}.
}
\label{fig:LFcum_MW_diffICG}
\end{figure*}
\begin{figure} 
\includegraphics[width=0.49\textwidth]{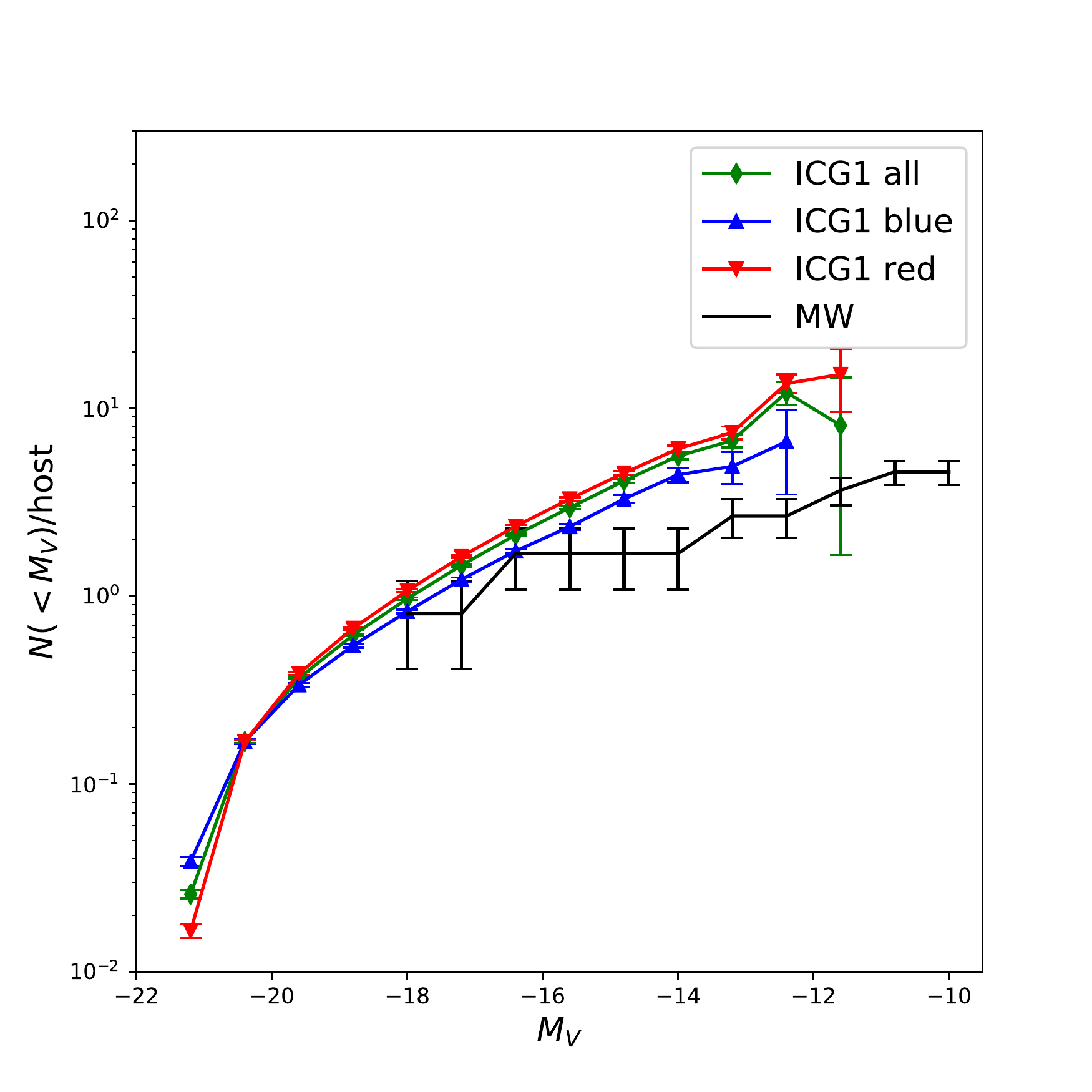}
\caption{Cumulative $V$-band LFs for satellite galaxies projected between 30 and 260~kpc to all, 
red and blue ICG1s with stellar mass in the range of $10.63<\log_{10}M_\ast/\msun<10.93$.
Errorbars are calculated in the same way as Figure~\ref{fig:LFcum_MW}.
}
\label{fig:LFcum_MW_ICGredblue}
\end{figure}
\begin{figure} 
\includegraphics[width=0.49\textwidth]{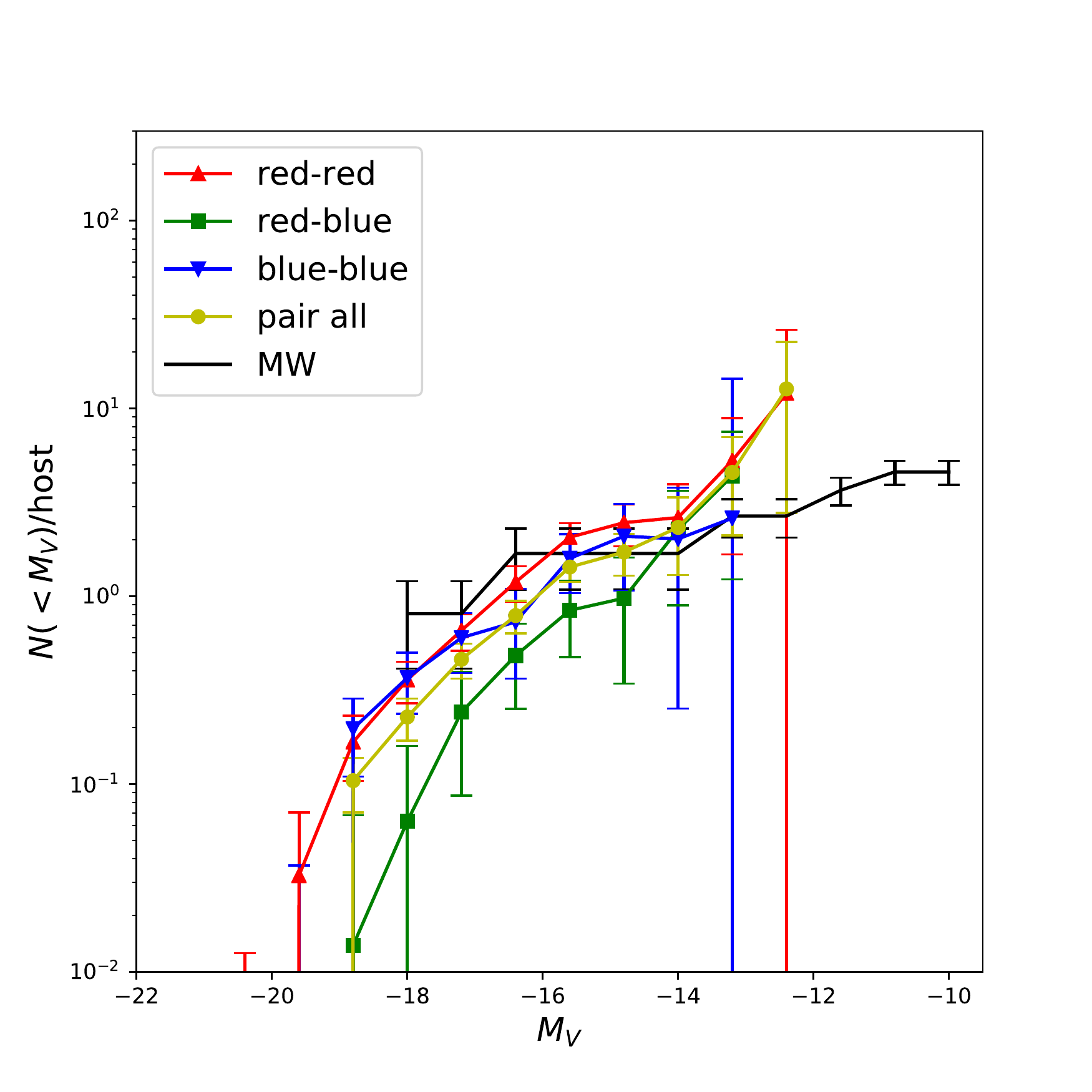}
\caption{Cumulative $V$-band LFs for satellite galaxies projected between 30 and 260~kpc to 
primary galaxies in pair. A few different colour combinations are adopted for the two primary galaxies,
i.e., ``red-red'', ``blue-blue'' and ``red-blue''. ``pair all'' is exactly the same as the yellow curve in 
Figure~\ref{fig:LFcum_MW_diffICG}. Primaries are required to have similar stellar mass to our MW, i.e., 
$10.63<\log_{10}M_\ast/\msun<10.93$, but the stellar mass of the other companion is not necessarily in 
this range and hence might not be used for this plot. All results are based on DECaLS. The black solid 
histogram shows the cumulative LF for MW satellite galaxies after projection and with the same inner 
radius cut of $r_p>30$~kpc. Errorbars are calculated in the same way as Figure~\ref{fig:LFcum_MW}. 
}
\label{fig:LFcum_MW_paircolor}
\end{figure}

According to the recent study by \cite{2015ApJ...806...96L}, the total stellar mass of our MW is about $(6.08\pm1.14)
\times 10^{10}\msun$. Centred on this value\footnote{The MW mass models provided by \cite{2011MNRAS.414.2446M} 
provide a slightly larger value of stellar mass, $6.43\pm0.63 \times 10^{10}\msun$, and as we will discuss later, 
this will not affect the conclusion of this paper.}, Figure~\ref{fig:LFcum_MW} shows satellite LFs around ICG1s 
in the stellar mass range of $10.63<\log_{10}M_\ast/\msun<10.93$. Results based on HSC, DECaLS and SDSS show very 
good agreement with each other, though HSC tends to show slightly higher amplitudes, and as have been discussed 
in the previous section, this is likely real (more details can be found in Appendix~\ref{app:blending}). 

All results tend to have higher amplitudes and steeper slopes than the MW satellite LF\footnote{The MW 
satellite count in each bin is not an integer. This is due to the projection effect (see Section~\ref{sec:method} 
for details).}, and the tension with the MW satellite LF is much more significant than the errorbars. 
The errors of the MW satellite LF are based on the 1-$\sigma$ scatters of 600 random projections, and thus 
the errors are smaller at the faint end, where the cumulative number of satellites becomes larger and less 
sensitive to the projected inner radius cut. Besides, it is very important to remember that the boot-strap 
errors cannot be used to quantify the intrinsic scatter.

With our statistical satellite counting and background subtraction methodology, it is difficult to 
directly infer the scatter, but we can estimate the scatter through the satellite LFs from numerical 
simulations. To select ICG1s in simulations, we project the simulation box along the $z$ direction, and 
assign each galaxy a ``redshift'' based on its coordinate and velocity along the projected direction. 
ICG1s can then be selected using the same isolation criteria as for real data. For each ICG1, its satellites
are counted within a 3-dimensional sphere\footnote{Satellite counts made in 3-dimensional coordinates 
can help to better capture the true scatter in satellite LFs. If the counts are made in projection, the 
scatter would also reflect the fluctuation in background counts, which we do not want to include. Also 
note satellite counts made in 3-dimensional coordinates would have slightly lower amplitudes than the 
counts made in projection, but the effect is only about 30\% (see Section~\ref{sec:sig} for more detailed 
discussions). } with radius of 260~kpc. Besides, we apply an inner radius cut of $r_p>30$~kpc perpendicular 
to the $z$ direction.

The green triangles and the shaded region in Figure~\ref{fig:LFcum_MW} show the averaged satellite
LF and the scatter for ICG1s selected from the 2015 L-Galaxies mock galaxy catalogue. At the bright end, 
the numbers of satellites around most primaries are either zero or one, and thus the scatter is not shown. 
The green triangles are slightly lower in amplitudes than the results based on real data, which is partly 
due to the difference between satellite counts made in projection and in 3-dimensional coordinates, but the 
overall agreement at fainter magnitudes is very good. The scatter, however, is very large. Despite the fact 
that the discrepancy between the MW satellite LF and our measurements of extra-galactic satellite LFs is 
significantly larger than the boot-strap errorbars, the discrepancy is much smaller than the scatter at 
$M_V<-14$, and is still marginally consistent with the scatter at $M_V\sim-13$. 

Figure~\ref{fig:LFcum_MW_fluxlimit} shows results based on HSC (left) and DECaLS (right), and we have tried a 
few different flux limits for each survey. Despite the difference in flux limits, the agreement is extremely 
good, indicating our satellite counting methodology works very well. The good agreement also proves the 
completeness of both surveys. However, we note that at $M_V>-15$, results based on fainter flux cuts tend 
to have slightly lower amplitudes than those with brighter cuts, though the differences are mostly smaller 
than the errors. This might indicate some small amount of incompleteness, due to, for example, failures in 
detecting low surface brightness satellites or mis-classifications of galaxies as stars. These small differences 
at the faint end, however, cannot violate the conclusions of this paper.

Now we start to compare MW satellites with the satellites around central primary galaxies selected in 
different ways. We will only show results based on DECaLS. This is because its footprint is larger than HSC 
and thus can include more galaxy pairs. Besides, the flux limit is deeper than SDSS, and thus the signal is 
better at faint ends. As we have explicitly checked, results based on HSC and SDSS are consistent. The left 
plot of Figure~\ref{fig:LFcum_MW_diffICG} shows the satellite LFs measured around ICG1s, ICG2s and primary 
galaxies in pairs. The green curve (ICG1s) is exactly the same as the one in Figure~\ref{fig:LFcum_MW}. As 
we have already investigated in Figure~\ref{fig:LFcum_diffICG}, the amplitude of satellite LFs around ICG2s 
is lower. The bright end cutoff also becomes more significant. As a result, the agreement with the MW satellite 
LF at the bright end is better, while the tension at fainter luminosities is smaller but still remains.

Similar to Figure~\ref{fig:LFcum_MW}, the prediction by the 2015 L-Galaxies model for satellite LF of ICG2s
agrees very well with the real data. At $M_V<-12$, the discrepancy between the MW satellite LF and the L-Galaxies 
model prediction is within the scatter. This is consistent with previous studies \citep[e.g.][]{2015MNRAS.454..550G,
2018MNRAS.476.1796S}, which also show that the mean/median satellite LFs predicted by simulations tend to have 
higher amplitudes than the MW satellite LF, while they still agree within the scatter. However, at the faint end 
($M_V>-12$), our measurements show that the difference between the averaged extra-galactic and the MW satellite 
LFs is larger than the scatter. In fact this is also revealed in Figure 2 of \cite{2015MNRAS.454..550G} and 
Figure 1 of \cite{2018MNRAS.476.1796S}, that the MW satellite LF is below those for most of the MW-mass systems in 
the simulation at $M_V>-12$. We will come back discussing the tension between extra-galactic and the MW satellite 
LF in Section~\ref{sec:sig}.

The yellow curve in Figure~\ref{fig:LFcum_MW_diffICG} shows the result for primary galaxies in pairs similar 
to the MW and M31. The LF shows stronger bright end cutoff. Compared with the MW satellite LF, the yellow 
curve has slightly lower amplitudes at the bright end, and higher amplitudes at $M_V>-15$. In fact, for 
primaries selected in these different ways, their averaged satellite LFs all tend to have steeper slopes 
and higher amplitudes at $M_V>-15$ than the MW satellite LF. 

It has been shown in previous studies that red isolated primaries have more satellites and are hosted by more 
massive dark matter haloes than blue primaries with the same stellar mass \citep[e.g.][]{2012MNRAS.424.2574W,
2014MNRAS.442.1363W,2016MNRAS.457.3200M,2019ApJ...881...74M}, and thus we further split our sample of primaries 
in pairs into red and blue populations. As has been mentioned in Section~\ref{sec:data}, this is achieved by 
a stellar mass dependent division of $^{0.1}(g-r)=0.065\log_{10}M_\ast/\msun+0.1$ over the colour-magnitude 
diagram of SDSS spectroscopic galaxies. The readers can find more details in \cite{2012MNRAS.424.2574W}. The 
satellite LFs of red and blue primaries in pair are shown as magenta and cyan curves in the right plot of 
Figure~\ref{fig:LFcum_MW_diffICG}. There is no significant difference between the satellite LFs around red 
or blue primary galaxies in pair, and the tension with the MW satellite LF is still present. Note that when 
cumulating satellite counts around one of the primary galaxies in the pair, we did not include any additional
requirements on the colour of the other primary. 

As a comparison, we also show in Figure~\ref{fig:LFcum_MW_ICGredblue} the satellite LFs around all, red and 
blue ICG1s. The number of red and blue MW-mass ICG1s (46352 and 33488) is significantly larger than the number 
of red and blue primary galaxies in pair (893 and 738). Thus the errorbars are much smaller. Red ICG1s in 
Figure~\ref{fig:LFcum_MW_ICGredblue} tend to have more satellites, which is consistent with conclusions in 
previous studies. 

While the small sample size and large errorbars probably have prevented us from tracking any significant
differences between red and blue primaries in pair, we now move on to investigate satellite LFs for galaxy 
pairs with different colour combinations. This is shown in Figure~\ref{fig:LFcum_MW_paircolor} for red-red 
(red curve), blue-blue (blue curve) and red-blue (green curve) pairs. We also overplot the yellow curve 
for all primary galaxies in pair from Figure~\ref{fig:LFcum_MW_diffICG}. The numbers of galaxies in red-red, 
blue-blue and red-blue pairs are 524, 341 and 766, respectively. There are two interesting features. First
of all, different colour curves all have steeper slopes than that of the MW satellite LF, and are lower in 
amplitudes at the bright end and higher in amplitudes beyond $M_V\sim -15$ than the MW satellite LF. Moreover, 
despite the large errorbars and the failure of tracking down any significant difference in the previous
Figure~\ref{fig:LFcum_MW_diffICG}, now we can clearly see that red-red and blue-blue pairs both have more 
satellites than red-blue pairs (see Section~\ref{sec:sigblueblue} for more details about the significance). 

Under the standard framework of cosmic structure formation, red galaxies formed early and grew fast at 
early stages, which then triggered feedback prohibiting late-time star formation activities, while their 
host dark matter halo and satellite populations keep growing through accretion. As a result, at fixed 
halo mass, the stellar mass of red galaxies tends to be smaller than that of blue galaxies, because red 
galaxies have stopped forming stars\footnote{Late-time dry mergers can contribute to the growth of stellar 
mass while still keep red galaxies quiescent, but considering the peak stellar to dark matter mass ratio 
\citep[e.g.][]{2010MNRAS.404.1111G}, the accreted stellar mass is much less than the amount of accreted 
dark matter. }. 

Therefore, red galaxies having more satellites can be explained under the standard cosmological model, and as 
have been checked by \cite{2012MNRAS.424.2574W} \citep[also see, e.g., ][]{2014MNRAS.442.1363W,2019ApJ...881...74M}, 
the trend can be reproduced by modern galaxy formation models. It is thus straight-forward to understand why 
red-red galaxy pairs can have more satellites than primaries in pair with other colour combinations. However, 
it is still puzzling why galaxies in blue-blue pairs tend to have more satellites than red-blue pairs in 
Figure~\ref{fig:LFcum_MW_paircolor}, given the fact that blue isolated galaxies do not show such a trend. 
In Section~\ref{sec:disc}, we provide more discussions, including a comparison with the prediction by 
Illustris TNG-100.

\section{Discussions}
\label{sec:disc}

\subsection{Comparison with Illustris TNG-100}
\label{sec:tng}
\begin{figure} 
\includegraphics[width=0.49\textwidth]{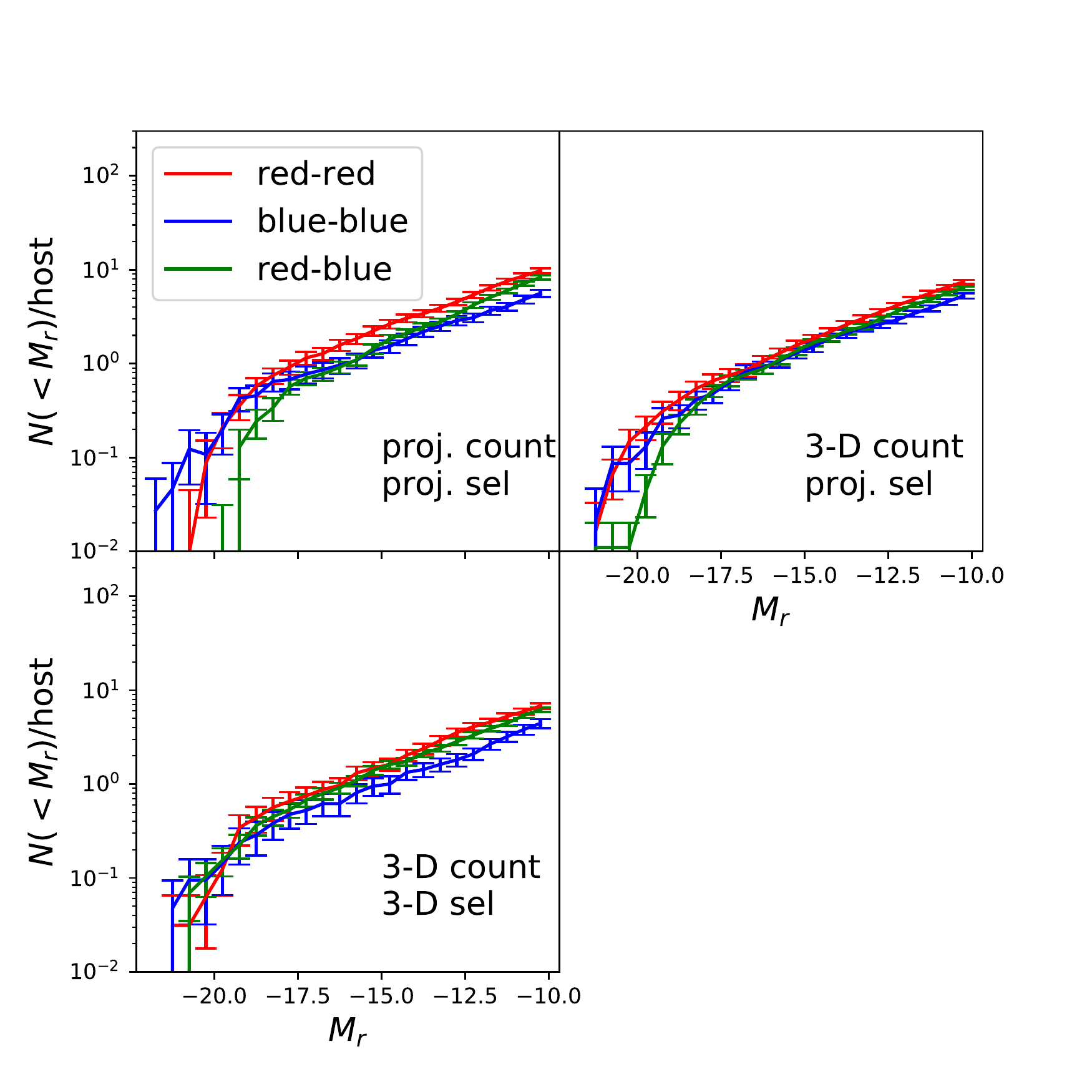}
\caption{Cumulative $r$-band LFs for satellite galaxies around MW-mass primary galaxies in pair from 
Illustris TNG-100. {\bf Top Left:} Primaries are selected in projection, and satellite counts are made in 
projection (30~kpc~$<r_p<$~260~kpc), with statistical fore/background subtractions. {\bf Top Right:} Primaries 
are selected in projection, and satellite counts are made in 3-D (30~kpc~$<r<$~260~kpc). {\bf Bottom Left:} 
Primaries are selected in 3-D, and satellite counts are made in 3-D as well. Errorbars are the 1-$\sigma$ 
scatters of 100 boot-strap subsamples. 
}
\label{fig:LFcum_MW_paircolor_tng}
\end{figure}

\begin{figure} 
\includegraphics[width=0.49\textwidth]{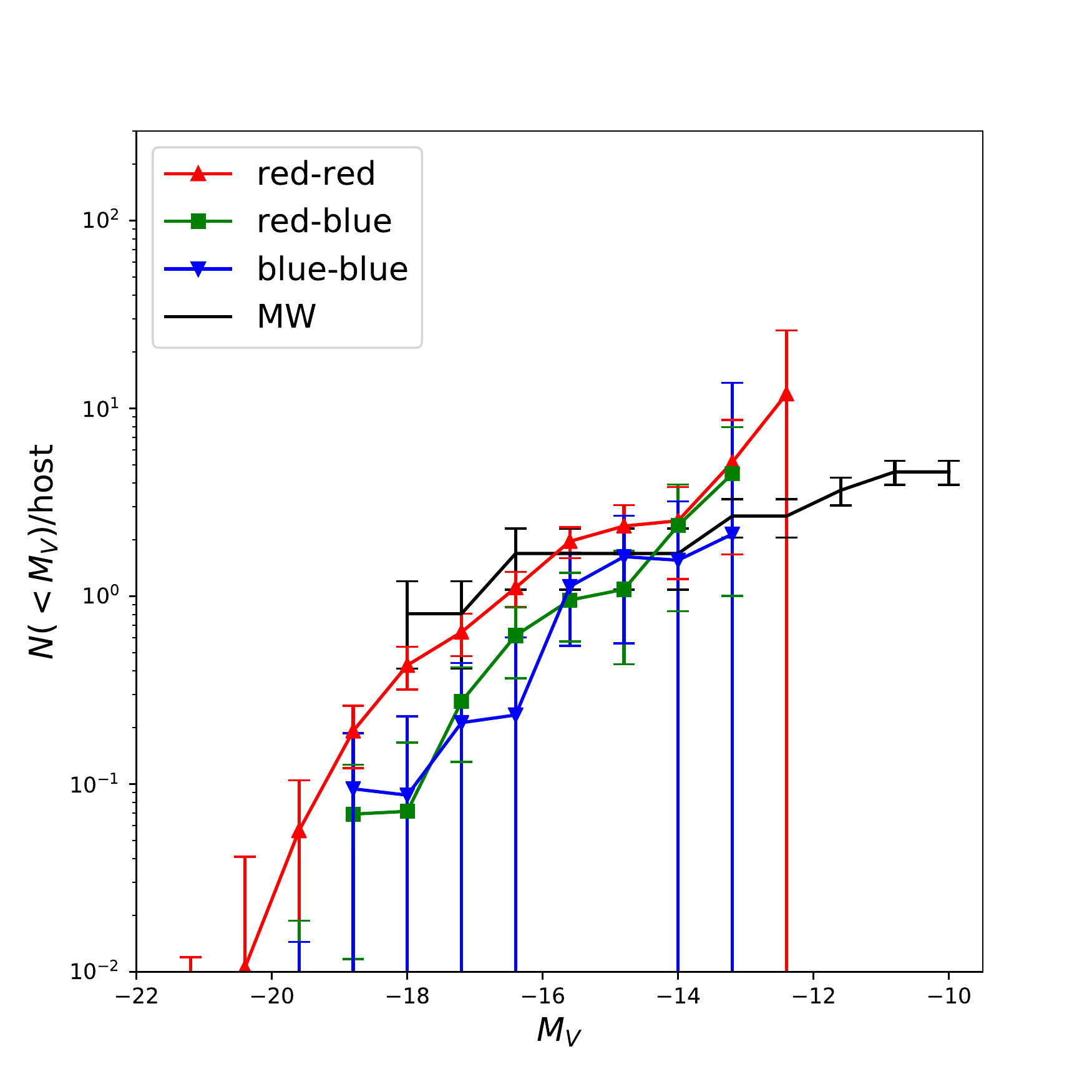}%
\caption{Cumulative $V$-band LFs for satellite galaxies projected between 30 and 260~kpc to MW-mass 
primary galaxies in pair. Galaxy pairs are selected in the same way as those in previous figures, 
but the flux limit is chosen as $r<16.7$. Black solid histogram shows the cumulative LF for detected 
MW satellite galaxies after projection and with the same inner radius cut of $r_p>30$~kpc. Errorbars 
are calculated in the same way as Figure~\ref{fig:LFcum_MW}. 
}
\label{fig:LFcum_MW_paircolor_test}
\end{figure}

We have shown that the averaged satellite LF of blue-blue primary galaxies in pair have higher amplitudes 
than that of primaries in pair with red-blue colour combinations. Here we start our investigation by checking 
whether modern numerical simulations can reproduce such a trend. We use the publically released Illustris 
TNG-100 halo and subhalo catalogues for the analysis. We choose to use TNG-100 instead of the L-Galaxies
model for our purpose here, because the colour distribution predicted by TNG-100 is in better agreement with 
real data, and there are not enough blue-blue MW-mass galaxy pairs in the 2015 L-Galaxies model.

To select primary galaxies in pair and in analogy to those in SDSS/DR7, we project the simulation box along $x$, $y$ 
and $z$-directions, and each galaxy can be assigned a ``redshift'' based on its coordinate and velocity along the 
projected direction. Galaxy pairs are then selected with exactly the same isolation criteria as those introduced 
in Section~\ref{sec:data}. Besides, we also select galaxy pairs in 3-dimensional coordinates, by applying the 
selection along the direction perpendicular to the line of sight to the 3-dimensional distance separations of 
galaxies in the simulation. We denote the two selections by ``proj. sel'' and ``3-D sel''. 

Satellites counts are made in projection or in 3-dimensional coordinates. The counts made in projection is analogous 
to real observation. For each projection, we first count companions projected within a cylinder, including those 
fore/background galaxies. The fore/background counts are then subtracted statistically through the counts around 
random points. The counts of all primaries and in all three projected directions ($x$, $y$ and $z$) are cumulated 
and averaged in the end. On the other hand, when counting satellites in 3-dimensional coordinates, we simply draw 
a sphere centred on each primary and count companions within the sphere. Satellites counted in the two different 
ways are denoted as ``proj. count'' and ``3-D count''. 

The satellite LFs are calculated directly from the absolute magnitudes of galaxies in TNG-100. We did not construct 
and use light-cone mock catalogues, and thus the direct projection of simulation box cannot fully represent the 
depth in background counts of a given flux limit. However, as we have very carefully checked and explicitly shown in 
the Appendix of \cite{2012MNRAS.424.2574W}, results based on directly projecting the simulation box and based 
on full light-cone mock catalogues are very similar to each other, though the latter is nosier due to its much 
larger background level. 

Figure~\ref{fig:LFcum_MW_paircolor_tng} shows the satellite LFs of primary galaxies in pair from TNG-100. Similar 
to Figure~\ref{fig:LFcum_MW_paircolor}, we try different colour combinations for the galaxy pair. The methods of 
primary selection and satellite counting are indicated by the text in each panel. Note that due to the resolution 
limit of TNG-100, the satellite LFs beyond $M_V\sim -16$ tend to be flattened, but we think this will not affect 
our comparison unless the satellite distributions around red and blue primaries are affected differently. 

In all three panels of Figure~\ref{fig:LFcum_MW_paircolor_tng}, satellite LFs around blue-blue galaxy pairs 
do not show higher amplitudes than other colour combinations at $M_r>-17.5$. At the bright end, the blue 
curves tend to show higher or comparable amplitudes as the green curve, which is more similar to what we 
see in real data. Except for the bright end, TNG-100 does not reproduce similar trends as the real observation. 

When the selection is made in projection, the number of primary galaxies in red-red, blue-blue and red-blue 
pairs are about 61, 46 and 91 (for $x$, $y$ and $z$ directions added together), out of which 53, 44 and 89 
are true halo central galaxies, respectively. The numbers are small, but is already enough for us to check
the trend, and at least in the simulation, the purity of galaxy pairs is high, and our results seem unlikely 
to have been affected by contamination of satellites in our sample of primary galaxies. 

\subsection{Why are there more satellites around blue-blue primary galaxy pairs?}
\label{sec:sigblueblue}

We estimate the significance of the difference between satellite LFs around blue-blue and red-blue galaxy
pairs by assuming a $\chi^2$ distribution
\begin{equation}
   \chi^2= 
    \sum_{i,j} D_i\times [C^{-1}]_{ij}\times D_j, 
    \label{eqn:significance}
\end{equation}
where $D_{i}$ means the difference between the $i$-th data point of the blue and green curves in Figure~\ref{fig:LFcum_MW_paircolor}.
$C^{-1}$ is the inverse of the covariance matrix, and is contributed by the covariance of both measurements, i.e.,
$C=C_\mathrm{blue-blue}+C_\mathrm{red-blue}$. The estimated significance is 2.70-$\sigma$. If the detection is genuine 
and robust, some new physical mechanisms beyond the merging origin of satellites under the standard cosmological context 
have to be proposed, for example, tidal dwarf galaxies. 

It has been discovered and reported as the 1-halo ``Galactic Conformity'' phenomenon that blue galaxies also 
tend to have bluer satellites with stronger star formation and more cold gas supplies, mainly because they 
are hosted by less massive dark matter haloes than red galaxies with the same stellar mass\citep[e.g.][]{
2006MNRAS.366....2W,2006ApJ...638L..55Y,2010MNRAS.409..491K,2012MNRAS.424.2574W}. The richer cold gas supplies 
seem to support the formation of tidal dwarf galaxies. However, it is hard to explain why in Figure~\ref{fig:LFcum_MW_diffICG}, 
if we do not include any restriction on the colour of the other primary galaxy in the pair, blue primaries do 
not have more satellites. It is thus still puzzling why there are more satellites around blue-blue galaxy pairs 
than red-blue galaxy pairs in real data. The mechanism might be related to the large scale environment, 
based on the fact that we only see such signals when both galaxies in the pair are blue. 

We have also checked the robustness of our results by repeating the calculation using galaxy pairs selected with 
different criteria. In the previous figures, galaxy pairs are selected by requiring that, centred on the middle 
point of the pair, all other companions projected within 800~kpc and within seven times virial velocity of the 
more massive galaxy in the pair and along the line of sight should be at least one magnitude fainter. We have tried 
to vary the projected separation to 1,500~kpc and the magnitude gap to 0.5. With the new selection, the sample size 
is decreased by about 1/3, and the measurements still show that there are more satellites around blue-blue galaxy 
pairs than red-blue pairs, but the significance drops to 1.79-$\sigma$. 

However, when we fix the selection to be the same as that of Figure~\ref{fig:LFcum_MW_paircolor}, but change the flux 
limit from $r<17.7$ to $r<16.7$, we fail to see significant difference between the amplitudes of blue and green curves. 
This is shown in Figure~\ref{fig:LFcum_MW_paircolor_test}. Thus the detection does not seem to be robust\footnote{We 
have carefully checked that our other conclusions are robust against changes in the flux limit of primaries.} against the 
variation in the flux limit or redshift range of primaries, as a brighter flux limit leads to a lower redshift range. 
This indicates our results are likely affected by large cosmic variations for the small number of galaxy pairs at low 
redshifts. Therefore, we avoid drawing a strong conclusion that blue-blue galaxy pairs have more satellites than red-blue 
pairs. Our results await more detailed follow-up studies by looking at, for example, galaxies pairs, their satellites and 
the evolution at higher redshifts.

\subsection{Is our MW special?}
\label{sec:sig}

As have been introduced, a few so-called challenges to the standard cosmological model were claimed, based on the observed 
properties of satellite galaxies in our MW \cite[e.g.][]{1999ApJ...522...82K,1999ApJ...524L..19M,2011MNRAS.415L..40B}. 
However, conclusions relying on the population of satellites in a single host galaxy might be unfair, as we don't know 
whether our MW is typical among galaxies of its type. This is why we looked at extra-galactic satellite systems in this 
study. Our investigations show that the averaged satellite LFs around MW-mass central primary galaxies selected in different 
ways tend to have more similar amplitudes to the MW satellite LF at the bright end, but the slopes at fainter magnitudes are 
significantly steeper, and the amplitudes beyond $M_V\sim-15$ are higher. 

Explicitly, at the bright end, ICG1s and ICG2s on average tend to have comparable number of satellites as our MW, 
while primary galaxies in pair tend to have a bit stronger bright end cutoff and lower amplitudes at $M_V<-16$ than 
the MW satellite LF. The average number of satellites with $M_V<-16$ are about 1.5 and 2.5 around ICG1s and ICG2s.
This is consistent with our MW, which has the LMC and SMC. It seems our results at the bright end are not quite 
consistent with the estimates by \cite{2011ApJ...733...62L}, in which the probability for isolated galaxies with 
luminosities similar to our MW and having two satellites about $\Delta M_V=2$ and $=4$ magnitudes fainter is 
reported to be only 3.5\%.

\cite{2011ApJ...733...62L}, however, adopted slightly more strict isolation criteria to select central primaries, 
and they counted satellites within a projected distance of 150~kpc. Their sample of MW-like primaries are selected 
according to luminosity. To ensure a fair comparison, we repeat our analysis by adopting the same selection of 
primaries as \cite{2011ApJ...733...62L}, and we use primaries with $-21.05<M_V<-20.75$ (or $M_{^{0.1}r}\sim 
-21.2$) and the same projected separation of 150~kpc to count satellites. Compared with ICG1s, the amplitude 
indeed drops, but the cumulative LF is still comparable to that of ICG2s at $-17<M_V<-15$, i.e., there are on 
average $\sim$1.5 such bright satellites with $M_V<-16$.

The fraction of dark matter haloes smaller than $M_{200}\sim1.25 \times10^{12}\msun$ and hosting both LMC and 
SMC-like satellites is also reported to be very low\footnote{\cite{2011MNRAS.414.1560B} looked for LMC-like 
subhaloes by also using the distance and orbital velocity information, which is responsible for the low 
possibility, because of the short travel time spent by the LMC at its current location, which is close to the 
orbital pericenter. In this study, we did not include any distance or velocity selection. } by \cite{2011MNRAS.414.1560B}, 
while the probability for more massive dark matter haloes hosting LMC and SMC-like satellites is significantly 
increased. According to recent studies, the virial mass of our MW is about $M_{200} \sim1\times10^{12}\msun$ 
\citep[e.g.][]{2019MNRAS.484.5453C,2020ApJ...894...10L}, and the readers can check \cite{2020SCPMA..63j9801W} 
for a review. Thus although the predicted probability of hosting both LMC and SMC-like satellites by numerical 
simulations is rare, it actually happens with our MW, unless the best-fit virial mass of our MW is under-estimated. 
On the other hand, the fact that we found 1.5--2.5 satellites with $M_V<-16$ perhaps indicate that the mean 
host halo mass for other MW-mass\footnote{Having similar stellar mass as the MW, not halo mass.} central 
primaries is larger than that of our MW. This likely implies that our MW deviates from the mean stellar 
mass versus halo mass relation of other galaxies, which have been pointed out in previous studies 
\citep[e.g.][]{2015MNRAS.454..550G,2020MNRAS.494.4291C}.

\cite{2011MNRAS.417..370G} and \cite{2012ApJ...760...16J} both reported a factor of two less extra-galactic 
satellites than the average of MW and M31 satellites with $M_V<-15$, which seem to be inconsistent with our 
results\footnote{Note the satellite LF measured by \cite{2012ApJ...760...16J} is significantly higher in 
amplitude than \cite{2011MNRAS.417..370G} in all luminosity bins of both primaries and satellites, despite 
the fact that they reported consistent results when comparing with MW and M31 satellites using primaries 
with $M_V\sim-21$. The inconsistencies in their two brighter primary bins of $M_r\sim-23$ and $M_r\sim-22$ 
are at least partly due to the difference in background subtraction, that \cite{2011MNRAS.417..370G} adopted 
a local background estimated from an outer annulus ring. However, the radius chosen for the annulus ring is 
well below the virial radius, which can result in significant over-subtractions. The disagreement in the 
faintest primary bin ($M_r\sim-21$), as checked through very detailed personal communications and careful 
one-to-one comparisons with Guo et al. (results not published), is mainly because both \cite{2012ApJ...760...16J}
and \cite{2012MNRAS.424.2574W} adopted K-corrections to $z=0.1$, while \cite{2011MNRAS.417..370G} adopted 
K-corrections to $z=0$. This results in a difference of $\sim$0.2--0.3 in $r$-band absolute magnitude, and 
thus with the seemingly same absolute magnitude bin, in fact different primaries were picked up. Note at 
the massive end, a small change in the stellar mass or luminosity of galaxies can lead to very quick changes 
in the host halo mass \citep[e.g.][]{2010MNRAS.404.1111G} or satellite abundance. For MW-luminosity primaries, 
the annulus ring adopted by \cite{2011MNRAS.417..370G} is larger than the virial radius, and as long as both 
studies did correct conversion from $M_{^{0.1}r}$ or $M_r$ to $V$-band magnitudes, their comparison with MW 
and M31 satellites should have consistent results.}. This is, however, mainly because the satellite LF averaged 
over MW and M31 \citep[][]{2008ApJ...686..279K} is boosted by the number of M31 satellites. M31 has 6 satellites 
with $M_V<-15$, while our MW only has two. This reflects the large diversity in the satellite LFs of different 
galaxies, and in the next subsection we will present the averaged satellite LF of more MW-mass systems in the 
Local Volume. Also note both \cite{2011MNRAS.417..370G} and \cite{2012ApJ...760...16J} detected $\sim$2 satellites 
with $M_V<-15$ of MW-luminosity primaries on average, consistent with our results. Given our careful analysis 
in this paper and the consistency and cross checks with \cite{2011MNRAS.417..370G} and \cite{2012ApJ...760...16J}, 
we think our results are robust.

The other four satellites contributing to the MW satellite LF between $M_V\sim-14$ and $M_V\sim-10$ are Fornax, Leo I, 
Sculptor and Sagittarius I, among which Sagittarius I is always excluded due to the 30~kpc inner radius cut. The MW 
satellite LF contributed by these objects has a much shallower slope than those extra-galactic satellite systems, which 
leads to significantly lower amplitudes at $-15<M_V<-12$. Though faint satellite candidates of the MW are continuously 
being discovered \citep[e.g.][]{2018PASJ...70S..18H,2019PASJ...71...94H}, and future surveys, such as LSST, are 
predicted to increase the number of faint MW satellites by a factor of two to ten, depending on model assumptions 
\citep{2018MNRAS.479.2853N,2019ARA&A..57..375S}, the number of such bright MW satellites with $M_V<-10$ is very 
unlikely to be significantly increased to match our measurements. Note the numbers of satellites around ICG1s and 
ICG2s with $M_V<-13$ are about ten and five, respectively. In addition, the extrapolated number of satellites with 
$M_V<-10$ around ICGs through the best-fit double Schecter function is about 33. Even if there is one more 
such bright satellite which sits behind the disc and is not yet discovered, the significance level can drop, but 
the discrepancy at the faint end cannot vanish.

The readers may wonder whether our satellite counts made in projection with statistical background subtraction 
may have been over-estimated, compared with the counts made in 3-dimensional coordinates. This is true because 
along the line-of-sight direction, the correlation between halo central galaxies and distant companions can persist 
out to very large distances, and companions with line-of-sight separations larger than the halo virial radius can 
still contribute to the signal. This explains why in Figure~\ref{fig:LFcum_MW_paircolor_tng}, the satellite LFs in 
the top left panel show slightly higher amplitudes than those in the top right panel. However, as have been carefully 
investigated in the Appendix of \cite{2012MNRAS.424.2574W}, by comparing results based on 3-dimensional coordinates 
and a full light-cone mock catalogue of galaxies, satellite counts made in projection are $\sim$30\% more than 
those made in 3-dimensional coordinates for MW-mass primaries, and the fraction is almost independent of the 
luminosity of satellites. This is thus far from enough to explain the difference between the averaged extra-galactic 
satellite LFs and the MW satellite LF we see in this study.

%The significance of the inconsistency between extra-galactic and the MW satellite LFs can be estimated through 
%Equation~\ref{eqn:significance} as well. Now $D_{i}$ is the difference between the $i$-th data point of the 
%averaged extra-galactic satellite LF and that of the MW satellite LF. The estimated significance values for central 
%primary galaxies selected in different ways are provided in Table~\ref{tbl:significance}. 

%%%%%%%%%%%%%%%%%%%%%%%%%%%%%%%%%%%%%%%%%%%%%%%%%%%
\begin{table}
\caption{The probabilities quantifying how common is the MW satellite LF compared with the satellite LFs 
of ICGs, based on the assumption of the Poisson distribution ($P_\mathrm{Poisson}$) or on the predictions
by the 2015 L-Galaxies model ($P_\mathrm{sim}$). }
\begin{center}
\begin{tabular}{lcc}\hline\hline
central primary & $P_\mathrm{Poisson}$ (30\% decrease) & $P_\mathrm{sim}$\\ \hline
ICG1 & $<$0.1\% & 1.2\% \\
ICG2  & $<$0.1\% & 1.5\% \\
\hline
\label{tbl:significance}
\end{tabular}
\end{center}
\end{table}
%%%%%%%%%%%%%%%%%%%%%%%%%%%%%%%%%%%%%%%%%%%%%%%%%%%

While the averaged extra-galactic satellite LFs show significant tensions with MW satellites, it is important to
quantify how common is our MW compared with other MW-mass systems by accounting for the intrinsic scatter. We cannot 
directly calculate the satellite LF for each individual primary galaxy with our method. Instead, we choose to estimate 
the percentile fractions of extra-galactic satellite systems which are more atypical than our MW with two alternative
approaches: i) assuming the underlying distribution of individual systems is Poissonian; ii) investigating 
the underlying scatter in satellite LFs using numerical simulations. 

In the first approach, we assume that the averaged extra-galactic satellite LFs are representative of the underlying 
true LF, and that the MW satellite system is a random draw from the parent distribution, which is Poissonian. We can 
estimate the probability in each luminosity bin assuming the Poisson distribution as

\begin{equation}
    P(k)=\frac{\lambda^k\exp{(-\lambda})}{k!},
\end{equation}
where $\lambda$ is the average number of extra-galactic satellites in a given luminosity bin, and $k$ is the number 
of MW satellites in the same bin. $k$ is required to be integers. The number of MW satellites might not be integers after 
accounting for the projection effect by averaging over 600 random directions for projection, and thus we calculate the 
probability for each direction at first and take the median probability in the end. The MW satellite counts are either 
zero or one in each differential luminosity bin. If assuming the satellite counts in different luminosity bins are 
independent from each other, i.e., ignoring the error correlation, we can estimate the joint probability of MW satellites 
under the assumption of Poisson distribution as
\begin{equation}
    P_\mathrm{total}=\prod_i P.CDF(k_i),
\end{equation}
where $k_i$ refers to the satellite counts in the $i$-th luminosity bin, which are either zero or one, and $CDF$ means 
the cumulative probability function value of $P(k)$ up to $k_i$. 

We then generate $10^5$ realisations by drawing from the random Poisson distribution of each magnitude bin, again assuming 
the averaged extra-galactic satellite LFs are the truth. The fractions of realisations that have a smaller probability than  
$P_\mathrm{total}$ are provided in Table~\ref{tbl:significance} for ICG1s and ICG2s, which we denote 
as $P_\mathrm{Poisson}$. $P_\mathrm{Poisson}$ are much smaller than 0.1\%. 
In this analysis, we have extrapolated the differential satellite LFs of ICG1s and ICG2s down to $M_V=-10.4$, based on the 
best-fit double Schecter functions. Note the data points in the three faintest bins are not used for the extrapolation 
because they are noisy, and the best-fit values are used instead to estimate  $P_\mathrm{Poisson}$. We have also manually 
decreased the amplitude of the satellite LFs for ICG1s and ICG2s by 30\%, to account for the difference between the global 
background subtraction and local or annular background subtractions (the over-estimates of satellite counts due to the 
correlated signal out to large distances, as discussed above).

$P_\mathrm{Poisson}$ depends on the extrapolations to the faint end, and as have been pointed out in previous studies, even 
at fixed halo mass, the true intrinsic scatter in satellite LFs can be larger than Poisson errors \citep[e.g.][]{2010MNRAS.406..896B,
2014MNRAS.445.1820C}. Thus in the second approach, we try to estimate the scatter and percentile fraction based on the satellite LFs 
from the 2015 L-Galaxies model. We choose to use L-Galaxies instead of TNG-100, because TNG-100 does not have enough resolution beyond 
$M_V\sim-16$. 

The predicted satellite LFs around ICG1s and ICG2s, and the associated scatters, have been shown in Figures~\ref{fig:LFcum_MW} 
and \ref{fig:LFcum_MW_diffICG}. Note instead of counting satellites in projection, the satellite counts are made within 3-dimensional 
spheres with radius of 260~kpc, to include the true intrinsic scatters without contamination by fluctuations in background counts. 
This is analogous to how MW satellite counts are made, and thus the L-Galaxies predictions can be directly compared with the MW 
satellite LF, without accounting for the slight over-estimates of satellite counts due to projection and global background subtraction. 
Also note the satellite LFs of ICGs based on real data, after being manually decreased by 30\% in amplitude, agree very well with the 
prediction by the L-Galaxies model, and thus we choose to ignore the small residual difference between the real data and simulation 
predictions.

At a given luminosity bin, $i$, we can estimate the fraction of systems in the model which have less or equal number of satellites 
than our MW, $f_i$. Again by assuming the satellite counts in each differential luminosity bins are independent from each other, the joint 
probability can be estimated as

\begin{equation}
    f_\mathrm{total}=\prod_i{f_i}.
\end{equation}

Then for each system in the model, we can estimate the same probability. The fractions of systems which have lower such 
probabilities than that of our MW are provided in Table~\ref{tbl:significance} for ICG1s and ICG2s, which we denote as 
$P_\mathrm{sim}$. $P_\mathrm{sim}$ values are 1.2\% and 1.5\% for ICG1s and ICG2s, respectively, which are still quite low
but are more likely to happen. Such low possibilities, in fact, are dominated by the few faintest luminosity bins. If we stop 
at the second faintest magnitude bin, $P_\mathrm{sim}$ would be 6.9\% and 8.0\% for ICG1s and ICG2s. Similarly, if stopping 
at the third faintest bin, $P_\mathrm{sim}$ would be 11.1\% and 12.7\% for ICG1s and ICG2s. If only including measurements 
at $M_V<-14$, $P_\mathrm{sim}$ can be as high as 24\% and 27\%. This is consistent with Figures~\ref{fig:LFcum_MW} and 
\ref{fig:LFcum_MW_diffICG}, i.e., the tension and the significance are dominated by the measurements at the faint end. 

The estimated probabilities based on the L-Galaxies model seem to suggest that although our MW system can be predicted 
by numerical simulations, it is not very common. At $-12<M_V<-10$, the number of satellites in our MW is less than those in 
most of the systems in numerical simulations. As have been mentioned in Section~\ref{sec:data}, the completeness of ICG1s 
is close to 90\%. This seems to suggest that among $\sim$90\% of central galaxies with stellar mass similar to our MW, 
only $\sim$1.2\% systems are more atypical than our MW. Of course, we should also bear in mind that the exact fraction 
can be model dependent. Despite the model dependence, our results are consistent with previous studies \citep[e.g.][]{
2015MNRAS.454..550G,2018MNRAS.479..284S} that at $M_V>-12$, the MW satellite LF is below most of the predicted satellite 
LFs of MW-mass systems by numerical simulations.

\begin{figure*} 
\includegraphics[width=0.9\textwidth]{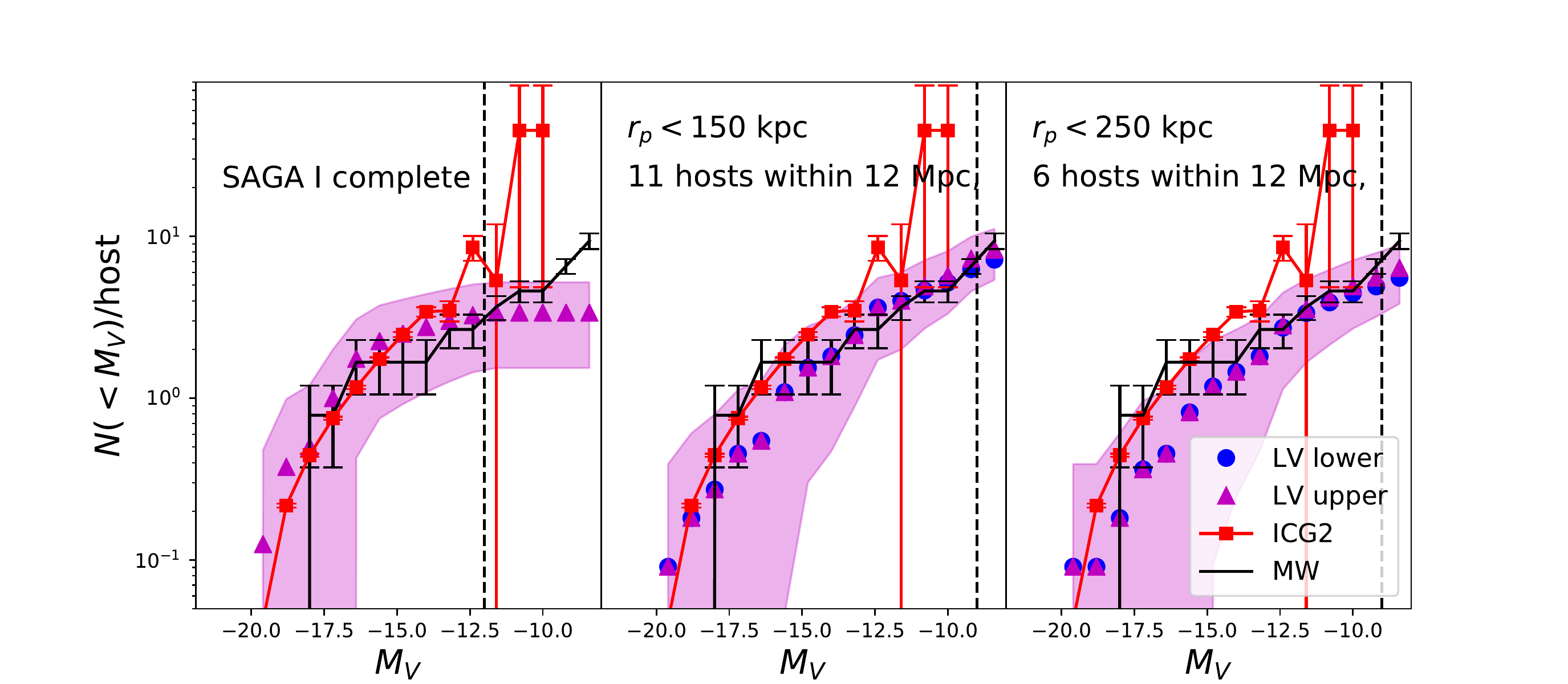}
\caption{A comparison between the MW satellite LF and the satellite LFs of other MW-mass galaxies within 40~Mpc in the 
Local Volume (LV).  {\bf Left:} The comparison with the averaged satellite LF of eight MW-mass galaxies between 20 and 
40~Mpc and from the SAGA survey (magenta triangles). Satellites are all confirmed and are complete out to $r_p\sim
300$~kpc to their central galaxies and down to $M_V\sim-12$. {\bf Middle:} The comparison with the averaged satellite 
LF of eleven galaxies within 12~Mpc. Satellites are complete out to $r_p\sim150$~kpc and down to $M_V\sim-9$. The blue 
dots and magenta triangles show the lower and upper limits when all unconfirmed satellites are treated as fore/background 
objects or as true satellites, respectively. {\bf Right:} Similar to the middle panel, but shows a subset of six systems 
whose satellites are complete out to $r_p\sim250$~kpc. In all panels, the black histograms are the MW satellite LF, whose 
errorbars are calculated in the same way as in previous plots. The magenta shaded regions are the Poisson errors 
for satellite LFs in the Local Volume, and this is not shown for blue dots, which have very similar errors. Red 
curves with errors are the averaged satellite LF of ICG2s, which is exactly the same as in Figure~\ref{fig:LFcum_MW_diffICG}, 
and the errorbars are the 1-$\sigma$ scatters of 100 boot-strap subsamples, which reflect the errors on the averaged LF. 
An inner radius cut of 30~kpc has been adopted for all. The vertical black dashed lines mark the magnitudes beyond which 
the LV satellites are incomplete. By going down to $M_V\sim-9$, five more MW satellites (Carina, Leo II, Sextans, Draco I 
and Ursa Minor) are included, in addition to those summarised in Table~\ref{tbl:MWsat}.
}
\label{fig:LFLV}
\end{figure*}

%%%%%%%%%%%%%%%%%%%%%%%%%%%%%%%%%%%%%%%%%%%%%%%%%%%
\begin{table*}
\caption{A sample of MW-mass galaxies in the Local Volume and their properties, including the distance, projected 
separation out to which satellites are complete, stellar mass, halo mass estimated from satellite dynamics or abundance 
matching, number of satellites between 30 and 150~kpc in projection, and the reference from which we take the information 
of their satellites.}
\begin{threeparttable}
\begin{center}
\begin{tabular}{lcccccccc}\hline\hline
Name      & $D$ & \tnote{1}\hspace{2mm}$r_\mathrm{comp}$ & $M_V$ &$M_\ast$ & $M_\mathrm{halo}$ & \tnote{2}\hspace{1mm}$N_\mathrm{sat}(<M_\mathrm{lim},$ & Source of   \\
      & [Mpc] & [kpc] &  & [$10^{10}M_\odot$] & [$10^{12}M_\odot$] & $30<r_p<150$~kpc) & satellites \\ \hline
\tnote{3}\hspace{4mm}M51       & 8.6 & 150 & $-$21.38 & $\sim$6 & --  & 0--3 & \cite{2020ApJ...891..144C,2020arXiv200602443C} \\
\tnote{3}\hspace{4mm}M64       & 5.3  & 300  & $-$20.20 & 4.9 & -- & 2 & in preparation \\
\tnote{3,8}\hspace{4mm}M81       & 3.69 & 250  & $-$21.10 & $\sim$5 & 4.9$\pm$1.4 & 13  & \cite{2013AJ....146..126C} \\
\tnote{3,8}\hspace{4mm}M101      & 6.52 & 300  &  $-$21.10 & $\sim$4 & 1.5$\pm$0.7 & 8 & \cite{2019ApJ...885..153B,2020ApJ...893L...9B} \\
                                 &      &      &           &         &             &   & \cite{2016AA...588A..89J} \\
                                 &      &      &           &         &             &   & \cite{2017ApJ...837..136D} \\
                                 &      &      &           &         &             &   & \cite{2020arXiv200602444C}  \\
\tnote{4}\hspace{4mm}NGC4565     & 11.9 & 150  & $-$21.80  & 7.6     & --          & 3--8 & \cite{2020ApJ...891..144C,2020arXiv200602443C} \\
\tnote{3,8}\hspace{4mm}NGC4258   & 7.2 &  150  & $-$20.94 & 5.1 & 3.2$\pm$1.0 & 5 & \cite{2020ApJ...891..144C,2020arXiv200602443C} \\
\tnote{5,8}\hspace{4mm}Centaurus A$\star$ & 3.77 & 200 & $-$21.04 & 8.13 & 6.7$\pm$2.1 & 15 & \cite{2019AA...629A..18M} \\
                                          &      &     &          &      &             &      & \cite{2007AJ....133..504K} \\
                                         &      &     &          &      &             &      & \cite{2019ApJ...872...80C} \\
                                          &      &     &          &      &             &      & \cite{2020arXiv200602443C} \\
\tnote{3}\hspace{4mm}NGC253    & 3.56 &  300  & $-$20.10 & 7.2 & -- & 1 & in preparation \\
\tnote{6,9}\hspace{4mm}NGC1023   & 10.4 &  200  & $-$20.90 & 7.7 & $\sim$6 & 10--12 & \cite{2020ApJ...891..144C,2020arXiv200602443C} \\
\tnote{3}\hspace{4mm}NGC5055   & 8.87 &  300  & $-$21.10 & 4.9 & -- & 7 & in preparation \\
\tnote{7}\hspace{4mm}NGC6744   & 8.95 &  300  & $-$21.62 & 6.6 & -- & 5--6 & in preparation \\
\tnote{10}\hspace{4mm}NGC5962   & 28.0 &  300  & -- & 3.31 & 1.35 & 1 & \cite{2017ApJ...847....4G} \\
\tnote{10}\hspace{4mm}NGC6181   & 34.3 &  300  & -- & 3.72 & 1.86 & 3 & \cite{2017ApJ...847....4G} \\
\tnote{10}\hspace{4mm}NGC5750   & 25.4 &  300  & -- & 3.39 & 1.20 & 1 & \cite{2017ApJ...847....4G} \\
\tnote{10}\hspace{4mm}NGC7716   & 34.8 &  300  & -- & 5.01 & 1.02 & 1 & \cite{2017ApJ...847....4G} \\
\tnote{10}\hspace{4mm}NGC1015   & 37.2 &  300  & -- & 3.72 & 1.12 & 1 & \cite{2017ApJ...847....4G} \\
\tnote{10}\hspace{4mm}PGC068743 & 39.2 &  300  & -- & 3.63 & 1.51 & 2 & \cite{2017ApJ...847....4G} \\
\tnote{10}\hspace{4mm}NGC2543   & 37.7 &  300  & -- & 4.37 & 1.07 & 1 & \cite{2017ApJ...847....4G} \\
\tnote{10}\hspace{4mm}NGC7541   & 37.0 &  300  & -- & 5.13 & 3.55 & 5 & \cite{2017ApJ...847....4G} \\
\hline
\label{tbl:LVhost}
\end{tabular}
\begin{tablenotes}
\item[1)] Projected separation out to which the observed satellites are complete.
\item[2)] We provide the number of satellites brighter than the limiting magnitude ($M_V\sim-12$ for SAGA galaxies and $M_V\sim-9$ for 
other more nearby galaxies) and projected between 30 and 150~kpc. In some systems, there are unconfirmed companions, and thus we provide 
the lower and upper boundaries. 
\item[3)] Stellar masses of M51, M64, M81, M101, NGC4258, NGC253 and NGC5055 are taken from the Spitzer Local Volume Legacy (LVL)
Survey \citep{2014MNRAS.445..899C}.
\item[4)] Stellar mass of NGC4565 is taken from the Spitzer Survey of Stellar Structure in Galaxies \citep{2010PASP..122.1397S}.
\item[5)] Stellar mass of Centaurus A$\star$ is taken from \cite{2014AJ....148...50K}.
\item[6)] Stellar mass of NGC1023 is estimated from the $K$-band absolute magnitude.
\item[7)] Stellar mass of NGC6744 is taken from \cite{2018PASA...35...15Y}.
\item[8)] Halo mass computed from orbital motions of satellites \citep{2014AJ....148...50K}.
\item[9)] Halo mass computed from satellite distribution \citep{2009MNRAS.398..722T}.
\item[10)] Halo mass calculated from the $k$-band absolute magnitude and abundance matching. 
\end{tablenotes}
\end{center}
\end{threeparttable}
\end{table*}
%%%%%%%%%%%%%%%%%%%%%%%%%%%%%%%%%%%%%%%%%%%%%%%%%%%

While our MW is the Galaxy which we can study in the most detail, it is not straight-forward to measure its 
exact position on the colour-magnitude diagram, due to the fact that we are within the MW and is difficult to 
directly observe its global properties, and the observation is often strongly affected by dust reddening. A  
study by \cite{2015ApJ...809...96L} picked up SDSS galaxies with similar stellar mass and star formation rates 
to our MW. After carefully correcting for Eddington bias and dust reddening, the photometric properties of these 
galaxies indicate that our MW locates at the ``green valley'', which is a region sparsely populated by galaxies 
between the bimodel red and blue clouds of galaxies in the colour-magnitude diagram. Though a few earlier studies 
provide bluer colour estimates, the measurement by \cite{2015ApJ...809...96L} is in good agreement with the novel 
measurement made by \cite{1986A&A...157..230V} using the Galactic background light taken by the Pioneer 10 spacecraft. 
Similar to our MW, the colour of M31 also locates at the green valley region \citep[e.g.][]{2011ApJ...736...84M}. 

\cite{2015ApJ...809...96L} estimated the rest-frame colour of our MW as $^0(g-r)=0.682_{-0.056}^{+0.066}$. If the 
K-correction is made to $z=0.1$, which is the standard adopted in this paper, the rest-frame colour of our MW is 
about $^{0.1}(g-r)=0.822$. Therefore, the colour of our MW indeed almost lies on the line of colour division for 
red and blue galaxies, i.e., $^{0.1}(g-r)=0.065\log_{10}M_\ast/\msun+0.1$. For our sample of primaries, we have 
repeated the calculation for primaries with $0.722<^{0.1}(g-r)<0.922$ and with concentration, $C=R_{90}/R_{50}$,
smaller than 2.6. Here the concentration parameter is adopted for a selection in morphology, so that the sample is 
dominated by low-concentration spiral galaxies. $R_{90}$ and $R_{50}$ are the 90\% and 50\% Petrosian radius, 
respectively. More details about the division by galaxy concentration can be found in \cite{2014MNRAS.443.1433D} 
and \cite{2019MNRAS.487.1580W}. The averaged satellite LF of primaries selected in this way, though quite noisy 
due to the small number of galaxies in the green valley region, still tends to have steeper slopes and higher 
amplitudes at the faint end than the MW satellite LF. 

A few early studies have investigated whether the global properties and scaling relations of our MW fits those of 
other galaxies, such as the Tully-Fisher relation \citep[e.g.][]{2006MNRAS.372.1149F}, the luminosity-velocity-radius 
relation \citep[e.g.][]{2016ApJ...833..220L}, star formation rate \citep[e.g.][]{2009A&A...505..497Y}, [Fe/H] 
abundance, angular momentum \citep[e.g.][]{2007ApJ...662..322H} and so on. Most of these studies report more than 
1-$\sigma$ of deviation from the main relation, and thus the global properties of our MW are also uncommon compared 
with the majority of other spiral galaxies. 

We conclude that the MW satellite LF is statistically atypical among other MW-mass galaxies. This is true unless 
the total stellar mass of our MW estimated by \cite{2015ApJ...806...96L}, as used in our analysis, is much higher 
than the true value, or the stellar mass of other primary galaxies estimated from their photometric colours is 
significantly wrong. Due to the difficulties of measuring the global properties of our MW from inside, 
we cannot rule out such possibilities, but so far this is the best estimate we can have. As we have mentioned 
before, another independent study by \cite{2011MNRAS.414.2446M} provides a slightly larger value of MW stellar 
mass, $(6.43\pm0.63) \times 10^{10}\msun$, which will lead to even higher amplitudes of extra-galactic satellite 
LFs at the faint end than that of our MW. 

Therefore, we comment that it is dangerous to use satellites within our MW to represent the entire satellite 
population around galaxies with similar properties, as our MW is special compared with not only the average 
properties of other MW-mass galaxies, but is also statistically uncommon if taking into account the scatter. 
Comparing the averaged satellite property distributions from numerical simulations only with our MW is unfair, and 
we cannot draw strong cosmological implications without quoting the statistical uncertainties. 

\subsection{Satellite luminosity functions in the local volume}

As we have mentioned in Section~\ref{sec:intro}, despite the fact that the satellite LFs of other host galaxies 
within 40~Mpc tend to show large diversities, it was found in previous studies that our MW satellite LF is 
typical among other MW-mass galaxies in the local Universe. Thus in this subsection, we further make a detailed 
comparison between the satellite LF of our MW and of other MW-mass galaxies in the Local volume (LV). 

A sample of 19 MW-mass galaxies in the LV is provided in Table~\ref{tbl:LVhost}. Eight systems are taken from 
the first stage paper of the SAGA\footnote{Very recently, the second stage SAGA results \citep{2020arXiv200812783M} 
have expanded the number of complete systems to 36.} survey \citep{2017ApJ...847....4G} out to $\sim$40~Mpc, and 
the satellites are complete down to $M_V\sim-12$. Eleven systems are within 12~Mpc, and the satellites are complete 
down to $M_V\sim-9$. Table~\ref{tbl:LVhost} includes a few rich systems, such as M81, Centaurus A$\star$ and NGC1023. 
Their dark matter haloes, as estimated from satellite dynamics, tend to be more massive than that of our MW 
($\sim 1\times 10^{12}M_\odot$, check \cite{2020SCPMA..63j9801W} for a review). Their stellar masses, on the other 
hand, are mostly within the mass range we adopt to select MW-mass galaxies ($10.63<\log_{10}M_\ast/\msun<10.93$), 
except for a few SAGA systems with $\log_{10}M_\ast<4\times10^{10}M_\odot$. The magnitude gaps between central 
primaries and the brightest satellites of these galaxy systems satisfy our selection of ICG2s. 

Figure~\ref{fig:LFLV} shows the comparison between the averaged satellite LFs of these LV systems and of our MW. 
Those observed by the SAGA survey and complete down to $M_V\sim -12$ are shown in the left panel. Those more 
nearby eleven systems within 12~Mpc are complete down to $M_V\sim -9$, while their satellites are complete 
out to different projected distances, $r_\mathrm{comp}$, but all systems are complete to 150~kpc. Thus in the 
middle panel we show the averaged satellite LF of all the eleven systems out to 150~kpc, and in the right panel, 
we show the averaged satellite LF for six out of the eleven systems, whose satellites are counted out to $r_p\sim$250~kpc.
We also overplot the averaged satellite LF of ICG2s. The averaged LV satellite LFs tend to be more similar to the 
MW satellite LF, which is consistent with the recently submitted second stage SAGA results \citep{2020arXiv200812783M}.

It seems the LF of satellites in our MW is more typical amongst the averaged satellite LF of MW-mass LV systems 
within 40~Mpc, but is more statistically uncommon amongst satellites of more distant MW-mass galaxies at redshift 
$z\sim0.1$. Such an inconsistency might be connected to the so-called ``local void'', that a large volume in the 
local Universe was reported to be deficient of galaxies \citep[e.g.][]{1988ngc..book.....T,2010Natur.465..565P}. 
The local void can be predicted by numerical simulations, but is not extremely common, as \cite{2014MNRAS.441..933X} 
reported that 11 out of 77 LG-like systems in the Millennium-II simulation have nearby low-density regions. 

\cite{2015ApJ...800..112G} reported that there are more satellites in filaments, and thus we can straight-forwardly 
think that central primary galaxies in over-dense regions tend to have more satellites. Similarly, we expect there 
are on average less satellites around central primaries in under-dense regions (or voids). Because the LV 
tends to be a ``void'' region, this can possibly explain why our MW satellites tend to be more typical in the LV, 
but the number of MW satellites with $M_V<-10$ tends to be less than the average of those around more distant 
galaxies. 

However, we note that our statistical satellite counting methodology is quite different from that for the LV 
satellites, which confirms satellites through the line-of-sight distances. In addition, though the magnitude gaps 
between primaries and their secondary satellites of the LV systems satisfy our selection of ICG2s, the projected 
and line-of-sight separations adopted in the selection of ICG2s might not exactly fit the condition of these LV 
systems. Thus although as have been checked by \cite{2012MNRAS.424.2574W} that the statistical satellite counts 
made in projection are $\sim$30\% more than those made in 3-dimensional coordinates for MW-mass primaries, 
more detailed comparisons of the two satellite counting methodologies, including a careful investigation on the 
selection effects of spectroscopically confirmed satellites in the LV, and possible differences in the selection 
of primaries are still necessary. We leave them to future studies. 

\section{conclusion}
\label{sec:concl}

In this study, we have measured the averaged satellite LFs of isolated central primary galaxies and galaxy pairs. 
Primary galaxies are selected from the SDSS spectroscopic Main galaxy sample. Satellites are constructed from 
the photometric source catalogues of HSC, DECaLS and SDSS. We compute the intrinsic luminosities of satellites 
following the method of \cite{2012MNRAS.424.2574W}, and fore/background source contaminations are statistically 
subtracted using the companion counts around random points. 

Despite the very different depth, resolution, footprint, mode of observation and steps of data reduction, satellite
LFs based on the three surveys agree very well with each other. Our measurements are also robust against 
the change in flux limits. We are not only able to extend previous measurements based on SDSS further down to the 
faint end by at most four magnitudes, but also we can measure the satellite LF around isolated central galaxies as
small as $9.2<\log_{10}M_\ast/\msun<9.9$ and $8.5<\log_{10}M_\ast/\msun<9.2$, which are smaller than most of the 
previous studies based on similar methods. Our measurements can thus reach isolated central galaxies as small as 
the LMC, which predict an average number of 3--8 satellites brighter than $M_V\sim-10$ of LMC-mass systems. 

We find both the bright end cutoff and the amplitude of satellite LFs over the whole luminosity range 
are sensitive to the magnitude gap between the central primary galaxy and its companions. A larger magnitude 
gap leads to more significant bright end cutoff by definition and also a decrease in the overall amplitude (or 
satellite abundance), indicating central primaries which have larger magnitude gaps than their secondary 
companions are hosted by less massive dark matter haloes. 

In previous studies, it was reported that red central galaxies tend to have more satellites and are hosted by 
more massive dark matter haloes than blue central galaxies with the same stellar mass \citep[e.g.][]{2012MNRAS.424.2574W,
2014MNRAS.442.1363W,2016MNRAS.457.3200M,2019ApJ...881...74M}. We fail to see such differences for red and blue 
primaries in pair, probably due to the small sample size and large errors, but by looking at galaxy pairs with 
different colour combinations, we find indications showing that galaxy pairs with red-red or blue-blue 
colour combinations have more satellites than red-blue galaxy pairs. However, such a trend for blue-blue pairs 
does not seem to be robust against changes in flux limits and might be affected by the sample variation at low 
redshifts, where the volume is small. And we fail to see similar trends in Illustris TNG-100. 

By selecting central primary galaxies having similar stellar mass to our MW ($10.63<\log_{10}M_\ast/\msun<10.96$), 
we find at the bright end ($M_V<-16$), isolated central galaxies have on average 1.5--2.5 satellites,
consistent with our MW. Satellite LFs around galaxies in pairs similar to the MW-M31 system tend to have 
slightly lower amplitudes at the bright end than the MW satellite LF.

We discover that the slopes of the averaged satellite LFs of both isolated central galaxies and galaxy pairs 
are steeper than that of MW satellites, and the amplitudes at $-15<M_V<-10$ are higher. The scatter 
of the satellite LFs predicted by numerical simulations can be quite large. Despite such a large scatter, 
the MW satellite LF at $M_V>-12$ still tends to be statistically uncommon. At such a bright magnitude 
range of MW satellites ($M_V<-10$), it is unlikely that there are other MW satellites remain undetected. 
%though previous studies have pointed out the possibility of incompleteness in bright and distant MW satellites 
%\citep[e.g.][]{2014MNRAS.439...73Y}

Therefore, we conclude that the LF of our MW satellites is atypical compared with those extra-galactic satellites
around MW-mass central primaries. Interestingly, a comparison with 19 MW-mass systems in the Local Volume ($<$40~Mpc) 
reveals that the MW satellite LF is more typical among other systems in the LV, possibly implying the LV is an under-dense 
region \citep[e.g.][]{1988ngc..book.....T,2010Natur.465..565P}. However, more detailed investigations on possible 
differences in the selection of primaries and the difference in satellite counting between systems in the nearby and 
the more distant Universe are still necessary in future studies.

Despite the much better agreement with LV systems, in order to have fair comparisons between real observations and 
theoretical predictions by modern numerical simulations, we can not use our MW to represent other MW-mass galaxies
in the Universe, and thus the observation of faint satellites in extra-galactic systems are crucial for proper 
cosmological implications. Future deep and wide photometric surveys, such as the LSST survey \citep{2008arXiv0805.2366I} 
and the Chinese Space Station Optical Survey Telescope (CSST) is very promising to further extend the study of 
extra-galactic satellites down to much fainter magnitudes, and our statistical approach of studying faint 
photometric satellites around spectroscopic central primaries can be straight-forwardly applied to these 
future observations.

\section*{Acknowledgements}
The Hyper Suprime-Cam (HSC) collaboration includes the astronomical communities of Japan and Taiwan, and Princeton University. The 
HSC instrumentation and software were developed by the National Astronomical Observatory of Japan (NAOJ), the Kavli Institute for 
the Physics and Mathematics of the Universe (Kavli IPMU), the University of Tokyo, the High Energy Accelerator Research Organization 
(KEK), the Academia Sinica Institute for Astronomy and Astrophysics in Taiwan (ASIAA), and Princeton University.  Funding was 
contributed by the FIRST program from the Japanese Cabinet Office, the Ministry of Education, Culture, Sports, Science and Technology
(MEXT), the Japan Society for the Promotion of Science (JSPS), Japan Science and Technology Agency  (JST), the Toray Science Foundation,
NAOJ, Kavli IPMU, KEK, ASIAA, and Princeton University.

This paper makes use of software developed for the Large Synoptic Survey Telescope. We thank the LSST Project for making their code
available as free software at  http://dm.lsst.org

This paper is based [in part] on data collected at the Subaru Telescope and retrieved from the HSC data archive system, which is 
operated by Subaru Telescope and Astronomy Data Center (ADC) at NAOJ. Data analysis was in part carried out with the cooperation 
of Center for Computational Astrophysics (CfCA), NAOJ.

The Pan-STARRS1 Surveys (PS1) and the PS1 public science archive have been made possible through contributions by the Institute 
for Astronomy, the University of Hawaii, the Pan-STARRS Project Office, the Max Planck Society and its participating institutes, 
the Max Planck Institute for Astronomy, Heidelberg, and the Max Planck Institute for Extraterrestrial Physics, Garching, The Johns 
Hopkins University, Durham University, the University of Edinburgh, the Queen’s University Belfast, the Harvard-Smithsonian Center
for Astrophysics, the Las Cumbres Observatory Global Telescope Network Incorporated, the National Central University of Taiwan, the 
Space Telescope Science Institute, the National Aeronautics and Space Administration under grant No. NNX08AR22G issued through the
Planetary Science Division of the NASA Science Mission Directorate, the National Science Foundation grant No. AST-1238877, the 
University of Maryland, Eotvos Lorand University (ELTE), the Los Alamos National Laboratory, and the Gordon and Betty Moore 
Foundation.

This project used data obtained with the Dark Energy Camera (DECam), which was constructed by the Dark Energy Survey (DES) 
collaboration. Funding for the DES Projects has been provided by the U.S. Department of Energy, the U.S. National Science 
Foundation, the Ministry of Science and Education of Spain, the Science and Technology Facilities Council of the United
Kingdom, the Higher Education Funding Council for England, the National Center for Supercomputing Applications at the 
University of Illinois at Urbana-Champaign, the Kavli Institute of Cosmological Physics at the University of Chicago, 
Center for Cosmology and Astro-Particle Physics at the Ohio State University, the Mitchell Institute for Fundamental 
Physics and Astronomy at Texas A\&M University, Financiadora de Estudos e Projetos, Fundacao Carlos Chagas Filho de Amparo,
Financiadora de Estudos e Projetos, Fundacao Carlos Chagas Filho de Amparo a Pesquisa do Estado do Rio de Janeiro, Conselho
Nacional de Desenvolvimento Cientifico e Tecnologico and the Ministerio da Ciencia, Tecnologia e Inovacao, the Deutsche 
Forschungsgemeinschaft and the Collaborating Institutions in the Dark Energy Survey. The Collaborating Institutions are
Argonne National Laboratory, the University of California at Santa Cruz, the University of Cambridge, Centro de Investigaciones
Energeticas, Medioambientales y Tecnologicas-Madrid, the University of Chicago, University College London, the DES-Brazil 
Consortium, the University of Edinburgh, the Eidgenossische Technische Hochschule (ETH) Zurich, Fermi National Accelerator 
Laboratory, the University of Illinois at Urbana-Champaign, the Institut de Ciencies de l'Espai (IEEC/CSIC), the Institut
de Fisica d'Altes Energies, Lawrence Berkeley National Laboratory, the Ludwig-Maximilians Universitat Munchen and the 
associated Excellence Cluster Universe, the University of Michigan, the National Optical Astronomy Observatory, the University
of Nottingham, the Ohio State University, the University of Pennsylvania, the University of Portsmouth, SLAC National Accelerator 
Laboratory, Stanford University, the University of Sussex, and Texas A\&M University.

BASS is a key project of the Telescope Access Program (TAP), which has been funded by the National Astronomical Observatories of 
China, the Chinese Academy of Sciences (the Strategic Priority Research Program ``The Emergence of Cosmological Structures'' Grant 
XDB09000000), and the Special Fund for Astronomy from the Ministry of Finance. The BASS is also supported by the External Cooperation
Program of Chinese Academy of Sciences (Grant 114A11KYSB20160057), and Chinese National Natural Science Foundation (Grant 11433005).

The Legacy Survey team makes use of data products from the Near-Earth Object Wide-field Infrared Survey Explorer (NEOWISE), which is 
a project of the Jet Propulsion Laboratory/California Institute of Technology. NEOWISE is funded by the National Aeronautics and Space 
Administration.

The Legacy Surveys imaging of the DESI footprint is supported by the Director, Office of Science, Office of High Energy Physics of the
U.S. Department of Energy under Contract No. DE-AC02-05CH1123, by the National Energy Research Scientific Computing Center, a DOE Office 
of Science User Facility under the same contract; and by the U.S. National Science Foundation, Division of Astronomical Sciences under
Contract No. AST-0950945 to NOAO.

WW is extremely grateful for very helpful suggestions by Marius Cautun.  WW also thank useful discussions with Naoki Yasuda, Jiaxin 
Han, Yaoyuan Mao, Chunyan Jiang, Quan Guo and Carlos Frenk. This work is supported by NSFC (12022307). WW gratefully acknowledge the 
support of the MOE Key Lab for Particle Physics, Astrophysics and Cosmology, Ministry of Education. This work is supported in part by World 
Premier International Research Center Initiative (WPI Initiative), MEXT, Japan, and MT is supported in part by JSPS KAKENHI Grant Numbers 
JP15H05887, JP15H05893, JP15K21733, and JP19H00677.

\section*{Data availability}
The data in plots of this paper can be shared on reasonable request to the corresponding author.

\bibliography{master}

%%%%%%%%%%%%%%%%%%%%%%%%%%%%%%%%%%%%%%%%%%%%%%%%%%%%%%%%
%% SECTION: APPENDIX                                  %%
%%%%%%%%%%%%%%%%%%%%%%%%%%%%%%%%%%%%%%%%%%%%%%%%%%%%%%%%
\appendix
\section{An inner radius cut to avoid source deblending mistakes}
\label{app:blending}

\begin{figure} 
	\includegraphics[width=0.49\textwidth]{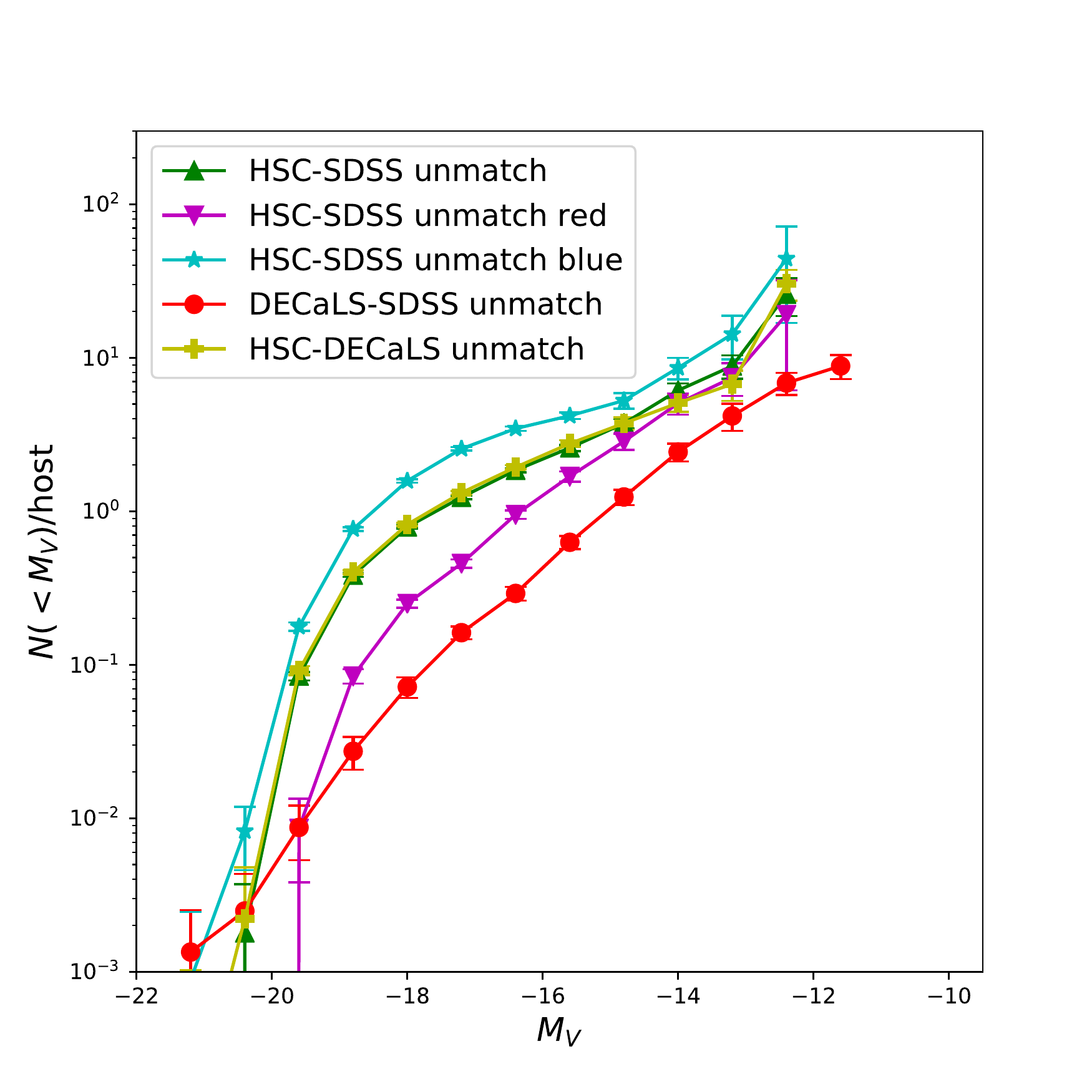}
	\caption{Green upper triangles connected by solid lines show the satellite LF based on extended sources detected 
	in HSC but not in SDSS, and projected within 260~kpc to ICG1s with $11.63<\log_{10}M_\ast/\msun<11.93$. No inner 
	radius cuts have been made. Magenta lower triangles/line and cyan stars/line show similar results, but are around 
	red and blue ICG1s in the same stellar mass range separately. We also show satellite LF based on extended sources 
	detected in DECaLS but not in SDSS as red dots/line, and the satellite LF using extended sources in HSC but not 
	in DECaLS as yellow plus symbols/line. These sources which are detected in one survey but not in the other show 
	very strong positive signals. Errorbars are based on the 1-$\sigma$ scatters of 100 boot-strap subsamples. 
	}
	\label{fig:blending}
\end{figure}

\begin{figure} 
	\includegraphics[width=0.49\textwidth]{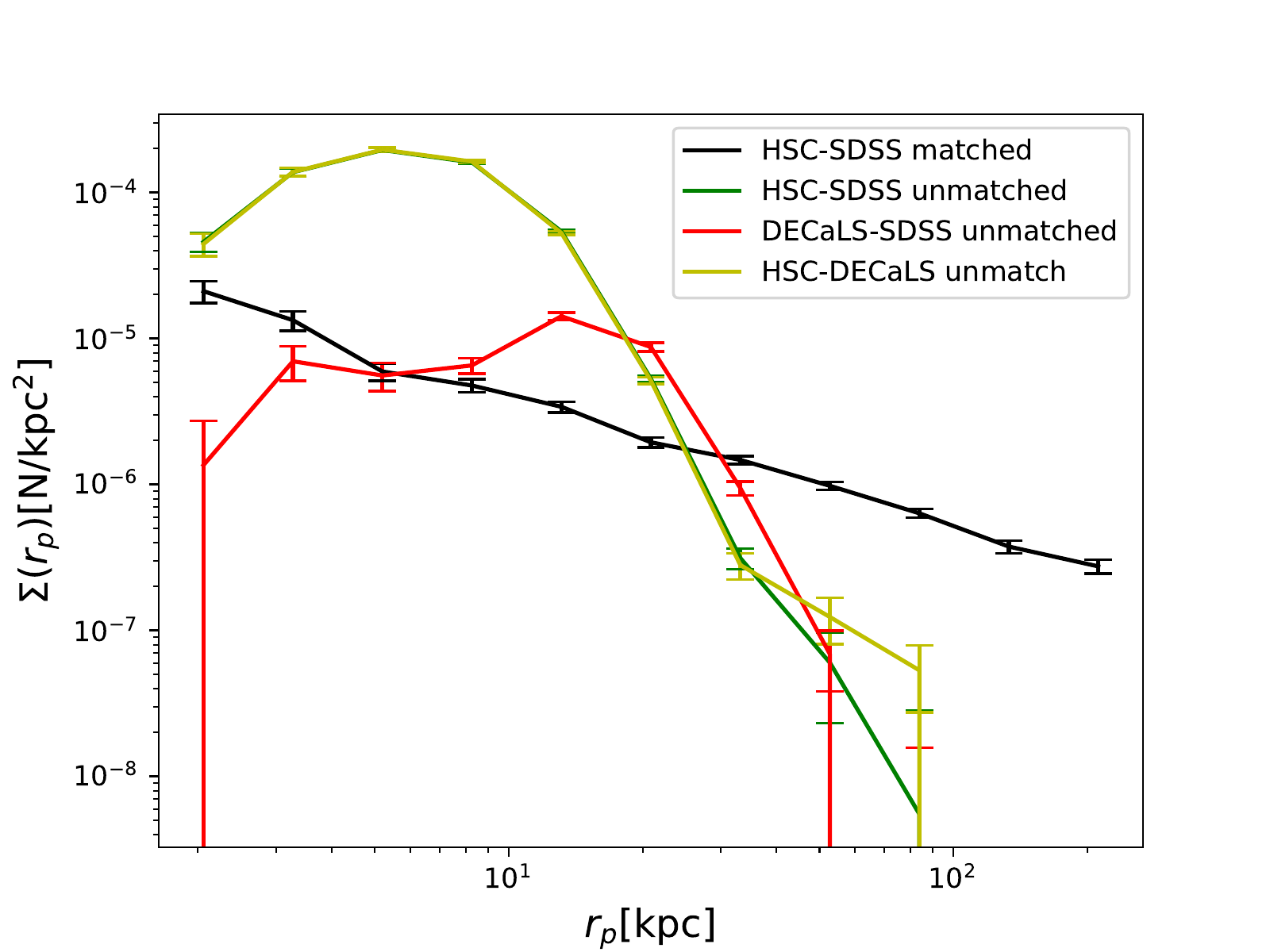}
	\caption{Green, red and yellow curves are the projected radial density profiles of extended sources detected 
	in HSC but not in SDSS, in DECaLS but not in SDSS, and in HSC but not in DECaLS, respectively. The profiles 
	are centred on ICG1s with $11.63<\log_{10}M_\ast/\msun<11.93$, and thus can be used to compare with symbols 
	and lines with the same colour in Figure~\ref{fig:blending}. The black curve shows the profile for extended 
	sources which are detected in both HSC and SDSS. Extended sources in both DECaLS and SDSS (or in both HSC 
	and DECaLS) have very similar projected radial density profiles as the black curve, and thus are not shown. 
	The sources existing in one survey but not in the other tend to peak in the inner 20--30~kpc region. Errorbars 
	are the 1-$\sigma$ scatters of 100 boot-strap subsamples. 
	}	
	\label{fig:blendingprof}
\end{figure}

\begin{figure*} 
\includegraphics[width=0.9\textwidth]{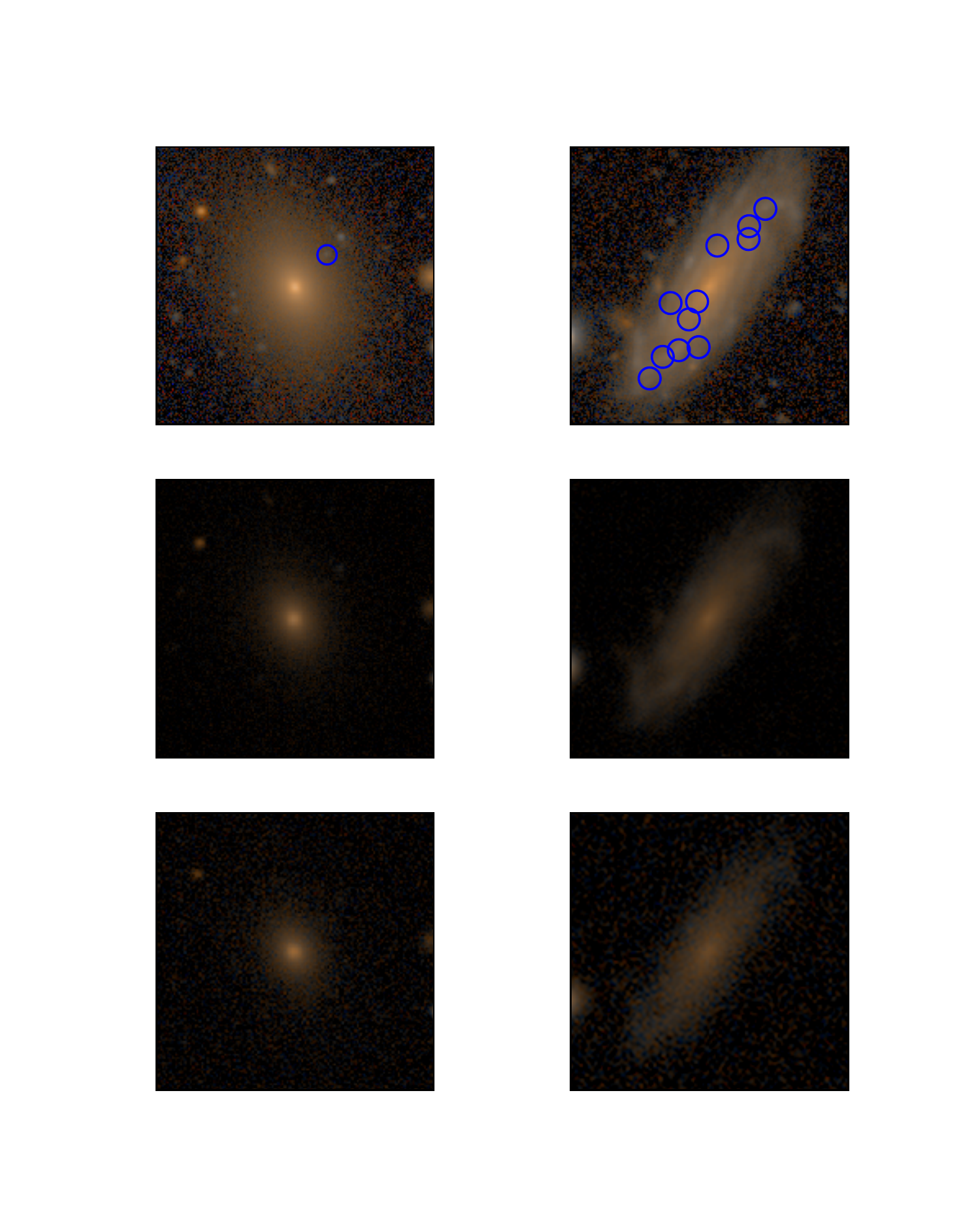}
\caption{Image cutouts in RGB from HSC (top), DECaLS (middle) and SDSS (bottom), and centred on two example galaxies 
(two columns). In the top panels, blue circles mark extended sources which are not detected in DECaLS and SDSS. The 
half edge lengths of these image cutouts are 40~kpc. The colour mapping is exactly the same for all six panels. 
Because DECaLS does not have $i$-band observations, we choose $r$, ($g+r$)/2 and $g$ as inputs for the RGB mapping. }
\label{fig:gal_color}
\end{figure*}

For results in the main text of this paper, we have applied an inner radius cut of $r_p>30$~kpc. In this Appendix we 
demonstrate the importance and the reason of choosing this inner radius cut by cross matching extended sources among 
HSC, DECaLS and SDSS. The matching also helps us to understand the difference among the three surveys. 

We match extended sources with $r<21$ in one of the deeper surveys (HSC, DECaLS and HSC) correspondingly to all 
extended sources in another shallower surveys (SDSS, SDSS and DECaLS). The matching is based on a searching radius 
of 1$\arcsec$. Figure~\ref{fig:blending} shows the satellite LFs based on sources in a deeper survey, which do not 
have a match within 1$\arcsec$ in a corresponding shallower survey. For the plot, all companions within 260~kpc 
to the central primaries are included without any inner radius cut. The signals are clearly positive. 

In order to investigate whether these sources detected in a deeper survey but not in another shallower survey 
are real, Figure~\ref{fig:blendingprof} shows their projected radial density profiles after fore/background 
counts estimated from random points are subtracted. Compared with the profile of sources which exist in both 
HSC and SDSS\footnote{Matched sources in both HSC and DECaLS or in both DECaLS and SDSS show very similar 
projected radial density profiles, and are hence not shown.}, unmatched sources tend to peak in very inner 
regions. The signals are still positive beyond 30~kpc, but drop very quickly with the increase in $r_p$.

We further show the colour images containing such unmatched sources on small scales and centred on two example  
primary galaxies in Figure~\ref{fig:gal_color}. The RGB figures are based on coadd images in $r$, ($g+r$)/2 and 
$g$-bands, and mapped to the RGB colour following the colour mapping of \cite{2004PASP..116..133L}. It is very 
clear that with the greater depth and higher image resolution, HSC reveals more extended low surface brightness 
emissions around the central galaxy and resolves more substructures, such as the spiral arms and star forming 
regions along the arms. In the top right panel, we mark the extended sources which are failed to be detected 
in both HSC and DECaLS. These sources clearly trace the spiral arms, indicating they are not real satellite 
galaxies, but instead, they are part of the central primary galaxy, which are mistakenly deblended as companion 
sources. Such deblending mistakes are more frequently happening in HSC than those in SDSS and DECaLS. However, 
there are also examples showing that HSC is able to detect some more real sources. For example, in the top left 
panel, the source which exists in HSC but not in the other two surveys does look like a real one.

In fact, we have looked at many image cutouts centred on our primary galaxies, and with different image sizes. 
Most of the sources close to the central galaxy, in a deeper survey but not in another shallower survey are 
faked detections due to deblending mistakes, and this often happens around bluer spiral galaxies, though 
there are also examples of real detections such as the one in the top left panel of Figure~\ref{fig:gal_color}. 

Magenta lower triangles/line and cyan stars/line in Figure~\ref{fig:blending} can prove and summarise what we 
have seen. The cyan stars connected by lines show satellite LFs based on extended sources in HSC but not in 
SDSS and around blue primaries. On the other hand, the magenta triangles connected by lines show similar results 
around red primaries. The amplitude of the cyan curve is much higher than the magenta one, indicating there are 
more such sources around blue galaxies. Blue star forming galaxies have more rich substructures, which are more 
likely to be mistakenly deblended as companions sources. 

Because such faked sources mostly exist within $r_p\sim$30~kpc according to Figure~\ref{fig:blendingprof}, we choose
an inner radius cut of 30~kpc for all the results in the main text of this paper. Such an inner radius cut, of course, 
would also exclude real sources on small scales, but by applying this cut to all of the three surveys and also to MW 
satellites after accounting for the projection effect (see Section~\ref{sec:method} for details), we can ensure that 
our comparisons among different surveys and the comparison with MW satellites are fair. Beyond 30~kpc, such sources 
still exist, but as the readers can see form Figure~\ref{fig:blendingprof}, even if all these sources are faked, they 
are subdominant and are unlikely to significantly affect our results. In fact, we have looked into such sources in HSC 
on larger distances to the central primaries, and have found examples that look like real galaxies. This can at least 
partly explain why satellite LFs based on HSC tend to have slightly higher amplitudes than those based on SDSS or DECaLS 
in the main text. 

Besides, the central primary may have another apparent massive companion galaxy close to it in projection, but it is in fact 
not physically associated because they have very different spectroscopic redshifts. There can be some faked sources around 
this massive companion galaxy, which are detected due to deblending mistakes, contributing to the number of unmatched sources.
We expect the probabilities for such cases to happen are equivalent around real central primaries and around random points. 
As a result, they should have been subtracted by using satellite counts around random points and thus are not expected to 
affect our results. However, if a true late-type satellite galaxy, that has a projected separation larger than 30~kpc 
to the central primary and is mistakenly deblended into multiple sources, the amplitude of our measured satellite LFs can 
be increased in a wrong way. We cannot rule out such possibilities and it is not trivial to estimate the exact fractions, 
but we believe our conclusions are unlikely to have been significantly affected, based on the fact that HSC is only slightly 
higher in amplitude than both DECaLS and SDSS.

\begin{figure} 
	\includegraphics[width=0.49\textwidth]{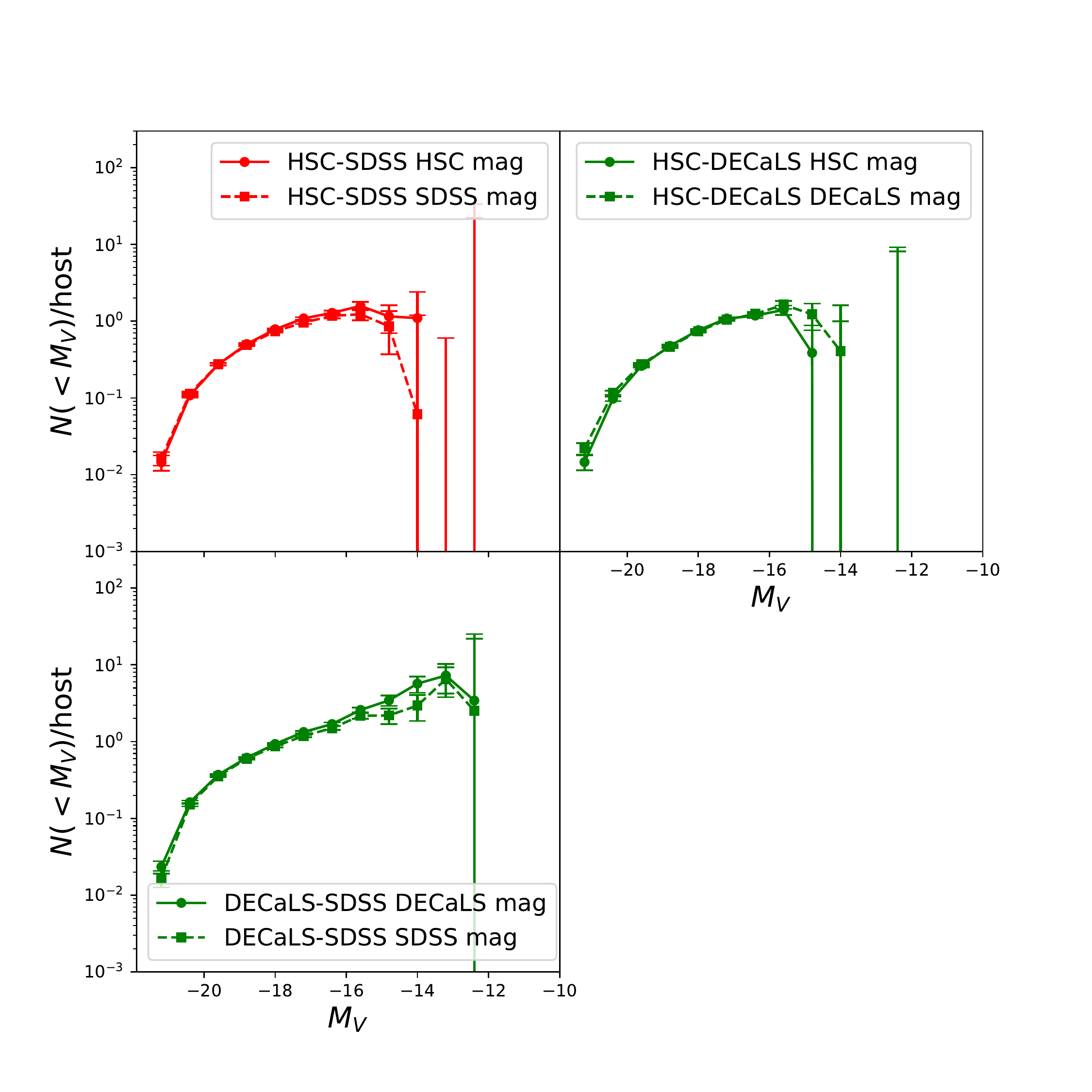}
	\caption{Satellite LFs based on matched extended sources between HSC and SDSS, between HSC and DECaLS, and 
	between DECaLS and SDSS. For each set, two different types of magnitudes taken from either of the two surveys
	are used. 
	}
	\label{fig:testmag}
\end{figure}

In addition to the unmatched sources, we also use the matched sources between different surveys to investigate 
whether the different magnitudes and filter systems defined in HSC, DECaLS and SDSS can lead to significant discrepancies. 
This is shown in Figure~\ref{fig:testmag}. The three panels correspond to matched extended sources in HSC-SDSS, HSC-DECaLS 
and DECaLS-SDSS, respectively. In each panel, the solid and dashed curves are results based on magnitudes from either 
of the two surveys. We can see some small differences at faint magnitudes, but the differences are far from being large 
enough to affect the conclusions of this paper.

\end{document}